\documentclass[a4paper,11pt]{article}
\usepackage{jheppub} 

\usepackage[english]{babel}

\usepackage{bbold}

\usepackage[utf8]{inputenc}
\usepackage{graphicx}
\usepackage{amssymb}
\usepackage{amsmath}
\usepackage{xcolor}
\usepackage{hyperref}
\usepackage{multirow}
\usepackage{slashed}
\usepackage{subcaption}
\usepackage{bm}
\usepackage{soul}

\renewcommand{\vec}[1]{\boldsymbol{#1}}
\newcommand{\be}{\begin{equation}}
\newcommand{\ee}{\end{equation}}
\newcommand{\ba}{\begin{eqnarray}}
\newcommand{\ea}{\end{eqnarray}}
\newcommand{\la}{\label}

\newcommand{\nn}{\nonumber \\}

\newcommand{\fQ}{f_{\mathcal Q}}

\newcommand{\amu}{a_\mu}

\newcommand{\ahvp}{a_\mu^{\mathrm{HVP}} }
\newcommand{\ahvpsib}{a_{\mu,SIB}^{\text{HVP}}}

\newcommand{\gpv}[1]{\left[\mathcal{G}(#1)\right]_\Lambda}

\newcommand{\atotvio}{a_\mu^{\text{HVP}, 38}}
\newcommand{\atotvioem}{a_{\mu,em}^{\text{HVP}, 38}}

\newcommand{\atotvioemlow}{a_{\mu,em}^{\text{HVP}, 38, \text{low}}}
\newcommand{\atotvioemhigh}{a_{\mu,em}^{\text{HVP}, 38, \text{high}}}
\newcommand{\atotviosib}{a_{\mu,SIB}^{\text{HVP}, 38}}
\newcommand{\alovio}{a_\mu^{\text{HVP},  38}}
\newcommand{\atotviotpta}{a_{\mu,(2+2)a}^{\text{HVP}, 38}}
\newcommand{\atotvioconn}{a_{\mu,(4)}^{\text{HVP}, 38}}

\newcommand{\alovioswap}{a_\mu^{\text{HVP},  83}}

\newcommand{\astrange}{a_\mu^{{\rm HVP},s}}

\newcommand{\atotisoS}{a_\mu^{\rm HVP,88}}

\newcommand{\dmK}{\Delta M_K}
\newcommand{\dmKem}{\Delta M_K^{em}}
\newcommand{\dmKel}{\Delta M_K^{elast}}
\newcommand{\dmKinel}{\Delta M_K^{inel}}
\newcommand{\dmKphys}{\Delta M_K^{phys}}

\DeclareMathOperator{\im}{\mathrm{Im}}

\begin{document}
\preprint{MITP-25-038}

\author[a]{Dominik~Erb,}
\affiliation[a]{PRISMA$^+$ Cluster of Excellence \& Institut f\"ur Kernphysik,
Johannes Gutenberg-Universit\"at Mainz,
D-55099 Mainz, Germany}
\emailAdd{domerb@uni-mainz.de}

\author[b]{Antoine~G\'erardin,}
\affiliation[b]{Aix-Marseille Universit\'e  , Universit\'e de Toulon, CNRS, CPT, Marseille, France}

\author[a,c]{Harvey~B.~Meyer,}
\affiliation[c]{Helmholtz~Institut~Mainz,
Staudingerweg 18, D-55128 Mainz, Germany}

\author[d]{Julian~Parrino,}
\affiliation[d]{Universit\"at Regensburg, Fakult\"at f\"ur Physik, Universit\"atsstraße 31, 93040 Regensburg, Germany}

\author[e]{Volodymyr~Biloshytskyi,}
\affiliation[e]{Institut f\"ur Kernphysik,
Johannes Gutenberg-Universit\"at Mainz,
D-55099 Mainz, Germany}

\author[e]{and Vladimir~Pascalutsa}

\title{
Isospin-violating vacuum polarization in the muon~$(g-2)$ 
 with SU(3) flavour symmetry from lattice QCD
}

\abstract{We compute the isospin-violating part $\atotvio$ of the hadronic-vacuum-polari\-zation (HVP) contribution to the muon $(g-2)$ in lattice QCD at the SU$(3)_{\rm f}$-symmetric point where $M_\pi=M_K\simeq416$ MeV.
All diagrams involving internal photons are evaluated in coordinate space, employing a Pauli-Villars-regulated photon propagator with a cutoff scale $\Lambda$ well below the lattice cutoff. The counterterm $(m_u-m_d)$, whose $\Lambda$ dependence is consistent with the expected logarithmic behaviour, 
is calibrated using the experimental kaon mass splitting as input.
The bare electromagnetic contribution at fixed $\Lambda$ is compared to  a phenomenological estimate based on the kaon-loop and pseudoscalar-pole contributions to the forward light-by-light amplitude.
An extension of these calculations to
physical pion and kaon masses appears promising.
}

\date{30.05.2025}

\maketitle

 \section{Introduction \label{sec::Intro}}

All charged particles with spin carry an intrinsic magnetic moment, proportional to the spin, with the `$g$-factor' as the proportionality constant. The leading prediction for the $g$-factor from a quantum relativistic description of an elementary charged particle with any spin is $g=2$. However, further quantum effects modify this value, giving rise to the `anomalous magnetic moment', 
 which for leptons ($\ell = e, \mu, \tau$) is denoted as $a_\ell=(g-2)/2$. 
 Precise experimental determinations of these quantities have, for a long time, been serving as one of the most stringent tests of the Standard Model (SM) of particle physics. 

The  level of precision achieved by the direct measurements of $a_\mu$ is presently reaching an absolute uncertainty of $22\times 10^{-11}$ or 0.19\;ppm \cite{Aguillard_2023,Muong-2:2021ojo,Muong-2:2006rrc}. 
Achieving this precision on the theory side involves very sophisticated
calculations in the SM. Their compilation by a broad consortium of experts has resulted in the `Muon $g-2$ Theory Initiative' White Papers (WPs) \cite{Aoyama:2020ynm,Aliberti:2025beg}. 
According to the 2025 WP, the SM prediction for the anomalous magnetic moment of the muon currently has a precision of $62\times 10^{-11}$, and is in good agreement with the experimental world average.

\begin{figure}[t]
\centering
\includegraphics[width=0.75\linewidth]{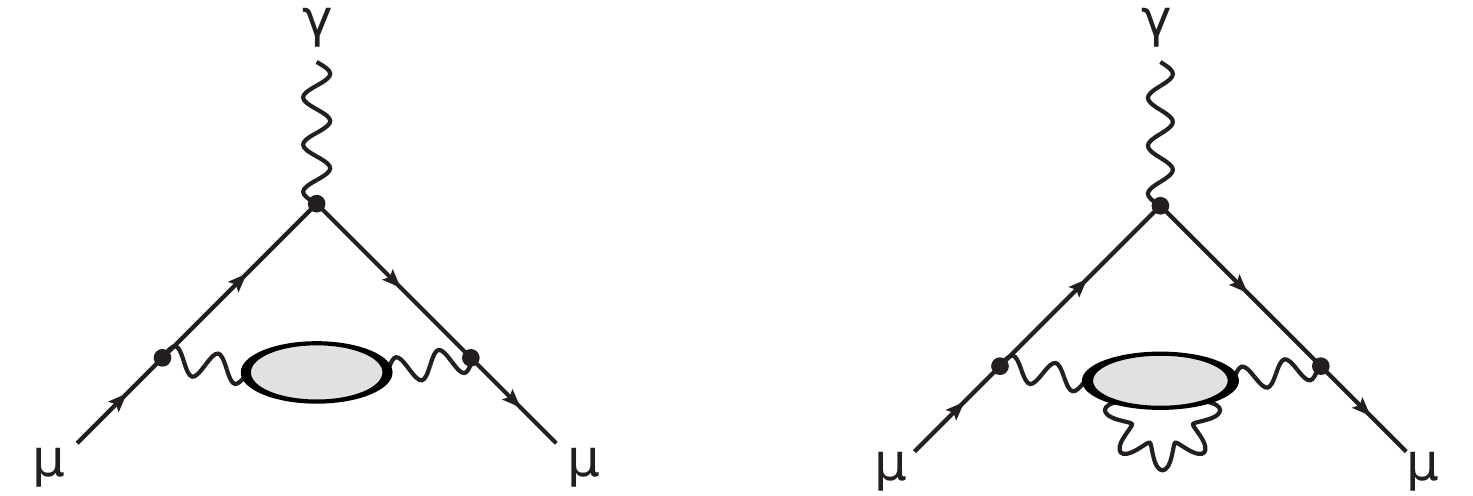}
  \caption{
Left: hadronic vacuum polarization contribution to the muon $(g-2)$. Right: electromagnetic correction to the latter. The blobs represents the non-perturbative contribution of QCD degrees of freedom.}
  \label{fig:HVPdiagrams}
\end{figure}

The strong-interaction (hadronic) contributions dominate the theory uncertainty. The leading hadronic contribution, the hadronic vacuum polarisation $\ahvp$ illustrated in Fig.~\ref{fig:HVPdiagrams}, carries by itself an absolute uncertainty of $61\times 10^{-11}$. Unlike the first edition of the WP \cite{Aoyama:2020ynm}, the new edition \cite{Aliberti:2025beg} uses an average of lattice QCD evaluations of the HVP contribution, rather than a dispersive evaluation. The latter are currently being omitted due to strong tensions in the experimental $e^+e^-\to\pi^+\pi^-$ cross-sections at center-of-mass energies below 1\,GeV. Clearly, in order to fully profit from the current precision of the direct measurement of $a_\mu$, and its imminently expected update by the Fermilab collaboration~\cite{Aguillard_2023}, a further factor-of-four reduction in the uncertainty of the HVP contribution is called for, corresponding to a relative error on $\ahvp$ at the two-permille level.

It is standard for lattice QCD simulations to be performed with equal up and down quark masses ($m_u=m_d$) and in the absence of photons. However, given the ambitious level of the targeted precision for $\ahvp$, it is imperative to calculate the corrections to the HVP contribution due to the isospin-breaking (IB) effects. The state-of-the-art is to expand the latter in $(m_u-m_d)$ and $\alpha$ to first order. The electromagnetic correction to $\ahvp$ is illustrated in Fig.\ \ref{fig:HVPdiagrams}. We note that, while the perturbative treatment of isospin-breaking effects has a long tradition \cite{Cottingham:1963zz,Gasser:1974wd,Cirigliano:2002pv}, it was first introduced in lattice QCD in 2013 \cite{deDivitiis:2013xla}.
Several lattice calculations of these corrections due to IB effects have already been performed \cite{Blum:2018mom,Giusti:2019xct,Borsanyi:2020mff,Ray:2022ycg,Djukanovic:2024cmq,Parrino:2025afq,Altherr:2025kqw}, including one \cite{Borsanyi:2020mff} that encompasses all required quark-level diagrams, but the corrections are not yet sufficiently well understood to reduce the theoretical uncertainty on $\ahvp$ to the desired level.

The leading correction in $\alpha$ to HVP has a connection to the hadronic light-by-light contribution (HLbL), which is the second-largest hadronic effect in the muon $(g-2)$, recently addressed in the lattice calculations~\cite{Blum:2023vlm, Chao:2021tvp,Fodor:2024jyn,Kalntis:2024dyd}. Indeed the O($\alpha$) correction to $\ahvp$ is expressible in terms of the forward HLbL amplitude \cite{Knecht:2003kc,Biloshytskyi:2022ets}.
However, while the HLbL contribution is UV-finite, the internal photon leads to divergences in the HVP calculation. To address this issue, we will work in the framework proposed in Ref.\ \cite{Biloshytskyi:2022ets}, where the photon is implemented with a Pauli-Villars regulated propagator, in the continuum and infinite volume. Doing so, one avoids power-law finite-volume effects that arise in the widely used QED$_L$ scheme (see for instance~\cite{Hayakawa:2008an,Bijnens:2019ejw,Borsanyi:2020mff}).
We will furthermore use the covariant coordinate-space (CCS) representation \cite{Meyer:2017hjv} for obtaining the HVP contribution on the lattice.

Specifically, we calculate the contribution to $\ahvp$ of the isospin-violating part $\langle j^{3}_\mu (x) j^{8}_\nu (0)\rangle$ of the electromagnetic-current correlator at the SU$(3)$-flavour symmetric point for three mass-degenerate quarks: up, down and strange. 
Clearly, this contribution vanishes in isospin-symmetric QCD\footnote{The currents $j_\mu^3$ and $j_\mu^8$ correspond to the  isovector, respectively isoscalar part of the electromagnetic current.}.
The task involves the calculation of the fully connected diagrams at the quark level, for which we also evaluate the (isovector) mass counterterm. This counterterm is fixed by the mass splitting of the kaon, and the combined result is UV finite. Additionally, one needs to include one quark-disconnected diagram, which is already UV-finite.
These calculations are performed on five different  gauge ensembles generated with $N_f=3$ dynamical quark flavours by the Coordinated Lattice Simulations (CLS) consortium~\cite{Bruno:2014jqa} at a pion and kaon mass around $416$\;MeV.

We start by defining the basic formalism as well as the different correlation functions to be computed in section \ref{sec::Master}, following which we present the lattice setup in section \ref{sec::LatSet}. The calculation of all relevant contributions is then split in three different sections. At first, we describe how we calculate the fully connected contributions (sec.\ \ref{sec::Conn}). Here, we also perform a crosscheck, by switching off the gluon interactions and comparing our result to a corresponding continuum calculation for the vacuum polarization contribution in pure QED.
In section \ref{sec:disco}, we compute the quark disconnected contribution, where we also compare our result to the prediction from the pseudoscalar meson exchange.
Then in section \ref{sec:counterterm} we determine the counterterm $(m_u-m_d)$ by  computing the kaon electromagnetic mass splitting on the lattice, using again the Pauli-Villars regulated photon propagator.
At the end of that section, we gather the three continuum-extrapolated quantities to build the renormalized  isospin-violating part of the HVP contribution in the muon $(g-2)$, still at a fixed value of the photon Pauli-Villars cutoff parameters $\Lambda$.
In a final step, we compare the lattice result extrapolated to $\Lambda=\infty$  with a phenomenological estimate, and conclude in section~\ref{sec::Concl}.
A number of technical aspects of the lattice calculation and the phenomenological models are provided in appendix.

 \section{Formalism} \label{sec::Master}

To calculate the hadronic vacuum polarization (HVP) contribution to the anomalous magnetic moment of the muon $a_\mu$, we employ the covariant coordinate-space (CCS) method, introduced in Ref.~\cite{Meyer:2017hjv}. This method is a proven alternative \cite{Chao:2022ycy} to the time-momentum representation (TMR) \cite{Bernecker:2011gh}, which is the most widely used to calculate this quantity on the lattice. In the CCS method, the HVP contribution is obtained from an integral over the space-time volume ($\int_z \equiv \int d^4z$) 
\begin{align}
\label{eq:ccs_lo}
    a_\mu^{\text{HVP}}=\int_z \ H_{\lambda \sigma}(z) G_{\lambda \sigma}(z), \qquad H_{\lambda \sigma}(z)&=-\delta_{\lambda \sigma} \mathcal{H}_1(|z|) + \frac{z_\lambda z_\sigma}{|z|^2} \mathcal{H}_2(|z|),
\end{align}
where $H_{\lambda \sigma}(z)$ is the CCS kernel
and $G_{\lambda \sigma}(z)=\langle j^{em}_\lambda(z) j^{em}_\sigma (0)\rangle_{\textrm{QCD}+\textrm{QED}}$ is the vector-vector correlator of the electromagnetic vector current
\ba 
j^{em}_\lambda (z) &=\frac{2}{3}\Bar{u}(z)\gamma_\lambda  u(z)  -\frac{1}{3}\Bar{d}(z)\gamma_\lambda  d(z) -\frac{1}{3} \Bar{s}(z)\gamma_\lambda  s(z).
\ea
In this work, we only consider the contributions of the up, down and strange quarks at the SU$(3)$ flavour symmetric point,  $m_u=m_d=m_s$.

Making use of partial integration and exploiting the fact that the electromagnetic current is conserved $(\partial_\lambda j_\lambda(z)=0)$, it is possible to add a total-derivative term  $\partial_\lambda [ z_\sigma g(|z|)] $ to the CCS kernel, without changing the result for $a_\mu^{\text{HVP}}$ in the continuum and infinite volume. 
Besides the transverse (`TV') kernel given in Eq.~\eqref{eq:ccs_lo}, we also define the traceless (`TL') kernel and a variant which is proportional to $z_\lambda z_\sigma$ (`XX') \cite{Ce:2018ziv, Chao:2022ycy}:
\begin{align}
    \label{eq:tl_kernel}
    H_{\lambda \sigma}^{TL}(z)&=\left(-\delta_{\lambda \sigma}+ 4\frac{z_\lambda z_\sigma}{|z|^2} \right) \mathcal{H}_2(|z|)\\
    \label{eq:xx_kernel}
    H_{\lambda \sigma}^{XX}(z)&=\frac{z_\lambda z_\sigma}{|z|^2} \left( \mathcal{H}_2(|z|) + |z| \frac{d}{d|z|} \mathcal{H}_1(|z|) \right)
\end{align}
The exact form of the two scalar weight functions $\mathcal{H}_1$ and $\mathcal{H}_2$ can be found in \cite{Meyer:2017hjv}. One may also use the rational approximations given in \cite{Biloshytskyi:2022ets}. In case of the disconnected diagram the `TL' version of the kernel is preferred over the `XX' one, due to its reduced finite size effects \cite{Parrino:2025afq}. For that reason all diagrams and values of the connected and disconnected contribution shown in this paper were calculated using the `TL' kernel if not stated otherwise. But nevertheless in the connected case both versions of the kernel give the same results within error. 

In order to generalize the expressions above in terms of the flavour structure, we introduce
the isovector~(3) and isoscalar~(8) parts of the electromagnetic~($em$)  charge matrix,
\ba 
\label{eq:charge_matrices}
\mathcal{Q}^{(3)} = \begin{pmatrix}
    1/2 & 0&0 \\
    0& -1/2 &0 \\
    0 & 0 & 0
\end{pmatrix}, \ 
\mathcal{Q}^{(8)} = \begin{pmatrix}
    1/6 & 0&0 \\
    0& 1/6 &0 \\
    0 & 0 & -1/3
\end{pmatrix}, \ 
\mathcal{Q}^{(em)} = \begin{pmatrix}
    2/3 & 0&0 \\
    0& -1/3 &0 \\
    0 & 0 & -1/3
\end{pmatrix} .\quad
\ea 
With the quark-flavour triplet $\vec{\psi}(x) = (u(x) , d(x) , s(x) )^T$,
we define the corresponding currents
\ba 
j_\lambda^3 (x) = \vec{\Bar{\psi}}^T(x) \gamma_\lambda \mathcal{Q}^{(3)} \vec{\psi}(x), \qquad j_\lambda^8 (x) = \vec{\Bar{\psi}}^T(x) \gamma_\lambda \mathcal{Q}^{(8)} \vec{\psi}(x), 
\ea 
such that $j^{em}_\lambda (x) = j^{3}_\lambda (x) + j^{8}_\lambda (x)$.
The correlator $G^{ab}_{\lambda \sigma}(z)=\langle j^{a}_\lambda(z) j^{b}_\sigma (0)\rangle_{\textrm{QCD}+\textrm{QED}}$ then leads to the sub-contributions
\begin{align}
    a_\mu^{\text{HVP},ab}=\int_z \ H_{\lambda \sigma}(z) \;G^{ab}_{\lambda \sigma}(z)
\end{align}
in the anomalous magnetic moment of the muon.

\begin{figure}[t]
	\centering
	\includegraphics[width=0.49\textwidth]{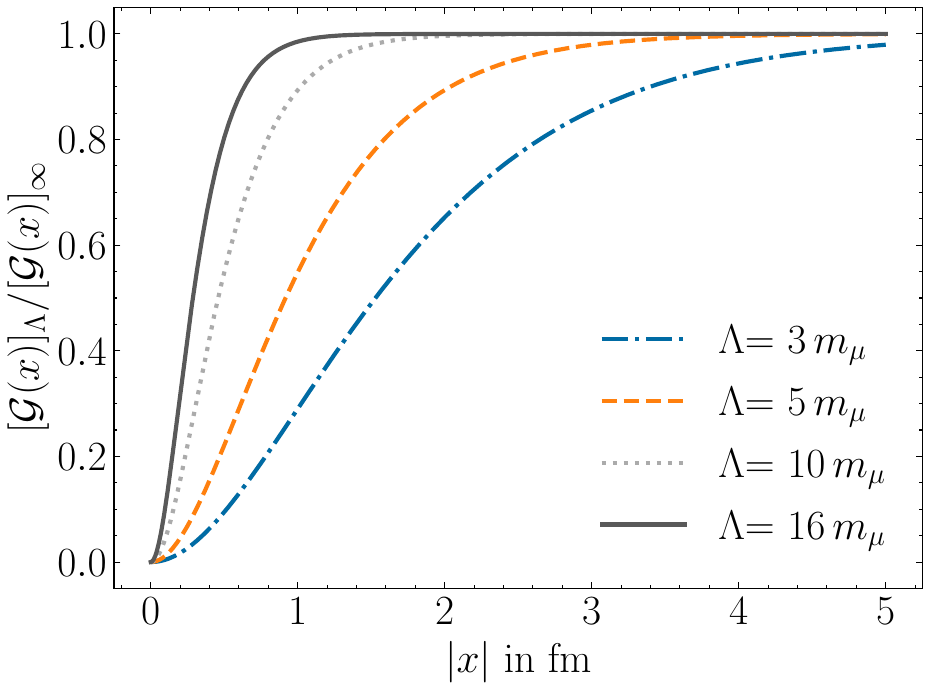}
	\caption{Ratio of the Pauli-Villars-regulated and non-regulated photon propagator for the values of $\Lambda$ used in later sections.}
    \label{fig::Master_PV_Comparison}
\end{figure}

Expanding the correlator to first order in $\alpha=\frac{e^2}{4\pi}$ around the isospin-symmetric QCD, one has
\ba 
\label{eq:vec_cor_expansion}
G^{ab}_{\lambda \sigma}(z)&=&\Big\langle j^{a}_\lambda(z)j^{b}_\sigma(0) \Big\rangle_\mathrm{QCD}
\\
&& -\frac{e^2}{2}\lim_{\Lambda\to\infty}\left\{\int_{x,y}\delta_{\nu \rho}\gpv{x-y} \langle j^{a}_\lambda(z) j^{em}_\nu(y) j^{em}_\rho(x) j^{b}_\sigma(0) \rangle_\mathrm{QCD}
    \;+C_T(\Lambda)\right\}
    \nonumber\\ && +\; O(\alpha^2)     .
    \nonumber
\ea 
The second term is the leading isospin-breaking correction, where we have introduced the Pauli-Villars (PV) regulated photon propagator in Feynman gauge $\delta_{\nu \rho}\gpv{x}$, with  \cite{Biloshytskyi:2022ets}
\ba
\label{eq:pv_photon_prop}
\gpv{x}  = \frac{1}{4\pi^2 |x|^2} - 2G_{\frac{\Lambda}{\sqrt{2}}}(x) +G_\Lambda(x).
\ea 
Here, $G_m(x) = {m K_1\left(m |x|\right)}/(4\pi^2 |x|)$ stands for the massive scalar propagator, with $K_1(x)$ being the modified Bessel function of the second kind. 
Fig.~\ref{fig::Master_PV_Comparison} compares the photon propagator with its Pauli-Villars-regulated version for various values of the cutoff $\Lambda$, often referred to as PV mass. The regulator ensures that the propagator is UV-finite. While UV-finiteness can be achieved already with a single subtraction term, we find that the regularization scheme in Eq.~\eqref{eq:pv_photon_prop} reduces cutoff effects more effectively and provides stronger suppression of short-distance contributions.

To obtain a physical result in the limit $\Lambda \rightarrow \infty$, it is necessary to specify an additional renormalization condition complementary to the scale setting procedure in isospin symmetric QCD  \cite{Bruno:2016plf,Strassberger:2021tsu,RQCD:2022xux}.
In this framework, one obtains a counterterm $C_T(\Lambda)$, which is calculated with the same PV regulator \eqref{eq:pv_photon_prop}.
In Eq.~\eqref{eq:vec_cor_expansion}, the counterterm cancels the UV-divergence that occurs when the vertices in the photon propagator $\gpv{x-y}$ approach each other, $x\rightarrow y$.

\begin{figure}[t]
	\centering
	\includegraphics[width=0.6\textwidth]{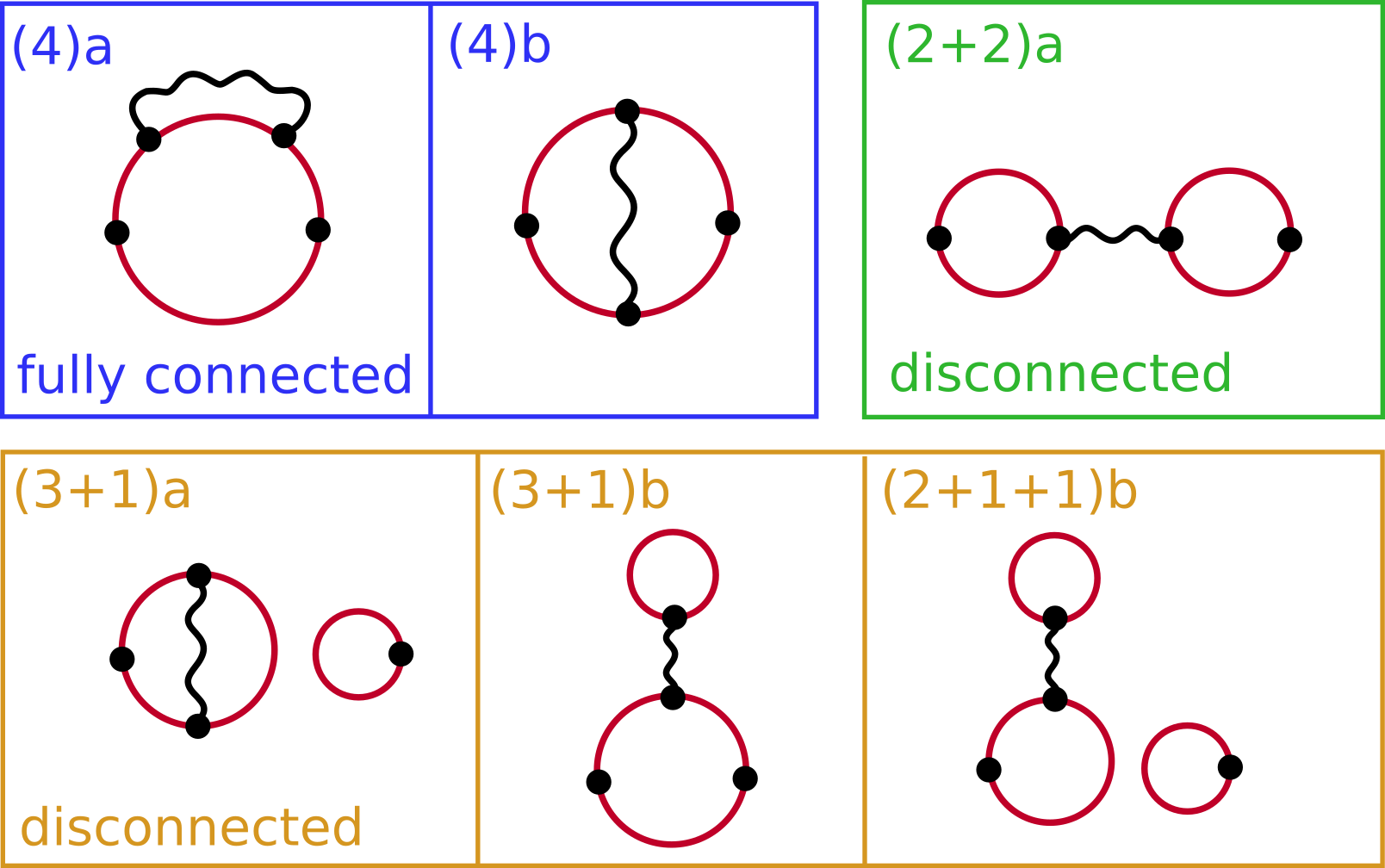}
	\caption{Feynman diagrams depicting all Wick contractions of the four-point function in Eq.~\eqref{equ::Master_4pt}. At the SU$(3)_{\rm f}$ symmetric point, the diagrams in the lower row vanish. }  
    \label{fig::Master_feynman}
\end{figure}

In the following, we focus on the $(3,8)$ component of the correlator,
$G^{3,8}_{\lambda \sigma}(z)$,
for which the leading-order contribution $ \alovio$ vanishes trivially in isospin-symmetric QCD. We therefore refer to it as the  \textit{isospin-violating correlator}. Inserting $G^{3,8}_{\lambda \sigma}(z)$
into Eq.~\eqref{eq:ccs_lo}, we obtain to order $O(\alpha^3)$
\begin{align} \label{equ::Master_amu38}
  \atotvio &= \lim_{\Lambda\to\infty} \atotvio(\Lambda),
  \qquad \atotvio(\Lambda) = \atotvioem(\Lambda) +C_T(\Lambda)\,,\phantom{_\Big|}
    \\
    \atotvioem(\Lambda)
    &=-\frac{e^2}{2}\int_{z,x,y}  H_{\lambda \sigma}(z) \delta_{\nu \rho}\, [\mathcal{G}(x-y)]_\Lambda \,\left\langle j^3_\lambda(z) j^{em}_\nu(y) j^{em}_\rho(x) j^8_\sigma (0) \right\rangle \,   .\label{equ::Master_4pt}
\end{align}
All possible Wick contractions contributing to the four-point function for the isospin-violating correlator in Eq.~\eqref{equ::Master_4pt} are shown in Fig.~\ref{fig::Master_feynman}.
In three-flavour theory, due to the property that the sum of the charges of the three light quarks is zero, the diagrams in the bottom row of Fig.~\ref{fig::Master_feynman} vanish at the SU$(3)$-flavour symmetric point. There, only the diagrams labeled with $(4)a$, $(4)b$ and $(2+2)a$ contribute. The former are discussed in section \ref{sec::Conn}, while the latter is discussed in section \ref{sec:disco}.

As a renormalization condition that fully specifies the counterterm, we require the kaon mass splitting
    $\dmK \equiv M_{K^+} - M_{K^0} $
 to assume its physical value  $\dmKphys = -3.934(20)$\;MeV \cite{PDG}. 
A short derivation, presented in Appendix \ref{sec:CTderivation}, leads then to the following expression for the counterterm,\footnote{By default, we use the non-covariant normalization of states, 
$\langle K_{\vec{p}}| K_{\vec{p}'} \rangle = (2\pi)^3 \delta^{(3)}(\vec{p}-\vec{p}')$. Covariantly normalized states will be denoted as $\bm{|K_{p} \rangle}$, such that $\bm{\langle K_{{p}}| K_{{p}'} \rangle }= (2\pi)^3 2E_{\vec{p}}\delta^{(3)}(\vec{p}-\vec{p}')$.} 
\begin{align}
  C_T(\Lambda) &= (\dmKphys-\dmKem(\Lambda))\; R_{38K}\;, \label{equ::Master_Counterterm} \phantom{\Big|}
    \\
        R_{38K} &=\frac{1}{ \langle K^+_{\vec 0}| \Bar{u}u-\Bar{d}d | K^+_{\vec 0} \rangle } \, \left. \frac{\partial \alovio}{\partial (m_u-m_d)}\right|_{m_u+m_d,m_s,g_0;\alpha=0}\\
        &=\frac{1}{2} \,
        \frac{\partial \ahvp}{\partial \dmK}\Big|_{\overline M_\pi,\overline M_K,\overline M_B}
        . \label{equ::HVP_MI_R38}
\end{align}
The quantity $\dmKem(\Lambda)$ is the bare electromagnetic mass splitting, computed with a photon propagator regularized at scale $\Lambda$.
The factor $R_{38K}$, as indicated by 
the second expression,\footnote{The factor $\frac{1}{2}$ here accounts for the fact that $\ahvp$ contains $\alovio$ and $\alovioswap$.}  expresses the response of the vacuum polarization contribution to the kaon-mass splitting at fixed isospin-averaged\footnote{The bar over the variables kept constant in Eq.\ (\ref{equ::HVP_MI_R38}) indicate that they are isospin-averaged.} pion, kaon masses and the scale-setting quantity (typically a baryon mass). It has no (leading) scheme dependence and
is a renormalization-group-invariant quantity that can be extrapolated to the continuum.

\subsection{Defining a strong isospin-breaking contribution \label{sec::SIB}}

In principle, a lattice computation of $\atotvio$ does not require a separation of strong and electromagnetic isospin-breaking effects. However,
it can be useful for interpretations in terms of phenomenological calculations
based on the Cottingham formula, where $\dmKem(\Lambda)$ is expressed via the kaon forward Compton amplitude. The Compton process is naturally split into the elastic and inelastic contributions and hence we replace  $\dmKem(\Lambda)$  in Eq.~\eqref{equ::Master_Counterterm} by $(\dmKinel(\Lambda) +\dmKel)$,
with the limit $\Lambda\to\infty$  taken `ahead of time' in $\dmKel$. Defining
\begin{align}
     \atotviosib &\equiv  (\dmKphys - \dmKel)\; R_{38K},
    \\
    C_T^{inel}(\Lambda) &\equiv -\dmKinel(\Lambda)\; R_{38K},
\end{align}
we may then write
\begin{align}
    \atotvio = \atotviosib+ \lim_{\Lambda\to\infty} \left\{ \atotvioem(\Lambda) + C_T^{inel}(\Lambda)\right\}\;,
\end{align}
such that the first term is interpreted as the strong isospin-breaking contribution and the second one as the electromagnetic contribution. While we do not make use of this separation in our lattice calculation, we use it to organize our phenomenological estimates.

One may further isolate an infrared-enhanced set of contributions to $\atotvioem(\Lambda)$ that are UV-finite, so that we finally have
\begin{align}\label{eq:model_schematic_final}
    \atotvio =  \atotviosib + \atotvioemlow + \lim_{\Lambda\to\infty} \left\{ \atotvioemhigh(\Lambda) + C_T^{inel}(\Lambda)\right\}\;.
\end{align}
The quantity $\atotvioemlow $ will be estimated by using a hadronic model involving the light pseudoscalar mesons of mass below 1 GeV. 
In doing so, the sum of the $\Lambda$-dependent terms in this last equation will be neglected, anticipating that, after the UV-divergence cancels, the remainder is small compared to the first two terms in Eq.\ (\ref{eq:model_schematic_final}), which are more long-distance dominated. As is well-known, and reviewed below, the (logarithmic) divergence stems from the self-energy of the light quarks, which in the continuum is proportional to the light-quark mass, and therefore has a very small prefactor.

 \section{Lattice setup} \label{sec::LatSet}
We perform the calculations on a subset of the ensembles generated by the CLS initiative \cite{Bruno:2014jqa, Bali:2016umi}, with $O(a)$ improved Wilson-Clover quarks and tree-level $O(a^2)$ improved Lüscher-Weisz gauge action. All our ensembles have degenerate up, down and strange quark masses, corresponding to a pion and kaon mass $m_\pi=m_K \sim 416$ MeV, more details are shown in Table~\ref{tab::LatSet_Properties}. 
To investigate the effect of the finite lattice volume, two of the ensembles have identical simulation parameters, but different volume. One of the ensembles, B450, has periodic boundary conditions in time while the other ones have open boundary conditions.

\begin{table}[b]
    \caption{Parameters of the employed CLS ensembles. The lattice spacing in physical units was extracted from \cite{Bruno:2016plf}. The pion mass values were taken from \cite{Ce:2022kxy}. The values for the VMD masses (see Eq.\ (\ref{equ::vmd_papram})) are from \cite{Chao:2020kwq}, and the $\hat{Z}_V$  values from Ref.\ \cite{Gerardin_2019}. All ensembles but B450 have open boundary conditions in time.}
    \label{tab::LatSet_Properties}
    \begin{tabular}{ c c c c c c c c c}
    \hline
        Id & $\beta$ & $(\frac{L}{a})^3 \times \frac{T}{a}$ & a [fm] & $m_\pi$ [MeV] & $M_{\text{VMD}}$ [MeV] & $m_\pi L$ & L[fm] & $\hat{Z}_V$\\ \hline
        H101 & 3.4 & $32^3 \times 96$ & 0.08636 & 416(4) & 921(13) & 5.8 & 2.8 & 0.71540 \\ \hline 
        B450 & 3.46 & $32^3 \times 64$ & 0.07634 & 415(4) & 942(25) & 5.1 & 2.4 & 0.72645 \\ \hline
        H200 & 3.55 & $32^3 \times 96$ & 0.06426 & 416(5) & 979(26) & 4.3 & 2.1 & 0.74030 \\
        N202 &  & $48^3 \times 128$ &  & 412(5) & 952(15) & 6.4 & 3.1 & \\ \hline
        N300 & 3.7 & $48^3 \times 128$ & 0.04981 & 419(4) & 1001(23) & 5.1 & 2.4 & 0.75912 \\ \hline
    \end{tabular}

\end{table}

\begin{table}[h!]
    \centering
    \caption{The number of used configurations for every ensemble and each part of the calculation. 
    }
    \label{tab::NumConfigs}
    \begin{tabular}{ l c c c c c }
    \hline
        Part & H101 & B450 & H200 & N202 & N300\\ \hline
        Connected & 200 & 200 & 200 & 120 & 120\\ \hline
        Disconnected & 182 & 200 & 200 & 200 & 200 \\ \hline
        Kaon mass splitting & 200 & 200 & 200 & 200 & 200\\ \hline
        Kaon mass derivative & 200 & 200 & 200 & 200 & 200\\ \hline
        HVP derivative & 2000 & 1600 & 2000 & 899 & 1540\\ \hline
    \end{tabular}
\end{table}

In order to improve our continuum extrapolations, we use two different discretizations for the vector current, the local $(l)$ and the point-split or conserved $(c)$ discretizations 
\begin{align}
    j^l_\nu(z)&=\Bar{q}_z\gamma_\nu q_z \\ 
    j^c_\nu(z)&= \frac{1}{2} \left[\bar{q}_{z+\hat{\nu}} (\gamma_\nu +1) U^\dagger_{\nu, z} \ q_{z} + \bar{q}_{z} (\gamma_\nu-1) U_{\nu, z} \ q_{z+\hat{\nu}}\right].    
\end{align}
$U_{\nu,z}$ is the gauge link in the direction $\nu$ associated with site $z$. No further improvements of the vector currents are implemented in this paper. \\
While the point-split vector currents do not need to be renormalized, a renormalization factor $\hat{Z}_V$ is required for the local currents. 
The renormalized local vector current is obtained from $j^{l,\text{ren}}_\nu(z)= \hat{Z}_V j^l_\nu(z)$. 
We absorb the combined renormalization factors for each local vector current in a global multiplicative constant. This constant always corresponds to $\hat{Z}_V$ raised to the power of the number of local vector currents $n_l$: $\mathcal{Z}= \hat{Z}_V ^{n_l} $. 
We extracted $\hat{Z}_V$ for each value of $\beta$  from Ref.\ \cite{Gerardin_2019}; the values are collected in Table \ref{tab::LatSet_Properties}. Apart from vector currents, pseudoscalar densities also appear in the course of our calculations, but their renormalization factors cancel and are therefore not needed.
Finally, we provide the level of statistics collected on each gauge ensemble for the different key observables in Table  \ref{tab::NumConfigs}.

\section{Connected electromagnetic contribution $\atotvioconn(\Lambda)$\label{sec::Conn}}
	\begin{figure}[ht]
		\centering
		\begin{subfigure}[t]{0.4\textwidth}
			\includegraphics[width=\textwidth]{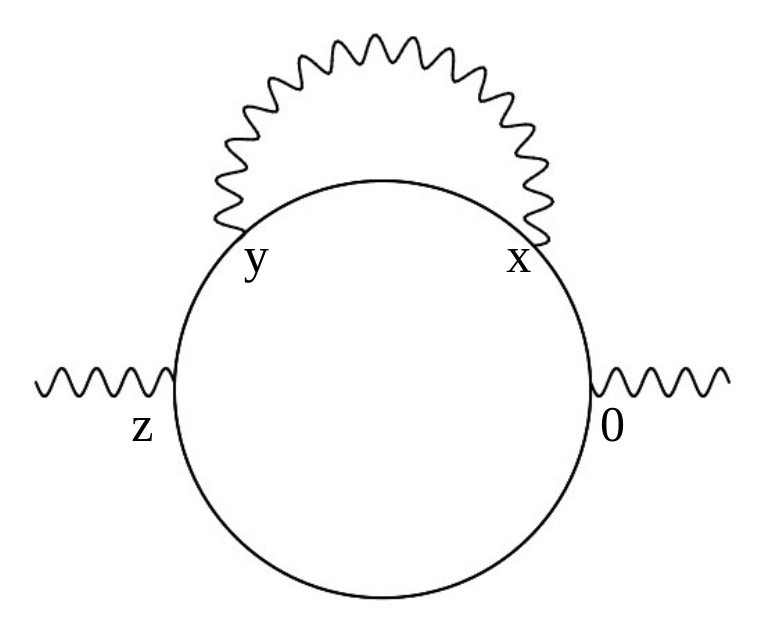}
			\caption{Diagram $(4)a$: the self-energy $\phantom{-}$ \mbox{diagram}, which contributes twice.}
   \label{fig::Conn_Feynman_Diagram_SE}
		\end{subfigure}~~~
		\begin{subfigure}[t]{0.4\textwidth}
			\includegraphics[width=\textwidth]{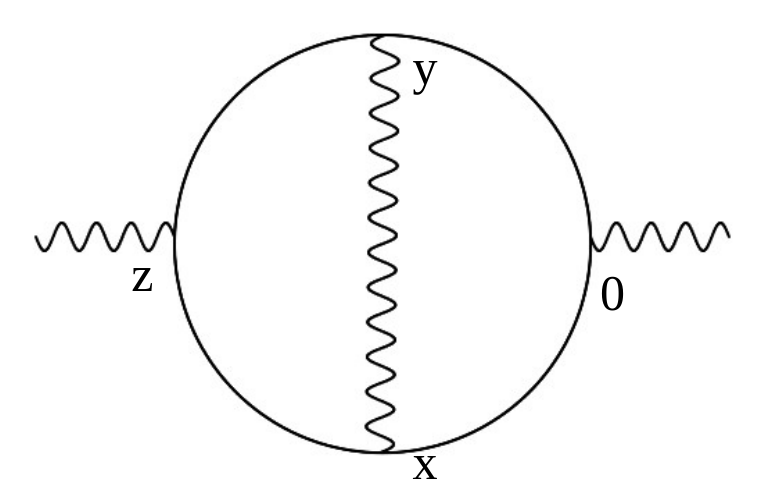}
			\caption{
            Diagram $(4)b$: the genuine 2-loop diagram.}
   \label{fig::Conn_Feynman_Diagram_2Loop}
		\end{subfigure}
  \caption{The fully connected diagrams which need to be calculated for $\atotvioconn$.
  }
  \label{fig::Conn_Feynman_Diagrams}
	\end{figure}

In this section, we describe the calculation of the connected contribution. 
Using Wick's theorem for the four-point function in Eq. \eqref{equ::Master_4pt} results in three different fully connected quark contractions, just as in the case of the calculation of the hadronic light-by-light contribution \cite{Chao:2020kwq}.
Two of these diagrams result in the same contribution. They correspond to the case depicted in Fig~\ref{fig::Conn_Feynman_Diagram_SE}, where the photon propagator can either be on the upper or the lower quark propagator. We refer to this diagram as the {\it self-energy} (SE) part,
\begin{align}
    C^{SE}_{\rho\nu\lambda\sigma}(x, y, z)&=  -2 \, \text{Re} \langle \text{Tr} [ S(0, x) \gamma_\rho S(x, y) \gamma_\nu S(y, z) \gamma_\lambda S(z, 0) \gamma_\sigma] \rangle_U\,.
\end{align}
Additionally, one needs to calculate the {\it genuine 2-loop} (2L) part 
\begin{align}
    C^{2L}_{\rho\nu\lambda\sigma}(x, y, z)&=  -2 \, \text{Re} \langle \text{Tr} [ S(0, y) \gamma_\nu S(y, z) \gamma_\lambda S(z, x) \gamma_\rho S(x, 0) \gamma_\sigma] \rangle_U.
\end{align}
The trace needs to be taken with respect to the Dirac and color indices, while $\langle ... \rangle_U$ denotes the expectation value over gauge configurations and $S(x,y)$ is the quark propagator where the sink is at $x$ and the source is at $y$.
In total the connected contribution yields
\begin{align}\label{equ::Conn_a_mu_1}
    \atotvioconn(\Lambda)= -\fQ^{(4)} \mathcal{Z} \;\frac{e^2} {2}\int_{z,x,y} \ &H_{\lambda \sigma}(z) \delta_{\nu \rho} [\mathcal{G}(x-y)]_\Lambda \\ &(2\, C^{SE}_{\rho\nu\lambda\sigma}(x, y, z) + C^{2L}_{\rho\nu\lambda\sigma}(x, y, z)). \nonumber
\end{align}
where the factor `$2$' for the SE part is due to the two possible positions of the photon propagator.
The charge factor $\fQ^{(4)}$ for the connected contribution is given by the trace over the product of charge matrices $\fQ^{(4)}= \text{Tr} \{ \mathcal{Q}^{(3)}\mathcal{Q}^{(em)}\mathcal{Q}^{(em)}\mathcal{Q}^{(8)} \}= \frac{1}{36}$, using Eq.~\eqref{eq:charge_matrices}.

For the implementation of Eq. \eqref{equ::Conn_a_mu_1} in our lattice calculation, we perform the summation over the $x$ and $z$ vertex on the fly for fixed values of $y$. Specifically, Eq.~\eqref{equ::Conn_a_mu_1} is rewritten as
\begin{align}\label{equ::Conn_a_mu_2}   
    \atotvioconn(\Lambda) = -\fQ^{(4)} \mathcal{Z}\,\pi^2 {e^2}  a^8\int_0^\infty  d|y|  \,  |y|^3 \sum_{z,x}& H_{\lambda \sigma}(z) \,\delta_{\nu \rho}\,[\mathcal{G}(x-y)]_\Lambda 
    \\ & \quad(2\, C^{SE}_{\rho\nu\lambda\sigma}(x, y, z) + C^{2L}_{\rho\nu\lambda\sigma}(x, y, z)).  
    \nonumber
\end{align}
The integrand of Eq.~\eqref{equ::Conn_a_mu_2} is calculated on the lattice for different values of $y$ along the $(1, 1, 1, 1)$  direction. 
The integral over $|y|$ is then computed using the trapezoidal rule. The origin was assigned to the middle time slice of each lattice.
For the calculation of $C^{SE}$, it is sufficient to use two one-to-all propagators with their origins at the $y$ and 0 vertices. For $C^{2L}$, we additionally make use of a sequential propagator starting at the 0 vertex, going over the $x$ vertex and ending at the $z$ vertex.

As mentioned in the previous section, we use both local and conserved discretizations of the vector currents. This applies only to the vertices over which the sum is performed, i.e. the $x$ and $z$ vertices. The discretization scheme is denoted as XdZd, where the character `d' is replaced with `l' for a local current or `c' for a conserved current at the corresponding vertex.

\subsection{Bare QED two-loop vacuum polarization contribution at fixed $\Lambda$ from the lattice }

In order to test whether the proposed methodology is viable, we first perform the calculation of $\atotvioconn(\Lambda)$ on ensembles without the strong interaction, i.e.\ where the gauge field variables $U\in {\rm SU}(3)_c$ are set to unity on all lattice links. 
Owing to the use of the Pauli-Villars regulated photon propagator (as opposed to the regularization by the finite lattice spacing), we can directly compare the lattice results to continuum calculations (see Appendix~\ref{app::QED_Cont}).
This serves as a useful crosscheck for our methodology.

\begin{figure}[t]
    \centering
    \begin{subfigure}{0.49\textwidth}
        \includegraphics[width=\textwidth]{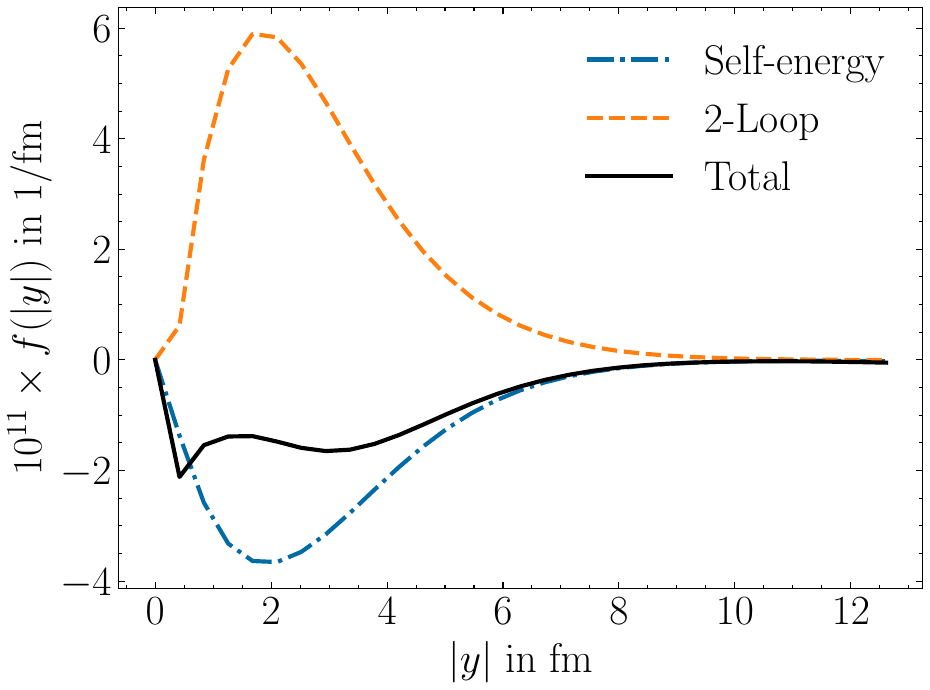}
    \end{subfigure}
	\caption{The integrand of Eq.~\eqref{equ::Conn_a_mu_2} for the self-energy (blue) and 2-loop (orange) part separately as well as combined to the total (black), given by 2$\times$SE+2L.   
    It is shown on the gluonless ensemble with $L=64$ with the XcZc discretization and the `TL' kernel. The lepton in the loop has mass $m_\ell=m_\mu$ and the Pauli-Villars mass is set to $\Lambda =3\, m_\mu$.}
    \label{fig::Conn_QED_Integrands}
\end{figure}

To this end, we employ eight $L^4$ lattices with different spacings, but with a constant volume of $m_\mu L=7.2$, with parameters provided in Table~\ref{tab::Conn_QED_Lattice_Parameters}.
We set the lepton mass $m_\ell$ appearing in the loop to be equal to the muon mass, $m_\ell=m_\mu$,
while the Pauli-Villars mass of the photon is set to $\Lambda=3\, m_\mu$ for the purpose of this test.
An example of the integrand of Eq.~\eqref{equ::Conn_a_mu_2} in the case of the lattice with $L=64$ is shown in Fig.~\ref{fig::Conn_QED_Integrands}. 
In this example, the currents at both the $x$ and $z$ vertices are conserved,
and the `TL' version of the CCS kernel is used.

\begin{figure}[h]
    \centering
    \begin{subfigure}{0.32\textwidth}
        \includegraphics[width=\textwidth]{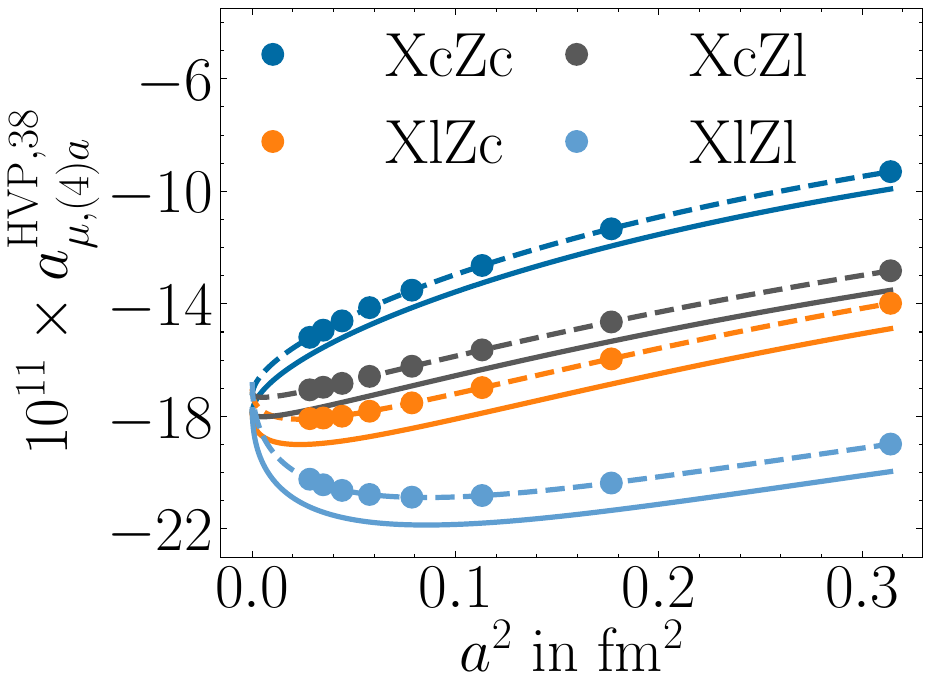}
        \caption{Self-energy}
        \label{fig::Conn_QED_Extrapolation_SE}        
    \end{subfigure}
    \begin{subfigure}{0.32\textwidth}
        \includegraphics[width=\textwidth]{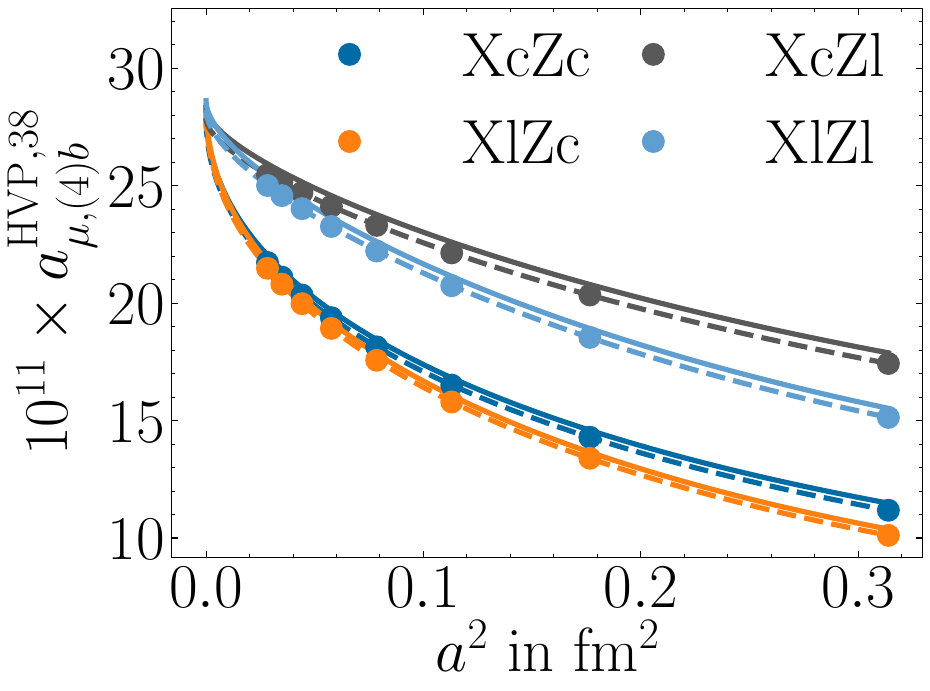}
        \caption{2-Loop}
        \label{fig::Conn_QED_Extrapolation_2L}
    \end{subfigure}
    \begin{subfigure}{0.32\textwidth}
        \includegraphics[width=\textwidth]{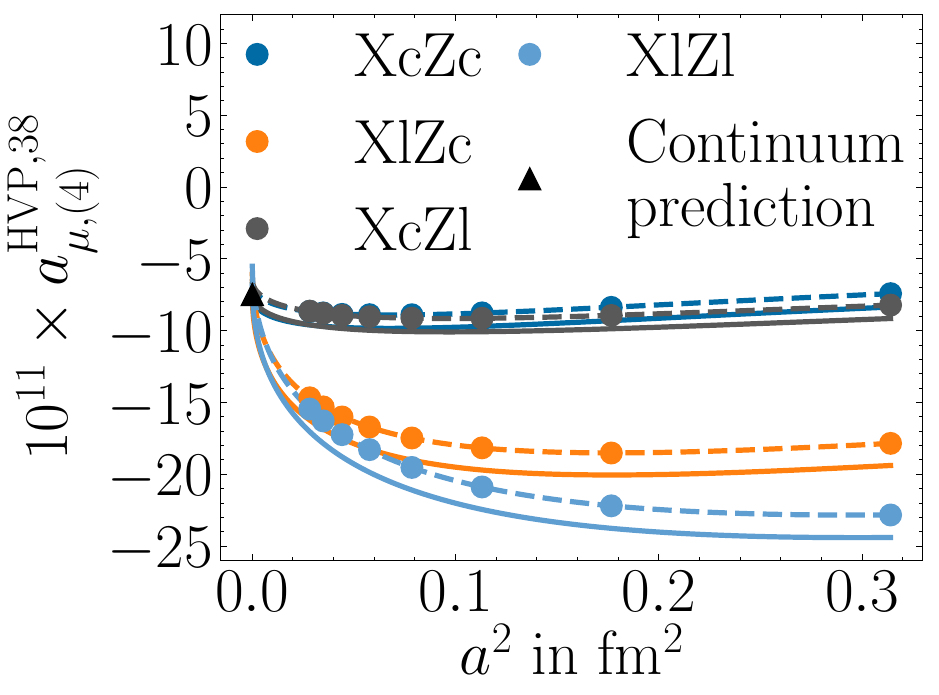}
        \caption{Total=2$\times$SE+2Loop}
        \label{fig::Conn_QED_Extrapolation_Total}
    \end{subfigure}

\caption{Continuum extrapolation of the self-energy and 2-loop parts and the total value for the `gluonless' lattices and $\Lambda=3\, m_\mu$. The different colors represent the different discretizations used for the current at the x and z vertices. Additionally, fits of the form given by Eq.~\eqref{equ::Conn_QED_Fit_Function} are shown. The dashed lines correspond to fits performed directly on the data points, while the solid lines represent the volume-corrected fits. The volume correction was calculated by doubling the volume of the coarsest lattices and adding the difference to the fit function. 
The figure showing the total result also includes the continuum prediction indicated by a black triangle, which was computed following the methodology outlined in Appendix~\ref{app::QED_Cont}.
}
\label{fig::Conn_QED_Extrapolation}
\end{figure}

The results for each ensemble using the four different discretizations are shown in Fig.~\ref{fig::Conn_QED_Extrapolation}. We perform a continuum extrapolation by fitting the data points with the function
\begin{align}
    f_{fit}(a)=c_0 + c_1\, a + c_2\, a^2 + c_3 \, a^3 \label{equ::Conn_QED_Fit_Function}.
\end{align}

\begin{table}[t]
	\centering
	\caption{Parameters of the gluonless ensembles. For each lattice the time extend is the same as $L$, i.e. they have a volume of $L^4$.}
	\label{tab::Conn_QED_Lattice_Parameters}
	\begin{tabular}{c | c c c c c c c c}
		\hline
		
		$L$ & 24  & 32 & 40 & 48 & 56 & 64 & 72 & 80 \\ \hline
		$am_\mu$ & 0.3 & 0.225 & 0.18 & 0.15 & 0.13 & 0.1125 & 0.1 & 0.09\\ \hline  
	\end{tabular}
\end{table}

\begin{table}[t]
	\centering
	\caption{Results of the volume corrected continuum extrapolations in Fig.~\ref{fig::Conn_QED_Extrapolation}. The values are given in units of $10^{-11}$. The errors are from the fit. The expected value of the total result is $-7.499 \times 10^{-11}$, as obtained using the methodology outlined in appendix~\ref{app::QED_Cont}.
    }
	\label{tab::Conn_QED_Results}
	\begin{tabular}{c c c c c}
		\hline
		 & XlZl & XcZl & XlZc & XcZc \\ \hline
		Total & $-6.90(15)$ & $-7.36(16)$ & $-7.44(18)$ & $-7.56(18)$ \\ \hline
		2-Loop & $\ 28.62(11)$ & $\ 28.32(8)$ & $\ 28.42(9)$ & $\ 28.18(7)$ \\ \hline
		Self-Energy & $-17.76(11)$ & $-17.84(11)$ & $-17.93(13)$ & $-17.87(12)$ \\ \hline
			
	\end{tabular}

\end{table}

Of special interest is the extrapolation of the total quantity in Fig.~\ref{fig::Conn_QED_Extrapolation_Total}. For both cases where the current at the $x$ vertex is conserved the extrapolation is flat without much curvature. Based on this observation the XcZl and XcZc discretization schemes are the ones used in the QCD calculation later on. This plot also shows the value resulting from the previously mentioned continuum prediction (see Appendix \ref{app::QED_Cont}) as a black triangle. This value 
 was obtained completely independently from the calculations described here and lies where all of the lines of the different discretizations meet. The results of the continuum extrapolation of the individual diagrams for different discretizations  are given in Table \ref{tab::Conn_QED_Results}. The discretization for which both the currents at the $x$ and $z$ vertices are conserved yields a result especially close to the value from the continuum prediction.

\subsection{The connected contribution $\atotvioconn$ at $m_\pi=m_K\simeq 416\,$MeV}
\begin{figure}[t]
    \centering
    \begin{subfigure}{0.49\textwidth}
        \includegraphics[width=\textwidth]{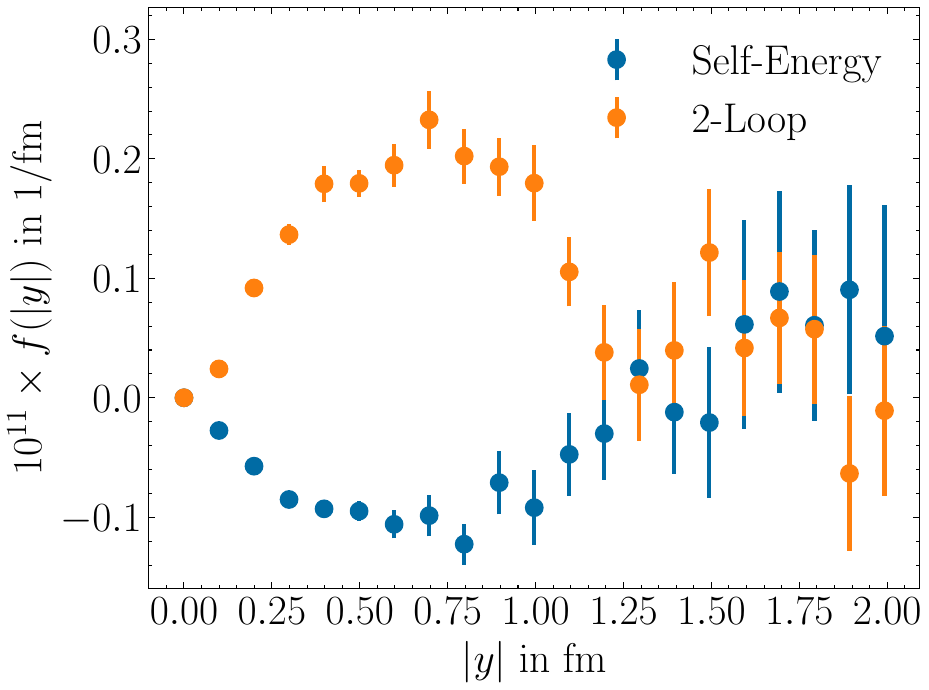}
        \caption{Self-energy and 2-Loop}
        \label{fig::Conn_QCD_Integrands_SE_2L}
    \end{subfigure}
    \begin{subfigure}{0.49\textwidth}
        \includegraphics[width=\textwidth]{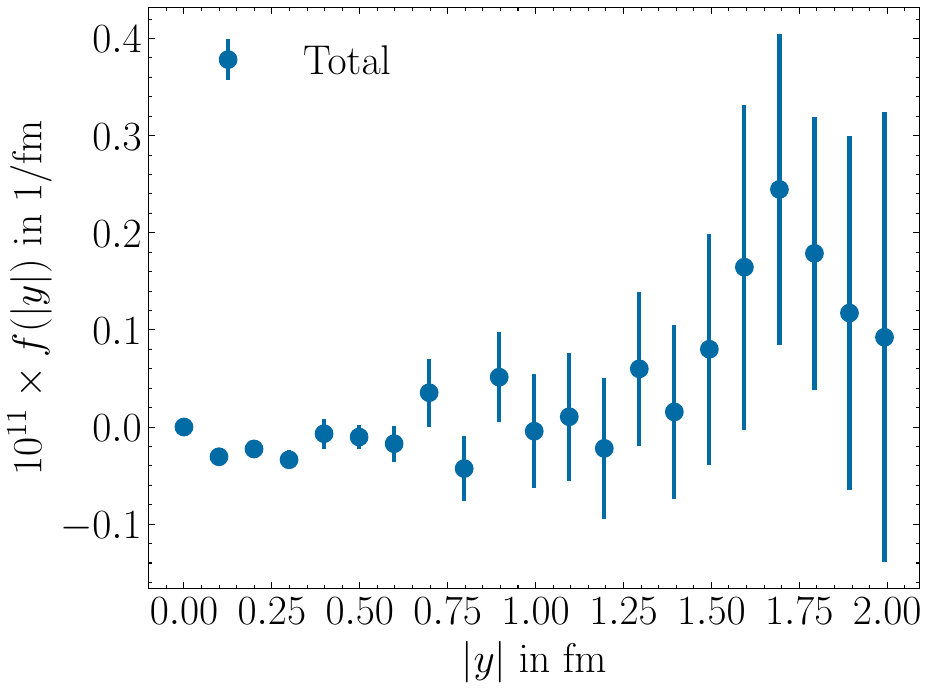}
        \caption{Total=2$\times$SE+2Loop}
        \label{fig::Conn_QCD_Integrands_Total}
    \end{subfigure}
\caption{The integrand of Eq.~\eqref{equ::Conn_a_mu_2} for the self-energy and 2-loop part separately on the left side as well as combined on the right side. It is shown on the N300 ensemble with the XcZl discretization and the `TL' kernel. The Pauli-Villars mass is set to $\Lambda =5\, m_\mu$.}
\label{fig::Conn_QCD_Integrands}
\end{figure}

The successful crosscheck on the `gluonless' ensembles against the continuum calculation confirmed the viability of the methodology. Now, the same principles shall be used for the calculation on the ensembles introduced in section \ref{sec::LatSet}. The integrands of the different parts on the N300 ensemble can be seen in Fig.~\ref{fig::Conn_QCD_Integrands}. For these plots the XcZl discretization and a Pauli-Villars mass of $5\, m_\mu$ is used. While the 'gluonless' calculation seen in Fig.~\ref{fig::Conn_QED_Integrands} already showed a cancellation between the self-energy and 2-loop part, the cancellation is much more severe for the QCD calculation. 
The combined integrand in Fig.~\ref{fig::Conn_QCD_Integrands_Total} is much smaller than the individual contributions from the self-energy and 2-loop diagrams depicted in Fig.~\ref{fig::Conn_QCD_Integrands_SE_2L}. This worsens the signal-to-noise ration of the total result compared to the contributions of the individual diagrams with the former being consistent with 0 starting at around $|y|=1$ fm. This behavior is seen for all ensembles, discretizations and CCS kernels we investigated.

Based on this observation, the integrand in Eq.~\eqref{equ::Conn_a_mu_2} is only evaluated up to a cut in $|y|$, which was chosen for each ensemble individually. For N300 this value is $|y|=1.45$ fm.
Another interesting observation is that the values of the integrands are more than an order of magnitude smaller than in the QED case and also much more short ranged, even before the cancellation between the self-energy and 2-loop part happens. This means one can also expect a much smaller value for the continuum extrapolated result.

\begin{figure}[t]
    \centering
    \begin{subfigure}{0.49\textwidth}
        \includegraphics[width=\textwidth]{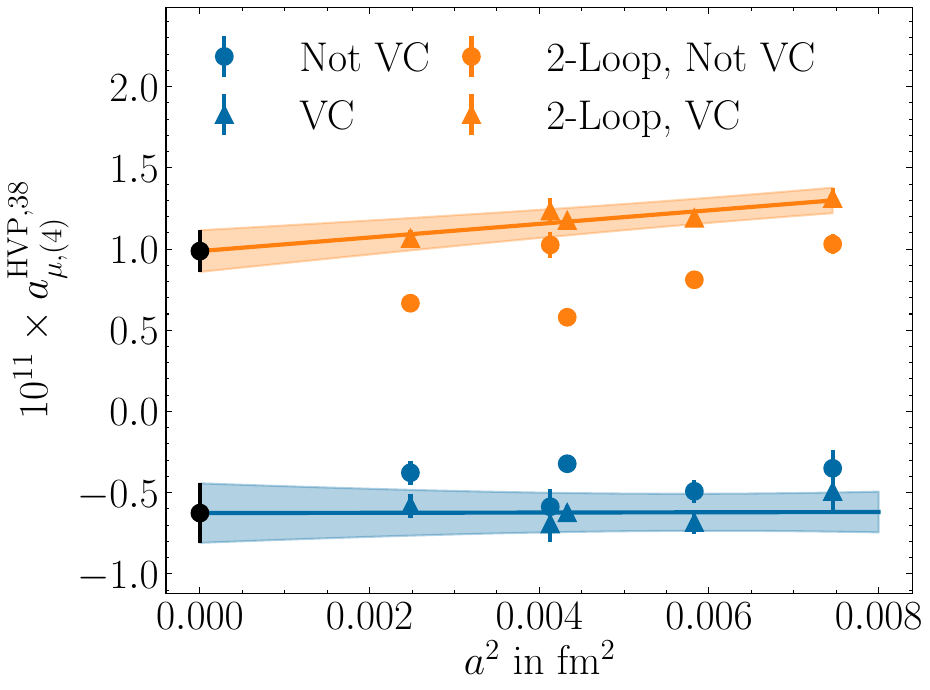}
        \caption{Self-energy and 2-Loop}
    \end{subfigure}
    \begin{subfigure}{0.49\textwidth}
        \includegraphics[width=\textwidth]{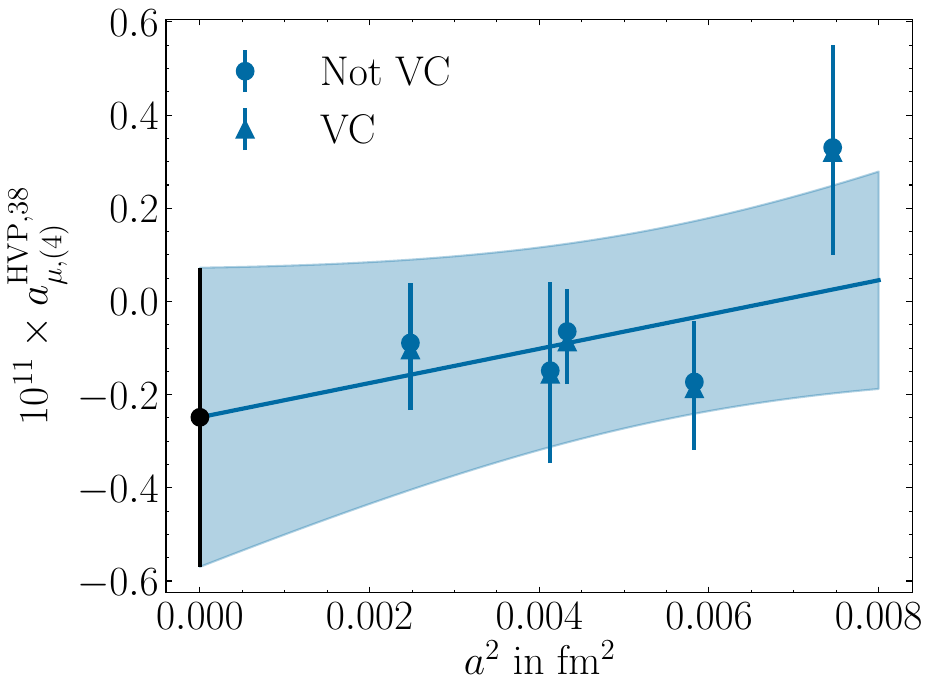}
        \caption{Total=2$\times$SE+2Loop}
    \end{subfigure}
\caption{Continuum extrapolation of the self-energy and 2-loop parts on the left side as well as the total value on the right side for the QCD ensembles. The Pauli-Villars mass is set to $\Lambda = 16\, m_\mu$. H200 is shown with a slight offset to higher $a^2$ to make it distinguishable from N202. Eq.~\eqref{equ::Conn_QCD_Fit_Function} is used as an ansatz for the fit function. The dots are the results from each ensemble, while the triangles are the same results, but with the volume term of the fit function subtracted without adjusting the error bars. The straight lines are the fits to these volume-corrected points. The black dots are the results of the continuum extrapolation.}
\label{fig::Conn_QCD_Cont_Extra}
\end{figure}

This continuum extrapolation of the different parts for a Pauli-Villars mass of $16\, m_\mu$ can be seen in Fig.~\ref{fig::Conn_QCD_Cont_Extra}. The fit function used in this case was modified compared to the `gluonless' case, because there are now several lattice volumes involved and fewer data points overall,
\begin{align}
    f_{fit}(a, m_\pi L)=c_0 + c_1 \, a^2 + c_2 \, e^{-\frac{m_\pi L}{2}}. \label{equ::Conn_QCD_Fit_Function}
\end{align}

The expectation set by the plots of Fig.~\ref{fig::Conn_QCD_Integrands} holds true: 
while we clearly observe a signal for the self-energy part and the two-loop part, a vast cancellation leads to the total contribution vanishing within its uncertainty.
Another interesting observation from these plots is the behavior of the finite volume correction.
For the two parts separately the volume correction is important, but when added together it is not statistically significant. 

\begin{figure}[ht]
		\centering
		\begin{subfigure}[t]{0.49\textwidth}
			\includegraphics[width=\textwidth]{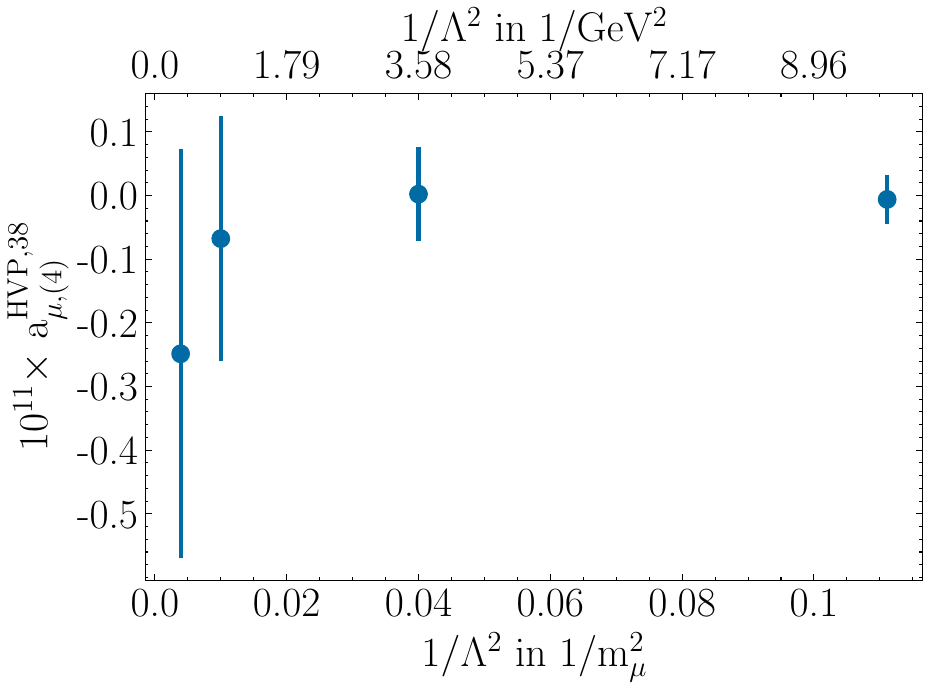}
		\end{subfigure}		

  \caption{Dependence on $\Lambda$ for the connected contribution. The coordinates of the data points are collected in Table~\ref{tab::Conn_Results}.}
  \label{fig::Conn_QCD_Lam_Extra}
\end{figure}

We finish this section by looking at the dependence on the PV-mass of the total connected contribution.
Table~\ref{tab::Conn_Results} lists the four calculated values together with their uncertainty and Fig.~\ref{fig::Conn_QCD_Lam_Extra} shows their dependence on $\Lambda$.
The values grow more negative for larger $\Lambda$ values, but the errors increase approximately at the same rate. This results in all values being consistent with zero within uncertainties, as previously noted.

\begin{table}[ht]
    \centering
    \caption{Continuum extrapolated values of $\atotvioconn$ for the different PV-mass values.}
	\begin{tabular}{c c c c c}
		\hline
		$\Lambda/m_\mu$ & 3 & 5 & 10 & 16 \\ \hline
		$m_\mu^2/\Lambda^2$  & 0.11 & 0.04 & 0.01 & 0.004 \\ \hline
		  $10^{11}\times \atotvioconn(\Lambda)\phantom{\Big|}$ & $-0.006(39)$ & $0.002(74)$ & $-0.068(193)$ & $-0.249(321)$\\ \hline
	\end{tabular}
    
    \label{tab::Conn_Results}
\end{table}

 \section{Disconnected electromagnetic contribution $\atotviotpta(\Lambda)$}
\label{sec:disco}
In addition to the fully connected diagrams shown in Fig.~\ref{fig::Conn_Feynman_Diagrams}, there is also one disconnected diagram contributing to $\atotvio$ in Eq.~\eqref{equ::Master_4pt} at the SU$(3)_{\rm f}$ symmetric point, which in the nomenclature of Ref. \cite{Djukanovic:2024cmq} is referred to as $(2+2)a$. This diagram is characterized by two disconnected valence quark loops that are connected by the photon propagator,
\ba\label{equ::tpta}
    \atotviotpta(\Lambda)
    =-e^2\fQ^{(2+2)a}\mathcal{Z} \int_{z,x,y} \ H_{\lambda \sigma}(z) \delta_{\nu \rho} \,[\mathcal{G}(x-y)]_\Lambda \,\left\langle \hat{\Pi}_{\lambda \nu}(z,x)\hat{\Pi}_{\rho \sigma}(y,0)\right\rangle_{U}.
\ea
The two-point correlation function using local vector currents is given by
\ba 
\label{eq:2ptf}
\Pi_{\mu \nu}(x,y) = -\text{Re}\Big(\text{Tr}\Big[S(y,x)\gamma_\mu S(x,y) \gamma_\nu  \Big] \Big).
\ea 
The vacuum expectation value needs to be subtracted in order to avoid double counting of the contribution, where the two QCD `blobs' are not interconnected,
\ba 
\label{eq:vev_subtr}
\hat{\Pi}_{\mu \nu}(x,y) = \Pi_{\mu \nu}(x,y)-\langle \Pi_{\mu \nu}(x,y) \rangle_{U}.
\ea
The charge factor for the `$38$' contribution is obtained from 
\ba 
\fQ^{(2+2)a}=\text{Tr} \{ \mathcal{Q}^{(3)}\mathcal{Q}^{(em)}\} \text{Tr} \{\mathcal{Q}^{(em)}\mathcal{Q}^{(8)} \} = \frac{1}{12}.
\ea 
A recent lattice QCD calculation of this contribution \cite{Parrino:2025afq} shows a significant contribution to the HVP from this particular diagram at the physical point. Here, we use the results from Ref. \cite{Parrino:2025afq} for the ensembles at the SU$(3)_{\rm f}$ symmetric point normalized to the correct charge factor for the `$38$' contribution. Since the $(2+2)a$ contribution is UV-finite \cite{Parrino:2025afq}, the Pauli-Villars mass $\Lambda$ is taken to be infinite in Eq. \eqref{equ::tpta}. 

\begin{figure}

    \begin{subfigure}{0.49\textwidth}
    \centering
    \includegraphics[width=1.\textwidth]{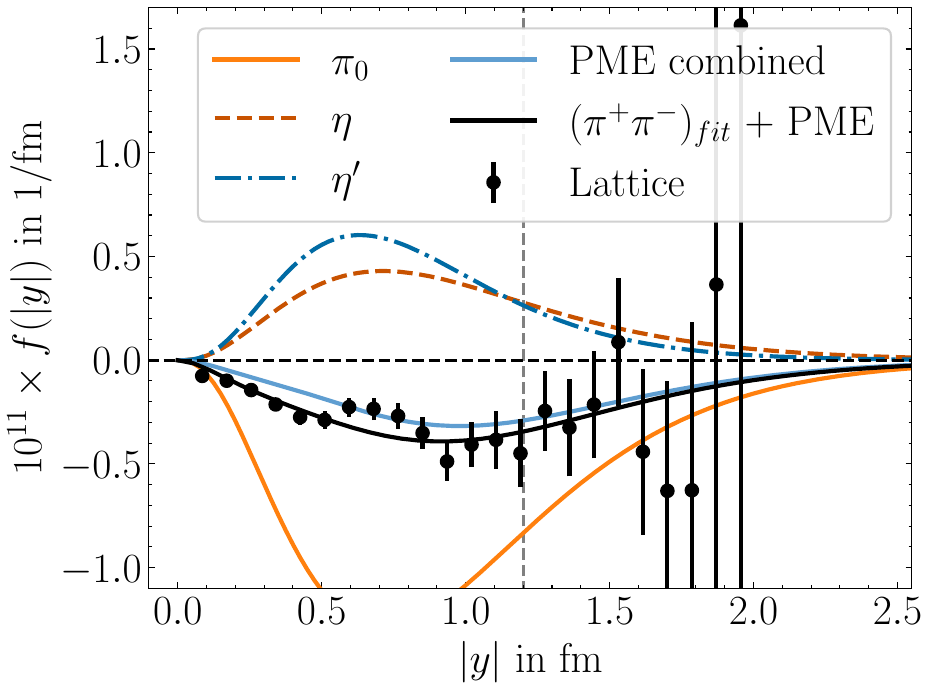}
    \caption{}
    \label{fig:tpta_N300}
\end{subfigure}
    \begin{subfigure}{0.49\textwidth}
    \centering
    \includegraphics[width=1.\textwidth]{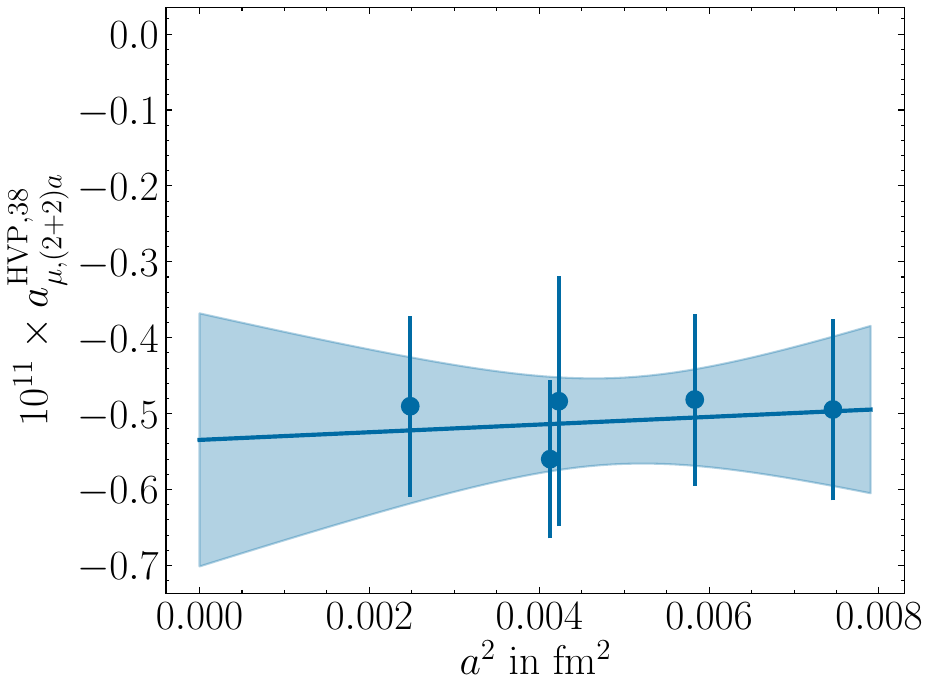}
    \caption{}
    \label{fig:tpta_continuum}
\end{subfigure}

\caption{(a) Integrand of Eq.~\eqref{equ::tpta} with respect to $|y|$ following the procedure in Ref.~\cite{Parrino:2025afq}, together with the prediction from the pseudoscalar mesons. The solid light blue curve is the sum of the contributions from $\pi^0$, $\eta$ and $\eta'$. For the black curve the charged pion contribution is fitted to the lattice data. The vertical dashed line represents the cut in $|y|$ from which we use the model to approximate the tail of the integrand.
(b) Continuum extrapolation of the results computed on the ensembles from Table~\ref{tab::LatSet_Properties}. H200 is shown with a slight offset to higher $a^2$ to make it distinguishable from N202. }
\end{figure}

A phenomenological description of the $(2+2)a$ contribution in terms of the exchange of neutral pseudoscalar mesons $\pi^0$, $\eta$, $\eta'$ and the charged pion loop was also discussed in Ref.\ \cite{Parrino:2025afq}. An important difference between the calculation at the physical point and the SU$(3)_{\rm f}$ symmetric point is the relative size of the charged pion loop and the pseudoscalar meson exchange (PME) contribution. While the former is dominant at the physical point, the hierarchy is swapped at the SU$(3)_{\rm f}$ symmetric point. In Fig~\ref{fig:tpta_N300}, the integrand with respect to $|y|$ of the Eq.~\eqref{equ::tpta} is shown after performing the integration over $x$ and $z$. We observe good agreement between the prediction from the pseudoscalar mesons and the lattice data. However, following the procedure from Ref.~\cite{Parrino:2025afq}, we also fit the contribution of the charged pion loop to the lattice data to approximate the tail of the integrand for $|y|>L/2$  by the total prediction from the model. From Fig.\ \ref{fig:tpta_N300}, one can see that the tail is totally dominated by the PME contribution, a fact that does not hold anymore when approaching the physical pion mass.

In order to remove the cut-off effects, we perform a simple continuum extrapolation for the results of the $(2+2)a$ contribution with a fit linear in $a^2$. The fit is depicted in Fig.~\ref{fig:tpta_continuum}. The continuum result for this contribution to the isospin-violating part of the HVP reads
\begin{equation}
     \atotviotpta= -0.53(17)\times 10^{-11}. \label{eq:tpta_Continuum}
    \end{equation}
 \section{Computing the counterterm \label{sec:counterterm}}
In order to obtain the counterterm according to Eq.~\eqref{equ::Master_Counterterm}, we split the calculation into three different parts. These are the electromagnetic kaon mass splitting, the light-quark mass derivative of the kaon mass and the light-quark mass derivative of the leading-order HVP. We discuss the calculation of the three quantities in that order. 

\subsection{The electromagnetic kaon mass splitting \label{sec::Kaon_MS}}
    \begin{figure}[ht]
		\centering
		\begin{subfigure}{0.4\textwidth}
			\includegraphics[width=\textwidth]{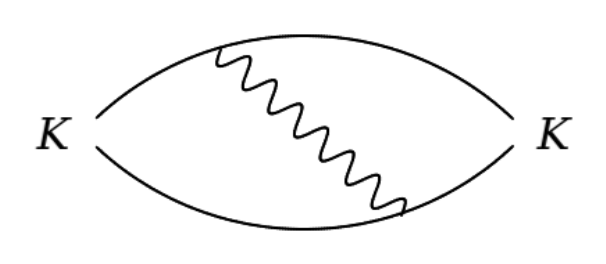}
			\caption{K1 diagram}
		\end{subfigure}
		\begin{subfigure}{0.4\textwidth}
			\includegraphics[width=\textwidth]{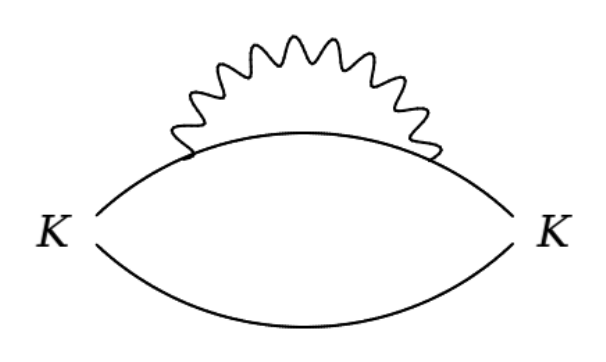}
			\caption{K2 diagram}
            \label{fig::Kaon_MS_Feynman_asymm}
   
		\end{subfigure}
  \caption{The Feynman diagrams of the leading order contributions to the e.m.\ mass splitting of the kaon. The bottom quark line corresponds to the strange quark. At the SU$(3)_{\rm f}$ symmetric point, only these two diagrams contribute to the mass splitting.}
  \label{fig::Kaon_MS_Feynman}

	\end{figure}
    
The first quantity that needs to be calculated for the counterterm is the leading electromagnetic contribution to the charged-neutral kaon mass splitting. From this point on, it will be abbreviated as the e.m.\ kaon mass splitting. At the SU$(3)_{\rm f}$ symmetric point, only the two diagrams in Fig.~\ref{fig::Kaon_MS_Feynman} contribute to the e.m.\ mass splitting; see for instance Ref.\ \cite{deDivitiis:2013xla}. These two diagrams can be interpreted as the self energy correction to the kaon mass. 

For large separation times between the two photon vertices, only intermediate kaon states contribute to this correction~\cite{Feng:2018qpx}, in other words, the elastic contribution dominates; the latter can be calculated analytically (see below). This naturally leads to the strategy of using the lattice calculation up to a maximum time separation between the two currents, while beyond that, we use the aforementioned analytic expression, directly in the continuum and with infinite volume~\cite{Biloshytskyi:2022ets}. This procedure avoids power-law finite-size effects, which would otherwise affect the four-point function at large separation times~\cite{Feng:2018qpx}.

In this section, we first discuss the calculation of the elastic part. Then the lattice calculation, which also includes the inelastic part, will be presented. 
Following that, we apply the Operator Product Expansion to predict the behavior of the e.m.\ kaon mass splitting at large PV masses. Finally, we compare this prediction to our results.

\subsubsection{Equivalent expressions for the e.m.\ correction to a hadron mass \label{sec::Kaon_MS_general}}
In the following, we review various equivalent expressions for the e.m.\ correction to a hadron mass.
We start by considering the leading-order electromagnetic corrections to the two-point function of a general hadron

\begin{align}
    C_2(z_0)=&\,\langle \Bar{h}(z_0)h(0)^\dagger \rangle_{\mathrm{QCD}+\mathrm{QED}} =\, C_2^{(0)}(z_0) +  C_2^{(2)} (z_0) +O(\alpha^2),
    \\
    C_2^{(0)}(z_0)=&\langle\Bar{h}(z_0)h(0)^\dagger \rangle_\mathrm{QCD},\\
    C_2^{(2)} (z_0)=&-\frac{e^2}{2} \int_{x,y}\delta_{\nu \rho} [\mathcal{G}(x-y)]_\Lambda \langle \Bar{h}(z_0) j^{em}_\rho (x) j^{em}_\nu (y) h(0)^\dagger \rangle_\mathrm{QCD}.       
\end{align}
In these equations $\Bar{h}(z_0)= \int d^3 z \ h(z)$ is the interpolator for a hadron at rest. The mass of the hadron can be extracted from this using the usual approach
\begin{align}
    M_h&=-\lim_{z_0\rightarrow \infty}\frac{d}{dz_0} \log C_2(z_0) = -\lim_{z_0\rightarrow \infty}\frac{d}{dz_0} \log \left( C_2^{(0)}(z_0) + C_2^{(2)} (z_0)\right)\\
    &=-\lim_{z_0 \rightarrow \infty} \frac{d}{dz_0} \left( \log  \left(C_2^{(0)}(z_0)\right) + \frac{C^{(2)}_2(z_0)}{C^{(0)}_2(z_0)}\right) + ... = M_h^{(0)} + \delta M_h.
    \nonumber
\end{align}
We are interested in the correction to the hadron mass $\delta M_h$, for which we now have the expression
\begin{align}
    \delta M_h=\frac{e^2}{2}\lim_{z_0 \rightarrow \infty} \frac{d}{dz_0}\frac{\int_{x,y} \delta_{\nu \rho} [\mathcal{G}(x-y)]_\Lambda \langle \Bar{h}(z_0) j^{em}_\rho (x) j^{em}_\nu (y) h(0)^\dagger \rangle }{\langle \Bar{h}(z_0)h(0)^\dagger \rangle}. \label{equ::Kaon_MS_DelM_h}
\end{align}
This formula corresponds to the use of the `summation method'~\cite{Maiani:1987by} to extract the ground-state matrix element.
By using the fact that only the relative position of the two vector currents to one another as well as their relative position to the hadron interpolators is of importance, we can rewrite the upper integrals in the following way:
\begin{align}
    &\int_{x,y} \delta_{\nu \rho} [\mathcal{G}(x-y)]_\Lambda \langle \Bar{h}(z_0) j^{em}_\rho (x) j^{em}_\nu (y) h(0)^\dagger \rangle \nonumber \\
    &=\int_{-\infty}^\infty dy_0 \ \left\langle \Bar{h}\left(\frac{z_0}{2}\right) \left( \int_x \delta_{\nu \rho}[\mathcal{G}(x)]_\Lambda \  j^{em}_\rho (x + y_0 \hat{e}_0) j^{em}_\nu(y_0\hat{e}_0) \right) \Bar{h}\left(-\frac{z_0}{2}\right)^\dagger \right\rangle.
\end{align}
For very large $z_0$ only the ground state of the hadron contributes to the four-point function. At that point there is translational invariance for an operator insertion between the $\Bar{h}^\dagger$ and the $\Bar{h}$ operator, which means the integral over $y_0$ simply yields a factor of $z_0$, and we can write
\begin{align}
    \delta M_h=\frac{e^2}{2} \lim_{z_0\rightarrow \infty}\frac{\int_{x} \delta_{\nu \rho} [\mathcal{G}(x)]_\Lambda \langle \Bar{h}(\frac{z_0}{2}) j^{em}_\rho (x) j^{em}_\nu (0) \Bar{h}(-\frac{z_0}{2})^\dagger \rangle }{\langle \Bar{h}(\frac{z_0}{2})\Bar{h}(-\frac{z_0}{2})^\dagger \rangle}. 
\end{align}
This expression corresponds to  the `mid-point' method used in hadron structure  calculations for extracting the ground-state matrix element. 
We now introduce 
\begin{align}
    f_{z_0}(x_0)&=\frac{e^2}{2} \frac{\int d^3x \  \delta_{\nu \rho} [\mathcal{G}(x)]_\Lambda \langle \Bar{h}(z_0) j^{em}_\rho (x) j^{em}_\nu (0) \Bar{h}(-z_0)^\dagger \rangle }{\langle \Bar{h}(z_0)\Bar{h}(-z_0)^\dagger \rangle}\\
    \overset{z_0\rightarrow \infty}{\rightarrow}    f(x_0)&=\frac{e^2}{2}\int d^3x \ \delta_{\nu \rho} [\mathcal{G}(x)]_\Lambda \langle h_{\vec{p}=0} | \text{T} \{ j^{em}_\rho (x) j^{em}_\nu (0) \} | h_{\vec{p}=0} \rangle,
    \nonumber
\end{align}
where we dropped the factor 1/2 of $z_0$ in the first definition since it does not matter in the limit. This finally leaves us with 
\begin{align}
    \delta M_h&= \int_{-\infty}^\infty dx_0 \ f(x_0)\,,
\end{align}
which is the form typically used in continuum  field-theoretic treatments.

\subsubsection{The elastic contribution \label{sec::Kaon_MS_elastic}}
We now shift our attention from a general hadron to the kaon.  Introducing the relativistically normalized kaon states, indicated by bold symbols,
\begin{align}
    \bm{\langle K_{\vec{q}}| K_{\vec{p}} \rangle} &= 2 E_{\vec{q}}\,  L^3 \, \delta_{\vec{q} \vec{p}}.    
\end{align}
we find an expression for the elastic contribution to the integrand $f(x_0)$ by inserting a complete set of one-particle states,
\begin{align}
    f_{elast}(x_0)&\overset{x_0\geq 0}{=}\frac{e^2}{2} \int d^3x \ \delta_{\nu \rho} [\mathcal{G}(x)]_\Lambda \frac{1}{L^3} \sum_{\vec{p}} \frac{1}{4E_{\vec{p}}M_K} \bm{\langle K_{\vec{0}}|} j^{em}_\rho (x) \bm{| K_{\vec{p}} \rangle \langle K_{\vec{p}}|} j^{em}_\nu (0) \bm{| K_{\vec{0}} \rangle}. \label{equ::elast_fin_1}
\end{align}
Because of translational invariance, the matrix elements can be rewritten as 
\begin{align}
    \bm{\langle K_{\vec{0}}|} j^{em}_\rho (x) \bm{| K_{\vec{p}} \rangle} &= \bm{\langle K_{\vec{0}}|} e^{Hx_0-i\vec{p}\cdot \vec{x}} j^{em}_\rho (0) e^{-Hx_0+i\vec{p}\cdot \vec{x}} \bm{| K_{\vec{p}} \rangle} \nonumber \\
    &=\bm{\langle K_{\vec{0}}|} j^{em}_\rho (0) \bm{| K_{\vec{p}} \rangle} \ e^{(M_k-E_{\vec{p}})x_0+i\vec{p}\cdot \vec{x}}. \label{equ::elast_mat_1}
\end{align}
In Euclidean space, the matrix element from the right-hand side of the last equation is connected to the form factor $F(-Q^2)$ by
\begin{align}
    \bm{\langle K_{\vec{p}}|} j^{em}_\rho(0) \bm{| K_{\vec{k}}\rangle}=-i(P_\rho + K_\rho) F\left(-(P-K)^2\right) \label{equ::elast_mat_2}
\end{align}
with the Euclidean on-shell four-momentum vectors $P=(iE_{\vec{p}},\ \vec{p})$ and $K=(iE_{\vec{k}},\ \vec{k})$. Substituting Eqs.~\eqref{equ::elast_mat_1} and \eqref{equ::elast_mat_2} into \eqref{equ::elast_fin_1} yields the following result for the elastic integrand, now valid for all values of $x_0$:
\begin{align}
    f_{elast}(x_0)=\frac{e^2}{4L^3} \int d^3x \ [\mathcal{G}(x)]_\Lambda  \sum_{\vec{p}} e^{(M_k-E_{\vec{p}})|x_0|+i\vec{p}\cdot \vec{x}} \ \frac{E_{\vec{p}}+M_K}{E_{\vec{p}}} F^2\left(-(P-K)^2\right). \label{equ::elast_fin_2}
\end{align}
In the limit of infinite volume, the expression simplifies to 
\begin{align}\label{eq:felast_infvol}
    f_{elast}(x_0)=\frac{e^2}{8\pi^2} \int_0^\infty dp \ &p^2 e^{(M_k-E_{\vec{p}})|x_0|} \ \frac{E_{\vec{p}}+M_K}{E_{\vec{p}}} F^2\left(-(P-K)^2\right) \ \times\\
    & \times \left[ \frac{e^{-|\vec{p}||x_0|}}{2|\vec{p}|} - \frac{e^{-\sqrt{|\vec{p}|^2 + \Lambda^2/2} |x_0|}}{\sqrt{|\vec{p}|^2+\Lambda^2/2}} +\frac{e^{-\sqrt{|\vec{p}|^2 + \Lambda^2} |x_0|}}{2\sqrt{|\vec{p}|^2+\Lambda^2}}\right]. \nonumber
\end{align}
It is important to note that the derivation above refers to the rest frame of the kaon, i.e. $K=(iM_K,\ \vec{0})$ in these equations.  We also remark that $\int_{-\infty}^\infty dx_0 \;f_{elast}(x_0)$ yields an expression for $\dmKel$ similar but not identical to the one obtained in~\cite{Stamen:2022uqh}, where the `elastic contribution' is defined via the forward Compton amplitude. The two expressions do however agree in the limit where the argument of the form factor is small.

The vector meson dominance (VMD) form factor 
\begin{align} \label{equ::vmd_papram}
    F_{\text{VMD}}(-Q^2)=\frac{1}{1+{Q^2}/{M^2_{\text{VMD}}}}
\end{align}
will be used to evaluate Eq.\ (\ref{eq:felast_infvol}). This turned out to be sufficient for describing the data, and appendix \ref{app::Kaon_FF} provides a more detailed discussion regarding the form factor. The VMD mass values for the different ensembles are collected in Table~\ref{tab::LatSet_Properties}.

\subsubsection{Lattice QCD calculation of $\dmKem$ \label{sec::Kaon_MS_inelastic}}
For the total contribution there are three parts which have to be evaluated on the lattice. Writing them in terms of propagators yields: 
\begin{align}
    C^{2pt}(x, y)&=-\text{Re} \langle \text{Tr} [ S^l(y, x) \gamma_5 S^s(x, y) \gamma_5] \rangle_U, \label{equ::Kaon_MS_2pt_function}\\
    C^{K1}_{\nu \rho}(x, y, z)&=-2\, \text{Re} \langle \text{Tr} [ S^l(y, z) \gamma_\nu S^l(z, x) \gamma_5 S^s(x, 0) \gamma_\rho S^s(0, y) \gamma_5] \rangle_U, \\
    C^{K2}_{\nu \rho}(x, y, z)&=-2\, \text{Re} \langle \text{Tr} [ S^l(y, 0) \gamma_\rho S^l(0, z) \gamma_\nu S^l(z, x) \gamma_5 S^s(x, y) \gamma_5] \rangle_U.
\end{align}
$S^s(x, y)$ and $S^l(x, y)$ are now specifically the strange and light-quark propagators, respectively, with their source at $y$ and sink at $x$. The two four-point functions represent the Feynman diagrams shown in Fig.~\ref{fig::Kaon_MS_Feynman},  while the two-point function represents the propagation of a kaon. For these diagrams we use sequential as well as doubly sequential propagators. This is because we cannot use the same trick as in the case of the connected contribution in section \ref{sec::Conn}, were we used rotational invariance to reduce one of the integrals over the vertices to an integral over just the absolute distance of this vertex to the origin. For the K1 diagram, we start with a one-to-all propagator from the origin and calculate two sequential propagators over the $x$ and $y$ vertices respectively. On the other hand, for the K2 diagram a double sequential propagator, which starts at the 0 vertex and goes over the $x$ and $y$ vertices, is needed.

As with the elastic contribution, we have to project the kaon to zero momentum, which means both $x$ and $y$ are restricted to a single time slice with $x_0=-t_s/2$ and $y_0=t_s/2$, and $t_s$ is the separation time between the creation and annihilation operators of the kaon. 
In terms of the diagrams defined above, making use of the property ${\rm Tr}\,\mathcal{Q}^{(em)}=0$, we can write down the e.m.\ kaon mass splitting in the following way,
\begin{align}\label{eq:dmKdiags}
    \dmKem(\Lambda) &= 
    \lim_{t_s\to\infty}
    \frac{e^2\,
    \fQ^{\dmKem}}{2} 
    a^4 \sum_{z_0=-\infty}^\infty
    \frac{\sum_{\vec{x}, \vec{y}, \vec{z}} \delta_{\nu \rho} [\mathcal{G}(z)]_\Lambda \, (C^{K2}_{\nu \rho}(x,y,z)-C^{K1}_{\nu \rho}(x,y,z) ) }{\sum_{\vec{x}, \vec{y}} C^{2pt}(x,y)}\, , 
    \\
    \fQ^{\dmKem} &= {\rm Tr}\left\{\lambda_3\,\mathcal{Q}^{(em)\,2} \right\}
    = \frac{1}{3}.
    \nonumber
\end{align}
In practice, the integral over $z_0$ is restricted over the interval $[-t_c, t_c]$, since the contributions for the parts with larger times than the cutoff time $t_c$ are calculated using the elastic part in infinite volume (Eq.\ (\ref{eq:felast_infvol})), as mentioned at the beginning of the section. Additionally, $\dmKem$ is calculated for multiple values of $t_s$, and an exponential fit to the resulting data is used to reconstruct the value for infinite separation times.

Another challenge that arises as part of this calculation is caused by the four-point function of K2 corresponding to the diagram in Fig.~\ref{fig::Kaon_MS_Feynman_asymm}. 
When the two vector currents of this contribution are on top of each other, i.e. if $z_0=0$, a large value of the integrand is generated. It can be shown that this corresponds to a logarithmic divergence in the $z_0$-integrand of Eq.~\eqref{eq:dmKdiags} caused by short-distance artifacts that are absent in the continuum. The value of the integral is still finite, but the continuum limit becomes more challenging. As a countermeasure, we have used the neighboring points of the integrand to extrapolate to $z_0=0$ via a linear fit. This method reduces the lattice artifacts of the observable considerably and improves the continuum limit. 

\subsubsection{Lattice results for $\dmKem$}
In the following, a value of 1\;fm is chosen for the cutoff time $t_c$, since this choice ensures that (a) the regime where only the elastic term contributes has been reached, (b) finite-size effects are expected to be small and (c) the currents are far enough from the temporal boundaries of the lattice so as not to be affected by them. All of the results presented here are indeed stable under small variations of $t_c$.
\begin{figure}[t]
		\centering
		\begin{subfigure}[t]{0.49\textwidth}
			\includegraphics[width=\textwidth]{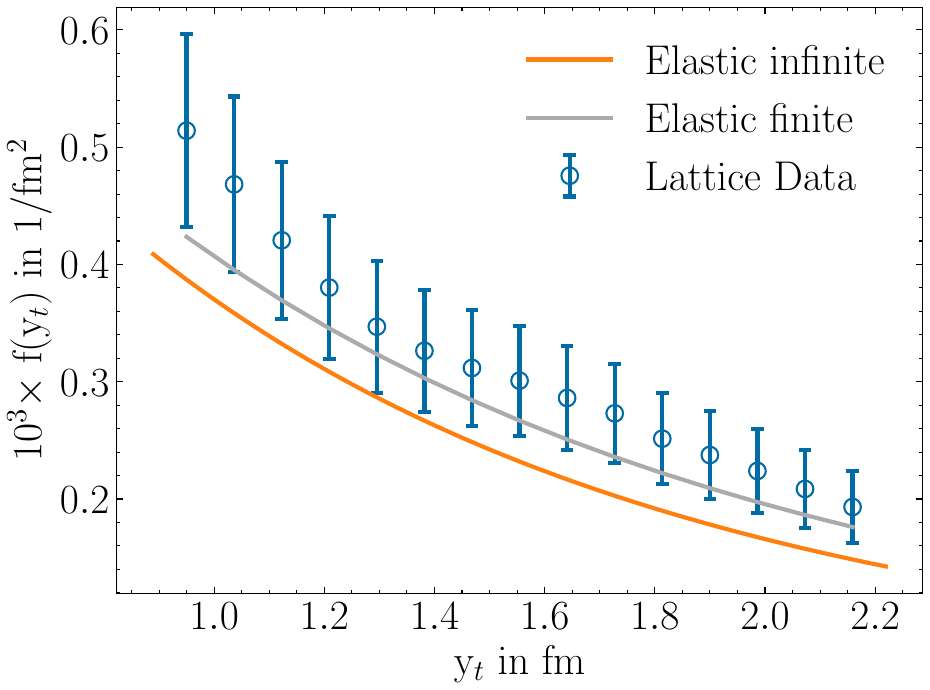}
			\caption{Elastic and lattice integrands.}  \label{fig::Kaon_MS_Method_x}
		\end{subfigure}		
		\begin{subfigure}[t]{0.49\textwidth}
			\includegraphics[width=\textwidth]{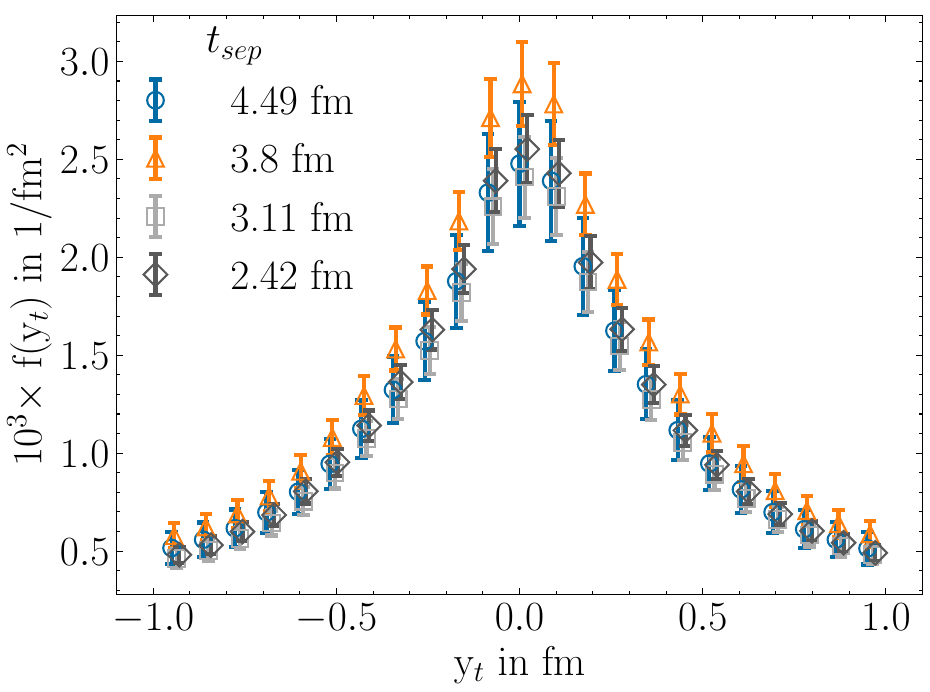}
			\caption{Data restricted to $y_t<1$ fm. The points are slightly to higher $y_t$ for visibility.}  \label{fig::Kaon_MS_Method_a}
		\end{subfigure}		
        \vspace{0.5cm}
        
        \begin{subfigure}[t]{0.49\textwidth}
			\includegraphics[width=\textwidth]{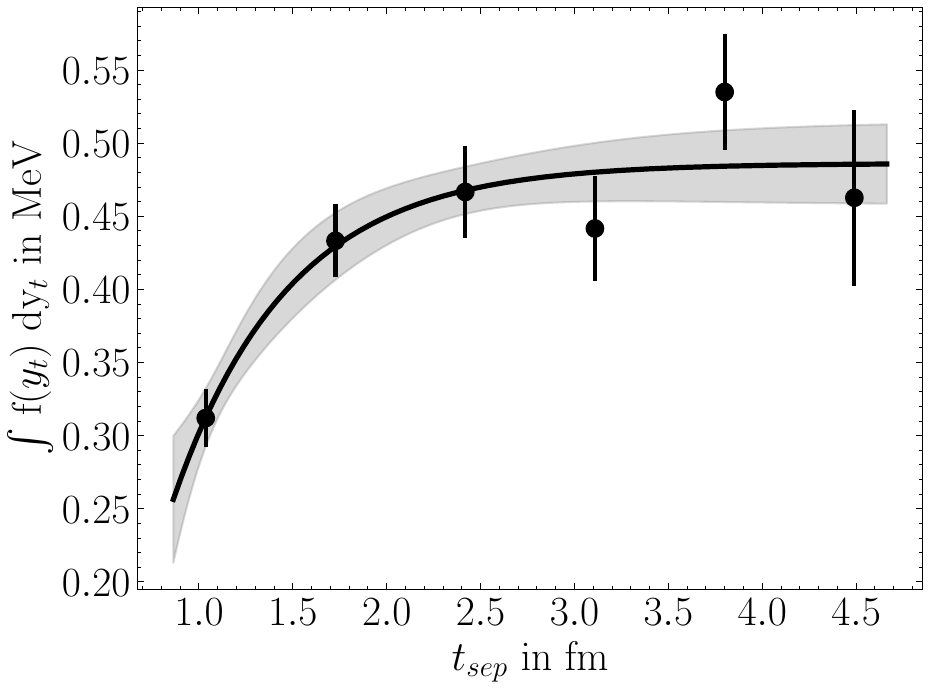}
			\caption{Extrapolation in $t_{sep}$ of restricted data.}  \label{fig::Kaon_MS_Method_b}
		\end{subfigure}
        \begin{subfigure}[t]{0.49\textwidth}
			\includegraphics[width=\textwidth]{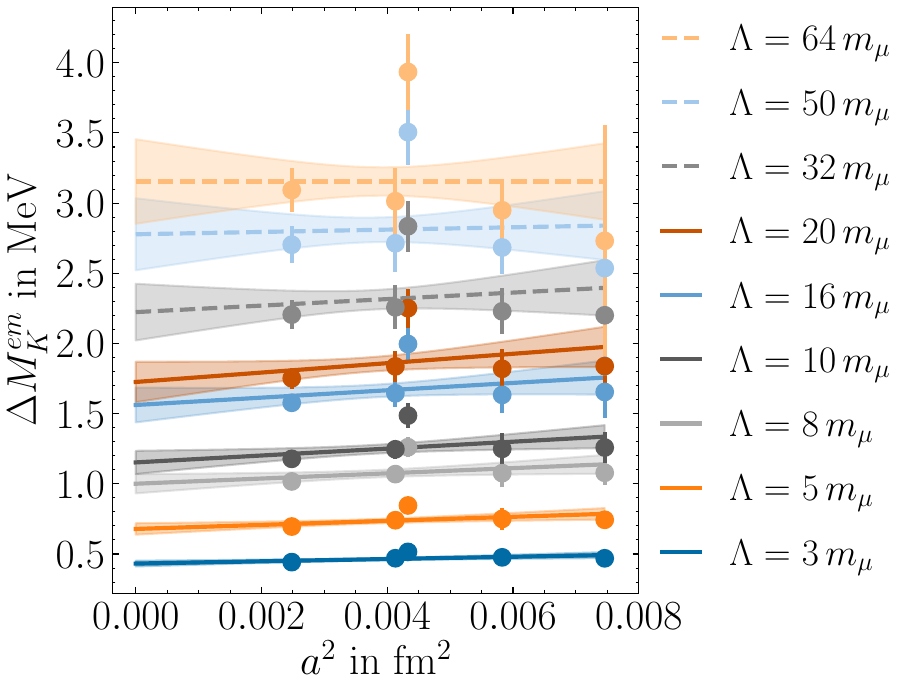}
			\caption{Continuum extrapolation.}  \label{fig::Kaon_MS_Method_c}
		\end{subfigure}
  \caption{
  (a) Comparison between the elastic part in finite and infinite volume derived in sec.~\ref{sec::Kaon_MS_elastic} and the integrand of Eq.~\eqref{eq:dmKdiags} calculated on the H101 ensemble for the largest separation time at a PV-mass of $5\, m_\mu$. 
  (b) The same integrand of H101 for different separation times with a $t_c$ value of 1 fm. (c) Extrapolation to infinite separation time of the integral over the integrands in (b).
  (d) Continuum extrapolation of the $\dmKem$ values for the different PV-masses. The points from H200 are shown with a slight offset to higher $a^2$ to make them distinguishable from the points of N202. The resulting values are reported in Table~\ref{tab::Kaon_MS_Cont}.   }
  \label{fig::Kaon_MS_Method}
\end{figure}

The three points raised above can be addressed using the first plot of Fig.~\ref{fig::Kaon_MS_Method}. It shows the integrand of Eq.~\eqref{eq:dmKdiags} for H101 at the largest separation time for large $y_t$ values as well as the elastic part, calculated with the formulas derived in section \ref{sec::Kaon_MS_elastic} for both the finite and infinite volume versions. The finite volume version is in good agreement with the data from the lattice. 

We have computed $f(x_0)$ for nine values of $\Lambda$ (see Table~\ref{tab::Kaon_MS_Cont}), since the incremental cost of including an additional value is small. In particular, beyond the ones that were used in computing 
$\atotvioem(\Lambda)$, we have included values as large as $64\, m_\mu\simeq 6.8\,$GeV in order to study the large-$\Lambda$ behavior. Note that our final result for $\atotvio$ only relies on $\Lambda$ values up to $16\, m_\mu$.

Plots (b) and (c) of Fig.~\ref{fig::Kaon_MS_Method} show an example of the methodology described previously on the H101 ensemble and with a PV-mass of $\Lambda=5\, m_\mu$.
The smallest two separation times are not shown in Fig.~\ref{fig::Kaon_MS_Method_a} in order to improve visibility. For the second-largest separation time we observe slightly larger results than for the other separation times; we interpret this as an upward statistical fluctuation.
In Fig.~\ref{fig::Kaon_MS_Method_c} the continuum limit of the kaon mass splitting value can be seen. For each value of the PV-mass, the ensembles show only very small variations, except for H200. It has significant differences when compared to the other ensembles. This can be explained by its comparatively small volume (see Table \ref{tab::LatSet_Properties}). Removing it from the extrapolation has no significant impact on the continuum result. For consistency of the data set, we still include this ensemble in the final analysis.
The continuum extrapolated values from these fits are reported in Table~\ref{tab::Kaon_MS_Cont}. The integral of the elastic part is already included for these values.

\begin{table}[ht]
	\centering
	\caption{Continuum extrapolated values of $\dmKem$. The continuum extrapolation can be seen in Fig.~\ref{fig::Kaon_MS_Method_c}.
    }
	\label{tab::Kaon_MS_Cont}
	\begin{tabular}{c c}
  \hline
		 $\Lambda/m_\mu$& $\dmKem$ [MeV] \\ \hline
		3 & $0.432 (21)$ \\ \hline
		  5 & $0.678 (41)$ \\ \hline
		  8 & $1.001 (69)$ \\ \hline
	\end{tabular}
 \hspace{0.4cm}
	\begin{tabular}{c c}
		\hline 
   $\Lambda/m_\mu$ & $\dmKem$ [MeV] \\ \hline
		10 & $1.153 (83)$ \\ \hline
		16 & $1.562 (124)$ \\ \hline
		20 & $1.726 (144)$ \\ \hline

 \end{tabular}
 \hspace{0.4cm}
	\begin{tabular}{c c}
		\hline 
   $\Lambda/m_\mu$ & $\dmKem$ [MeV] \\ \hline
		32 & $2.224 (202)$ \\ \hline
		50 & $2.778 (257)$ \\ \hline
		64 & $3.154 (302)$ \\ \hline

 \end{tabular}

\end{table}

The following section will discuss the behavior of the results for the kaon mass splitting in dependence of the PV-mass $\Lambda$.

\subsubsection{Large Pauli-Villars mass behavior of $\dmKem(\Lambda)$ \label{sec::Kaon_MS_Large_PV}}
Starting from Eq.~\eqref{equ::Kaon_MS_DelM_h} the divergent terms for the hadron mass splitting can be obtained by doing an Operator Product Expansion (OPE); see for instance section~4 of Ref.\ \cite{Biloshytskyi:2022ets}. This yields: 
\begin{align}
    \delta M_h(\Lambda)\overset{\Lambda \rightarrow \infty}{\underset{z_0\rightarrow \infty}{=}} c_m(\Lambda)\sum_f \mathcal{Q}_f^2m_f \frac{\partial M_h}{\partial m_f} - c_0(\Lambda) \left( M_h - \sum_f m_f \frac{\partial M_h}{\partial m_f} \right),
\end{align}
with 
\begin{align}
    c_m(\Lambda)&= \frac{3\alpha}{2\pi} \left(\log \frac{\Lambda}{\mu_{IR}} + O(\log \log\Lambda) \right)\\
    c_0(\Lambda)&= \frac{3\alpha}{2\pi} \left(\frac{1}{48 \pi^2b_0}\log \frac{\Lambda}{\mu_{IR}} + O(\log \log\Lambda) \right).
\end{align}
Calculating the mass splitting explicitly for the kaon results in the cancellation of many of the terms, leading to
\begin{align}
    \dmKem(\Lambda)&\stackrel{\Lambda\to\infty}{\sim} c_m(\Lambda) \fQ^{\dmKem}\,
    m_l \frac{\partial M_K}{\partial m_l}
    =\mathcal{C} \log \frac{\Lambda}{\mu_{IR}}. \label{equ::Kaon_MS_OPE_Prediction}
\end{align}
In the last step, all of the constant factors were summarized in the constant $\mathcal{C}$, which is approximately 0.12 MeV. The continuum limit for $m_l\frac{\partial M_K}{\partial m_l}$ was obtained using calculations from section \ref{sec::Kaon_MI}. Using Eq.~\eqref{equ::Kaon_MS_OPE_Prediction} the plausibility of the continuum extrapolated $\dmKem$ values in Table~\ref{tab::Kaon_MS_Cont} can be tested. 

\begin{figure}[t]
	\centering
	\includegraphics[width=0.5\textwidth]{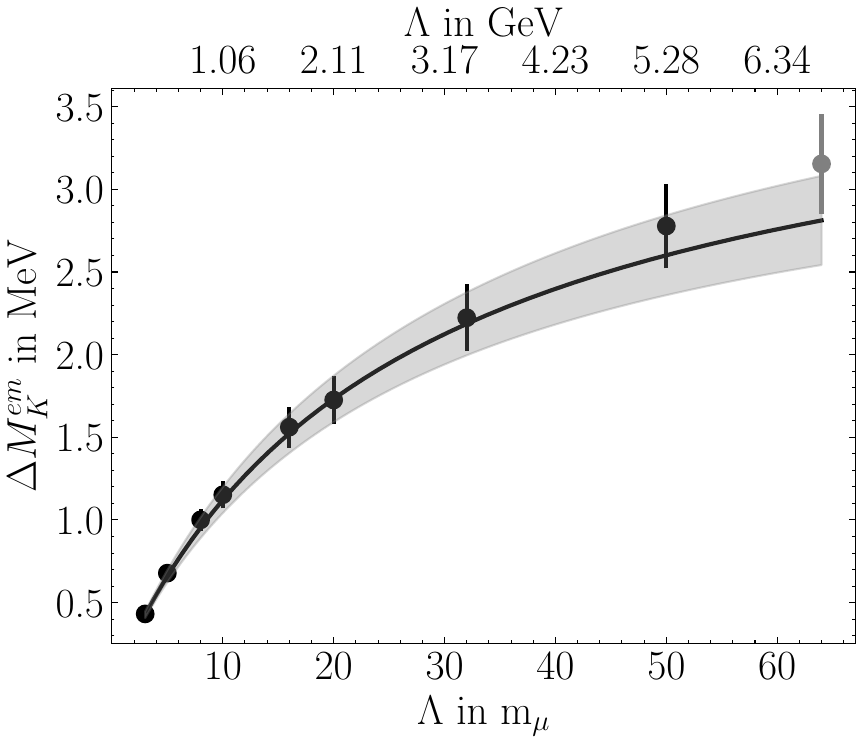}
	\caption{Plot of the continuum extrapolated values of $\dmKem$ in dependence of the Pauli-Villars mass. A fit to the data, where the fit function is given by Eq.~\eqref{equ::Kaon_MS_fit_func}, with its corresponding error bands is also shown. This fit function respects the expected behavior of $\dmKem$ as obtained by an Operator Product Expansion. The point with the largest PV-mass is marked as gray as it shows cutoff effects and is excluded from the fit.
    }
    \label{fig::Kaon_MS_PV_Extra}
\end{figure}

Fig.~\ref{fig::Kaon_MS_PV_Extra} shows the values from Table~\ref{tab::Kaon_MS_Cont} together with a fit. The fit function inspired by Eq.~\eqref{equ::Kaon_MS_OPE_Prediction} is given by 
\begin{align}
     f_{fit}(\Lambda)= c_0\, \frac{\Lambda}{\Lambda+c_1} + \mathcal{C}\, \log(\frac{\Lambda +c_2}{c_2}). \label{equ::Kaon_MS_fit_func}
\end{align}
The point with the largest PV-mass, $\Lambda=64 \, m_\mu$, was excluded from the fit, as we suspect it could be affected by cutoff effects. It is marked in gray in Fig.~\ref{fig::Kaon_MS_PV_Extra}.
Since the different values for $\dmKem$ are obtained by using a different parameter for $\Lambda$ in the calculation of the photon propagator, the results given in Table~\ref{tab::Kaon_MS_Cont} are highly correlated.
Because of this we use a correlation of 0.9 between all of the input parameters for the fit. 
This value comes from calculating the correlation of the separation-time extrapolated results on each ensemble; for all of them this correlation was consistently around 0.9.
The resulting $\chi^2/DOF$ of the fit is 0.92.
The factor in front of the logarithmic term in the fit function is set to the value which was predicted via the OPE. In addition to the logarithmic term, a second term, which is linear for small values of $\Lambda$ and constant for large values, has to be added. This fit function vanishes for $\Lambda=0$, which has to be the case since the photon propagator is zero in that limit.

We note that one can also reproduce the mass splittings with simpler fits. If only the first term of Eq.~\eqref{equ::Kaon_MS_fit_func} is used, the resulting fit has a $\chi^2/DOF$ of 0.84. Alternatively, it is also possible to use only the second term as a fit function. In that case $\mathcal{C}$ is determined by the fit, instead of setting it to 0.12 MeV. This ansatz results in a $\chi^2/DOF$ of 0.96. But the extracted value of $\mathcal{C}$ is 1.29(19) MeV, which is about one order of magnitude higher than the value predicted by the OPE.

The latter fit ansatz assumes that the dominant $\Lambda$-dependence of $\dmKem$ is given by the divergent term of the inelastic part. But appendix \ref{app::Pheno_Kaon_MS} shows that the elastic part $\dmKel$ still has a sizable dependence on the PV-mass. $\dmKel$ is defined as the part resulting from the elastic contribution to the forward Compton amplitude on the kaon. It goes from a value of 1.53 MeV for $\Lambda=16\,m_\mu$ to 2.4 MeV for infinite $\Lambda$. Comparing this to the continuum extrapolated value of $\dmKem$ in Table \ref{tab::Kaon_MS_Cont}, which is 1.562(124) MeV for $\Lambda=16\,m_\mu$, it becomes clear that our lattice results in this section are not precise enough to resolve the $\Lambda$-dependence of the divergent term underneath the corresponding dependence of the elastic contribution.

\subsection{The response of the kaon mass to an up/down quark mass splitting\label{sec::Kaon_MI}}
The electromagnetic kaon mass splitting from the previous section is part of the effort to fix the bare mass splitting of the light quarks in order to calculate the counterterm. The second part of this calculation is the mass insertion into the charged kaon propagator. For the ensembles with open boundary conditions, this calculation is relatively straightforward. Again, the two-point function from Eq.~\eqref{equ::Kaon_MS_2pt_function} is needed. Additionally, a three-point function has to be calculated,
\begin{align}
    C_{3pt}(x, y, z)=-\text{Re} \langle \text{Tr} [ S^l(y,z) \mathbb{1}S^l(z,x) \gamma_5 S^s(x,y)\gamma_5] \rangle_U.
\end{align}
The source and sink of the kaon are set to specific time slices and projected to zero momentum, with a setup similar to our calculation of the kaon mass splitting: $x_0=-t_s/2$ and $y_0=t_s/2$. Ideally the matrix element would be calculated with
\begin{align}
    \langle K^+ | \Bar{u}u-\Bar{d}d | K^+ \rangle = \lim_{t_s \rightarrow \infty} \frac{a^3\sum_{\vec{y},\vec{z}} C_{3pt}(x,y,z)}{\sum_{\vec{y}} C_{2pt}(x,y)} \Bigg|_{z_0=0}. \label{equ::Kaon_MI_Matrix1}
\end{align}
In practice there are two methods. For the first one, the expression of Eq.~\eqref{equ::Kaon_MI_Matrix1} is implemented as closely as possible by calculating the $n$-point functions for as large separation times as possible. Following this, the sum over the spatial components of $z$ is carried out and a constant fit to the central plateau in the coordinate $z_0$ is performed: the $z_0$ values entering the fit are chosen between the source and sink, sufficiently far away from $-t_s/2$ and $t_s/2$.

The second method instead uses a different approach. First the sum over the lattice data for different separation times is computed:
\begin{align}
    S(t_s)= a^4\sum_{z_0= -t_s/2+a}^{t_s/2-a} \frac{\sum_{\vec{y},\vec{z}} C_{3pt}(x,y,z)}{\sum_{\vec{y}} C_{2pt}(x,y)}. \label{equ::Kaon_MI_Summation1}
\end{align}
It is important to note that the sum goes over all $z_0$ values between the source and sink. This sum follows the relation \cite{koponen2022isovectoraxialformfactor}
\begin{align}\label{eq:summation_method_fit}
    S(t_s)=b+t_s\, \langle K^+ | \Bar{u}u-\Bar{d}d | K^+ \rangle + \mathcal{O}(e^{-\Delta \, t_s})
\end{align}
with $\Delta$ the relevant mass gap,
which means the matrix element can be obtained via a linear fit to the calculated values of $S(t_s)$. In the steps following this section, we use the values obtained from fitting the plateau to a constant, while the results of the summation method will serve as a cross-check for these values.

For lattices with periodic boundary conditions, which in the present calculations only affects the ensemble B450, an additional treatment of the two-point function becomes necessary due to wrap-around effects. In particular, at $t_s$ equal to half the time extent of the lattice, these effects result in an additional factor of two. To correct for this, an exponential fit that takes into account the periodicity is performed to the lattice data sufficiently far away from the source position of the kaon. 
The ratio between this exponential fit function and the equivalent function in infinite volume, where no wrap-around effects are present, can be expressed via
\begin{align}
    \frac{C^{fit}_{2pt}(x,y)}{C_{2pt}^{fit,\infty} (x,y)}\overset{t_s\to\infty}{\approx} 1+ e^{-M(T-2t_s)}.
\end{align}

The parameter $M$ is taken from the aforementioned fit and $T$ is  the time extend of the ensemble, which for B450 is $64 a$. The ratio is used to correct the normalizing factor for the matrix element in Eq.~\eqref{equ::Kaon_MI_Matrix1}. With the additional steps, B450 reaches the same precision as the other ensembles which can be seen in Table~\ref{tab::Kaon_MI_Results}.

\begin{figure}[ht]
		\centering
		\begin{subfigure}[t]{0.49\textwidth}
			\includegraphics[width=\textwidth]{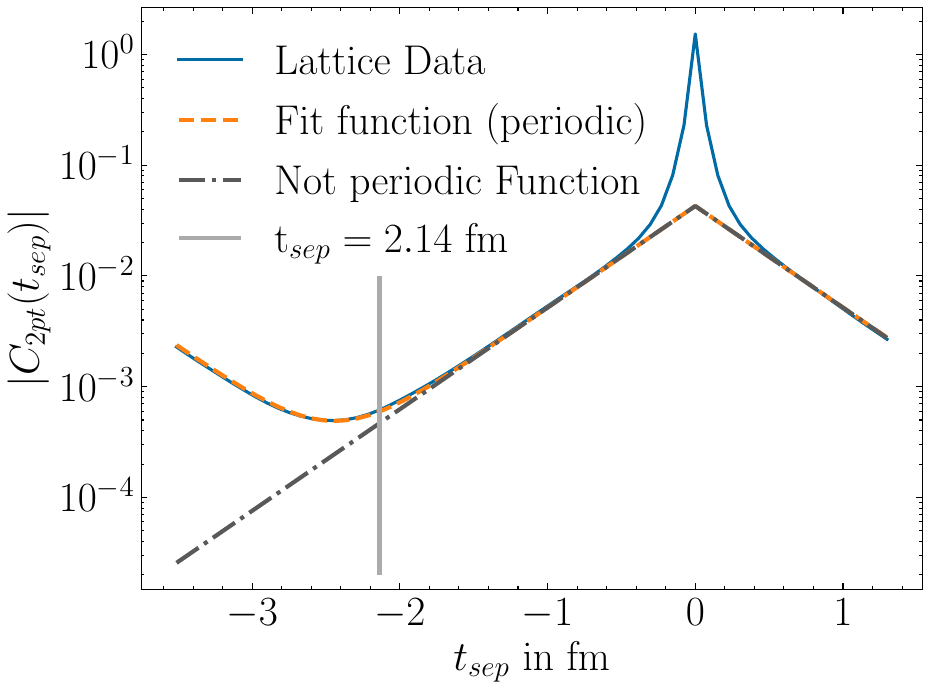}
			\caption{Calculation of the non periodic two-point function for B450.}  \label{fig::Kaon_MI_B450_2pt}
		\end{subfigure}		
        \begin{subfigure}[t]{0.49\textwidth}
			\includegraphics[width=\textwidth]{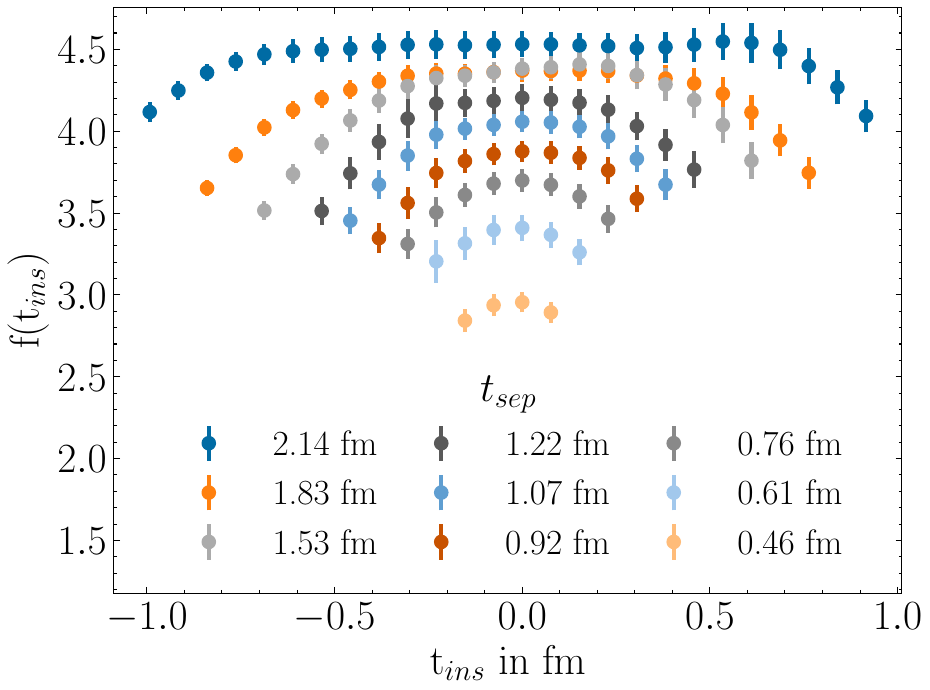}
			\caption{Lattice data for different separation times for B450.}  \label{fig::Kaon_MI_B450_Data}
		\end{subfigure}
  \caption{The data of B450 used for the fits seen in Fig.~\ref{fig::Kaon_MI_B450_Fits}. The step on the left side is only necessary for B450 since the other ensembles have open boundary conditions in time.}
  \label{fig::Kaon_MI_B450}
\end{figure}

\begin{figure}[ht]
		\centering
		\begin{subfigure}[t]{0.49\textwidth}
			\includegraphics[width=\textwidth]{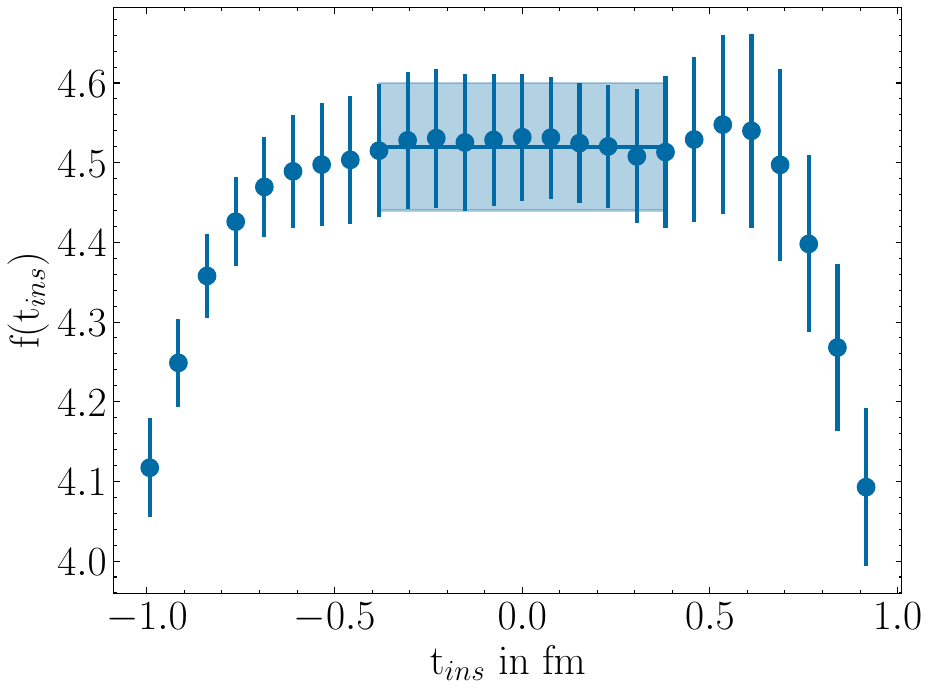}
			\caption{Constant fit.}  \label{fig::Kaon_MI_B450_Constant_Fit}
		\end{subfigure}		
        \begin{subfigure}[t]{0.49\textwidth}
			\includegraphics[width=\textwidth]{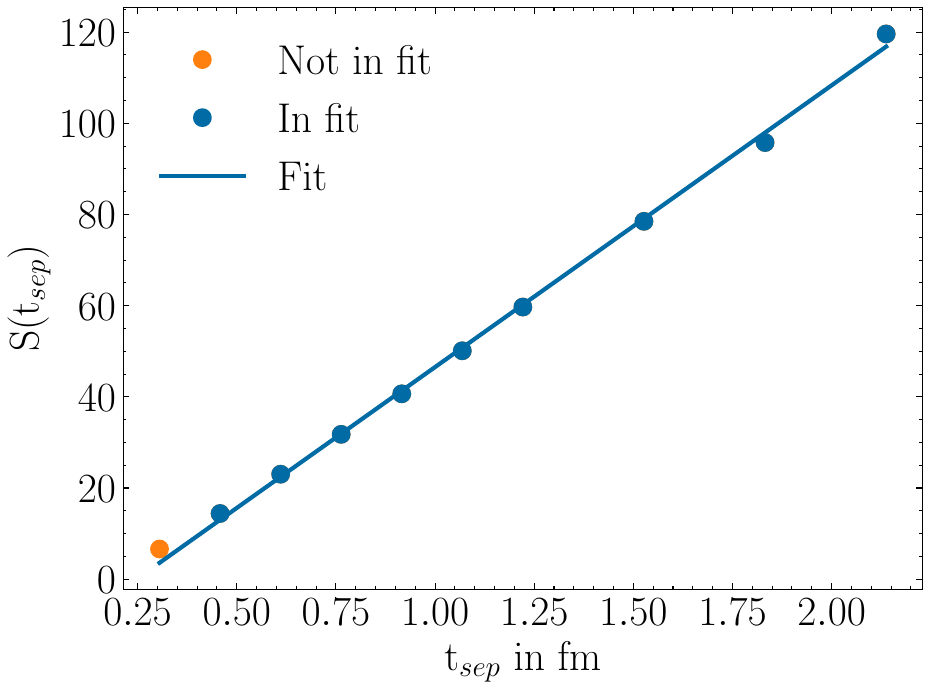}
			\caption{Summation method.}  \label{fig::Kaon_MI_B450_Summation_Method}
		\end{subfigure}
  \caption{Fits performed to extract the kaon isovector scalar matrix element from the lattice data of B450. (a) Constant fit for the largest calculated source-sink separation $t_s$, together with the error bands of the fit; see Eq.\ (\ref{equ::Kaon_MI_Matrix1}). (b) Fit of the summation method, Eq.\ (\ref{eq:summation_method_fit}), the kaon matrix element corresponding to the slope. The resulting values for both methodologies and all ensembles are collected in Table~\ref{tab::Kaon_MI_Results}.}
  \label{fig::Kaon_MI_B450_Fits}
\end{figure}

\subsubsection{Lattice results for $\partial\dmK/\partial(m_u-m_d)$ and comparison with chiral perturbation theory}

Fig.~\ref{fig::Kaon_MI_B450_2pt} shows the previously  described way of calculating the two-point function for the lattices with periodic boundary conditions. The difference between the periodic fit and the non periodic function is clearly visible at separation times, which are relevant for our calculation. On the right-hand side, Fig.~\ref{fig::Kaon_MI_B450_Data} shows the result for the ratio between the two- and three-point function for different values of $t_s$ for the ensemble B450. One can observe the trend that the data series approaches a plateau for larger and larger separation times. 
The fits for the two ways of extracting the matrix element can be seen in Fig.~\ref{fig::Kaon_MI_B450_Fits}.
The values obtained for all ensemble are provided in Table~\ref{tab::Kaon_MI_Results}.
There is very good agreement between the results obtained by the constant fit method and the summation method for all ensembles except B450. It has a bit less than $2\sigma$ discrepancy between the two values. Because of the periodic boundary conditions of B450, the maximal separation times are much more limited compared to the other ensembles. This is a challenge for both methods.
Based on the following two observations we continue using the result of the constant fit method for B450 without further modifications. Firstly, if one performs an extrapolation to infinite separation time with the data in Fig.~\ref{fig::Kaon_MI_B450_Data} one recovers up to rounding errors the result of the constant fit at the largest separation time. 
Secondly, the change in the results quoted in Table~\ref{tab::R_Ratio_Fits}
is one order of magnitude smaller than the errors, if one just changes to the summation method for B450. 
If the method is changed for all ensembles the results change even less.

\begin{figure}[t]
		\centering
		\begin{subfigure}[t]{0.49\textwidth}
			\includegraphics[width=\textwidth]{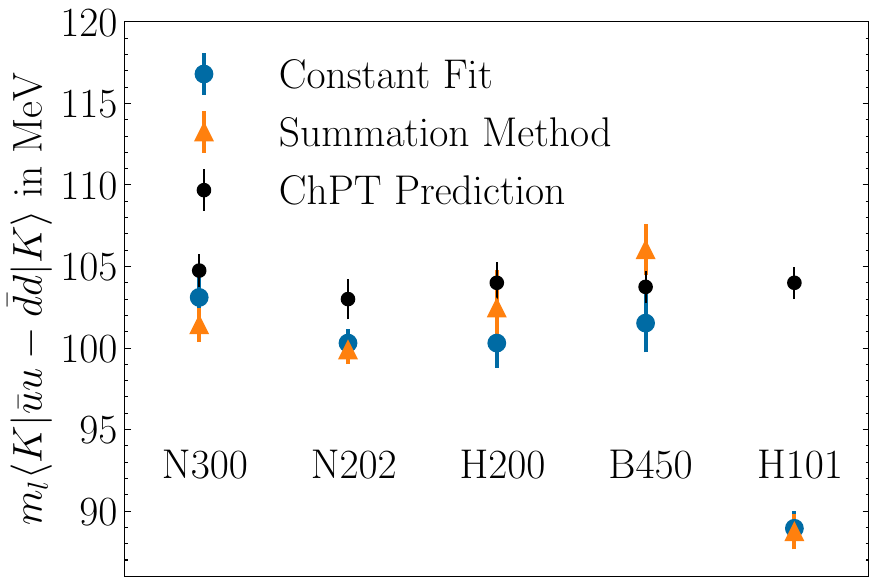}
		\end{subfigure}		

  \caption{Comparison between the ChPT prediction from Eq.~\eqref{equ::Kaon_MI_ChPT_Prediction} and the calculated values of the matrix element for each ensemble. }
  \label{fig::Kaon_MI_Check}
\end{figure}

\begin{table}[b]
	\centering
 \caption{Values of the light-quark mass derivative of the kaon mass for the different ensembles.} \label{tab::Kaon_MI_Results}
	\begin{tabular}{c c c c c c}
		\hline
		
	$\partial \Delta M_K/\partial(m_u-m_d)$	 & N300 & N202 & H200 & B450 & H101 \\ \hline
		Constant fit &  $4.90 (6)$  & $4.67(4)$ & $4.69(7)$ & $4.52(8)$ & $4.11 (5)$\\ \hline
		Summation Method & $4.82 (5)$ & $4.65 (4)$ & $4.77(11)$ & $4.72 (7)$ & $4.10 (5)$ \\ \hline
	\end{tabular}
\end{table}

One can use a leading-order chiral perturbation theory (ChPT) prediction in order to check the plausibility of the values calculated in this section. If $m_l$ is the average up/down quark mass, the prediction can be expressed as
\begin{align}
     m_l\;\frac{\partial\dmK}{\partial (m_u-m_d)}\approx \frac{M_\pi^2}{4M_K}. \label{equ::Kaon_MI_ChPT_Prediction}
 \end{align}
This particular quantity can also be obtained from our lattice results.
Until now, we have only used the bare (subtracted) versions of the light-quark masses $m_l$ and the operator~$(\Bar{u}u - \Bar{d}d)$. But for this consistency check we need their improved and renormalized versions. From \cite{Bhattacharya:2005rb} we get the following relation for the product of these two at the SU$(3)_f$-symmetric point:
\begin{align}
    [m_l  (\Bar{u} u - \Bar{d} d)]_{I, R} = r_m \, m_l (\Bar{u} u - \Bar{d} d) [1+ \mathcal{O}(am_l)].
\end{align}
We will not consider the $\mathcal{O}(am_l)$ effects, which means we only need to multiply our results with $r_m$. The necessary values of $r_m$ can be extracted from \cite{Heitger:2021bmg}. The comparison between the calculated values and the ChPT prediction are shown in Fig.~\ref{fig::Kaon_MI_Check}. The predictions and their respective uncertainties are calculated from the masses found in Table~\ref{tab::LatSet_Properties}. All ensembles except for H101 show agreement between the calculated values and the prediction. H101 is the ensembles with the largest lattice spacing, which means with the largest $\mathcal{O}(am_l)$ effects. 
From this we can conclude that our results of this part of the counterterm are plausible. 

\subsection{The light-quark mass derivative of the HVP \label{sec::HVP_MI}}
The last missing ingredient to obtain the counterterm is the light-quark mass derivative of the HVP. In comparison to the CCS method from sections \ref{sec::Conn} and \ref{sec:disco}, here, the time momentum representation (TMR) of the HVP is used \cite{Bernecker:2011gh}. Additionally, for the data in this section stochastic wall sources were used in order to increase statistics~\cite{Ce:2022kxy}. The master formula for this observable is
\begin{align}
\frac{\partial \atotvio }{\partial (m_u-m_d)}&=
\frac{1}{2} \frac{\partial \ahvp }{\partial (m_u-m_d)}
= \left( \frac{\alpha}{\pi}\right)^2 \int_0^{\infty} dx_0 \ w(x_0) G_{3pt}(x_0). \label{equ::HVP_MI_Master}
\end{align}
The kernel $w(x_0)$ is taken from the appendix of \cite{DellaMorte:2017dyu} and $G_{3pt}(x_0)$ is given by 
\begin{align}
    G_{3pt}(x_0)= -\frac{2}{3} \fQ^{3pt} a^7\sum_{i=1}^3 \sum_{\vec{x}} \sum_z  \text{Re} \langle \text{Tr} [ S(0,x) \gamma_i S(x,z) \mathbb{1} S(z,0)\gamma_i] \rangle_U. \label{equ::HVP_MI_3pt}
\end{align}
In this case the charge factor $\fQ^{3pt}=({1}/{2})\,{\rm Tr}\{{\cal Q}^{(3)} {\cal Q}^{(em)}{}^2\}=1/12$. This three-point function has large uncertainties for large values of $x_0$. However, it is possible to derive a relation between the three-point function and the two-point function\footnote{ At the SU(3)$_{\rm f}$ symmetric point, the expression 
$2\left( \frac{\alpha}{\pi}\right)^2 \int_0^{\infty} dx_0 \ w(x_0) G_{2pt}(x_0)$ yields the isoQCD contribution of the up, down and strange quarks to $\ahvp$.}
\begin{align}
    G_{2pt}(x_0)= \frac{1}{3} \fQ^{2pt} a^3\sum_{i=1}^3 \sum_{\vec{x}}  \text{Re} \langle \text{Tr} [ S(0,x) \gamma_i S(x,0)\gamma_i] \rangle_U, \label{equ::HVP_MI_2pt}
\end{align}
which can be calculated with higher accuracy ($\fQ^{2pt}=(1/2)\,{\rm Tr}\{ {\cal Q}^{(em)}{}^2\}=1/3$). For large $x_0$, and in our range of box sizes $L\lesssim 4$\,fm, the two-point function is approximated by
\begin{align}
    G_{2pt}(x_0)\overset{x_0\to\infty}{\approx} A\, e^{-M x_0}. \label{equ::HVP_MI_2pt_approx}
\end{align}
Keeping in mind that both $A$ and $M$ depend on the light-quark mass difference, Eq.~\eqref{equ::HVP_MI_2pt_approx} can be used to derive a similar approximation for the three-point function
\begin{align}
    G_{3pt}(x_0)=\frac{\partial G_{2pt}(x_0)}{ \partial (m_u-m_d)} \overset{x_0\to\infty}{\approx}(A' -B' x_0) e^{-M x_0}. \label{equ::HVP_MI_3pt_approx}
\end{align}

\begin{figure}[t]
	\centering
	\begin{subfigure}[t]{0.49\textwidth}
		\includegraphics[width=\textwidth]{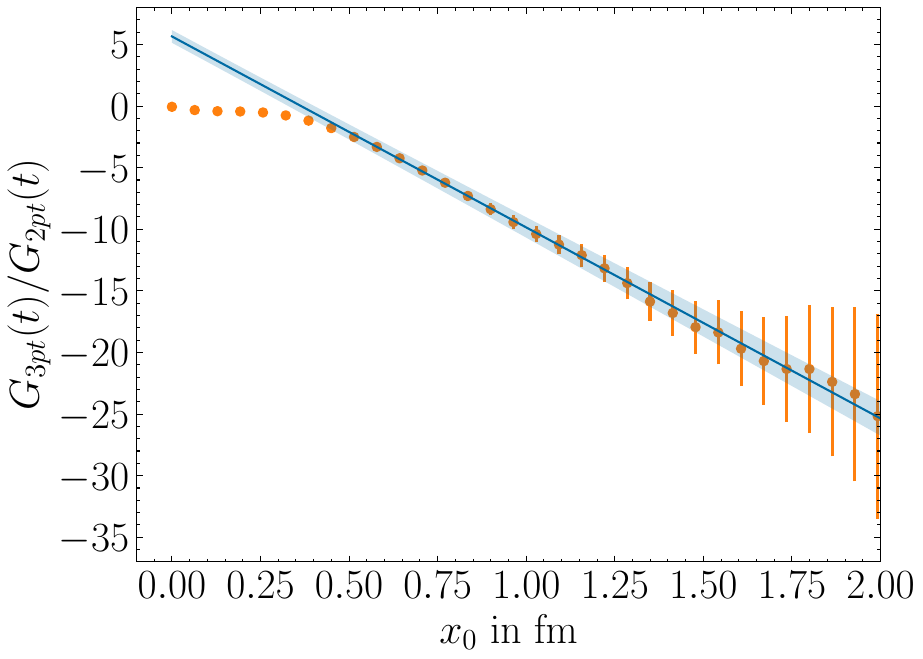}
  \caption{Linear fit to the ratio of the two- and three-point function.}
  \label{fig::HVP_MI_Ratio_Fit}
		\end{subfigure}
		\begin{subfigure}[t]{0.49\textwidth}
			\includegraphics[width=\textwidth]{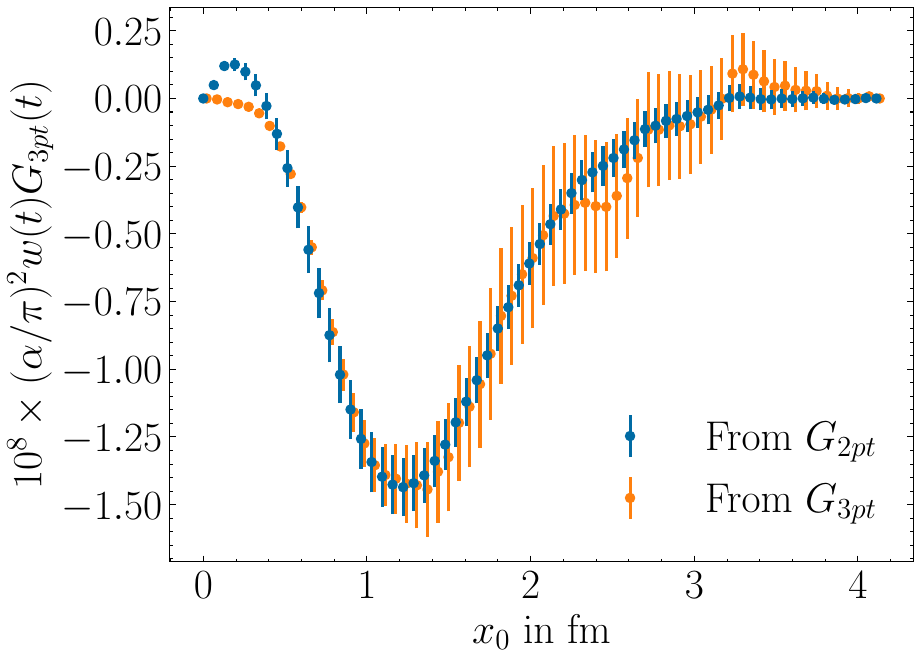}
      \caption{Comparison between the two methods of extracting the three-point function}   
  \label{fig::HVP_MI_Integrand_Compare}
		\end{subfigure} 
  \caption{(a) Example of the linear fit to the ratio of Eqs. \eqref{equ::HVP_MI_2pt_approx} and \eqref{equ::HVP_MI_3pt_approx} on N202. (b)~Integrand of equation \eqref{equ::HVP_MI_Master} for N202. The yellow dots were obtained by calculating the three-point function directly, while the blue dots were calculated form the two-point function, by multiplying it with the linear term from the fit on the left hand side. } 
  \label{fig::HVP_MI_Integrand}
 \end{figure}

Here, it is important to note that the exponents of the approximations are the same. So the values of the three-point function can be obtained from the two-point function by multiplying it with a first-order polynomial in $x_0$. The coefficients of this term can be obtained from a linear fit to the ratio of the three- and two-point functions.

This ratio together with a fit is shown in Fig.~\ref{fig::HVP_MI_Ratio_Fit} for the N202 ensemble. The upper limit of the fit range was set to 2 fm. Multiple models using different lower limits were considered. The weights by which they contribute were determined using the Akaike Information Criterion \cite{Bozdogan_1987, Kimura_2017}. The right-hand side of the same figure, Fig.~\ref{fig::HVP_MI_Integrand_Compare}, depicts the integrand of Eq.~\eqref{equ::HVP_MI_Master} on the same ensemble, both from the three-point function directly and from the two-point function using the fit from the left hand side. For very small values of $x_0$, where higher mass states are still important, there is a sizable discrepancy between the two methods. But already at 0.5 fm both methods have equal values within the error. At first the uncertainties of the values from the three-point function are smaller, but around 1 fm this switches and stays this way until the upper limit. The other ensembles have the same behavior. 

Based on these observations we will use a hybrid method for the results presented in this section and for the subsequent calculations. For the integrand below 1\;fm, the three-point function from Eq.~\eqref{equ::HVP_MI_3pt} is used directly. For $x_0$ values larger than that, we use the approximation from the two-point function of Eq.~\eqref{equ::HVP_MI_2pt}. The results are collected in Table~\ref{tab::HVP_MI_Results}.

Similar to section \ref{sec::Kaon_MI} the periodic boundary conditions in the time direction of B450 need to be accounted for. In comparison to the other ensembles the integral is not saturated when the maximal time extend is reached. We use the same methodology as before, where a fit respecting the periodicity is performed to the two-point function. This is then used to calculate the corresponding non-periodic version of the two-point function, which is fed into the integral. Here, the correction is small compared to the error, the value of the integral goes from $-6.74(36)$ 1/MeV to $-6.89(37)$ 1/MeV, but this change is still noticeable during the continuum extrapolation of $R_{38K}$, slightly reducing the $\chi^2$.

\begin{table}[t]
	\centering
 \caption{Values of the light-quark mass difference derivative of the HVP for the different ensembles, as well as the ratio $R_{38K}$ obtained using the kaon matrix element values from the constant fit.} 
 \label{tab::HVP_MI_Results}
	\begin{tabular}{l c c c c c}
		\hline
		
		 & N300 & N202 & H200 & B450 & H101 \\ \hline
		$10^{11} \times \frac{\partial\alovio}{\partial(m_u-m_d)} \phantom{\bigg|}$\,[1/MeV] &  $-7.37 (79)$ & $-9.32 (85)$ & $-7.05 (93)$ & $-6.89 (37)$ & $-7.58 (64)$\\
        $10^{11} \times R_{38K}$ [1/MeV] & $-1.50 (17)$ & $-1.99 (19)$ & $-1.50 (20)$ & $-1.52 (9)$ & $-1.84 (16)$\\ 
        \hline 
	\end{tabular}
\end{table}

\subsubsection{The ratio $R_{38K}$, and the strong IB contribution in the FLAG24 scheme}

An important physical quantity is the ratio $R_{38K}$ introduced in Eq.~\eqref{equ::HVP_MI_R38}.
 We recall that it expresses the response of the vacuum polarization contribution to a kaon mass splitting, at fixed isospin-averaged pion, kaon and scale setting quantity (typically a baryon mass).
The results from this section and section \ref{sec::Kaon_MI} can be used in order to calculate this ratio for each gauge ensemble. It is the ratio between the light-quark mass derivative of the HVP and the light-quark mass derivative of the kaon mass (see Eq.~\eqref{equ::HVP_MI_R38} and \eqref{eq:dmKdm}). The resulting values are given in Table~\ref{tab::HVP_MI_Results}.

\begin{figure}[t]
		\centering

			\includegraphics[width=0.49\textwidth]{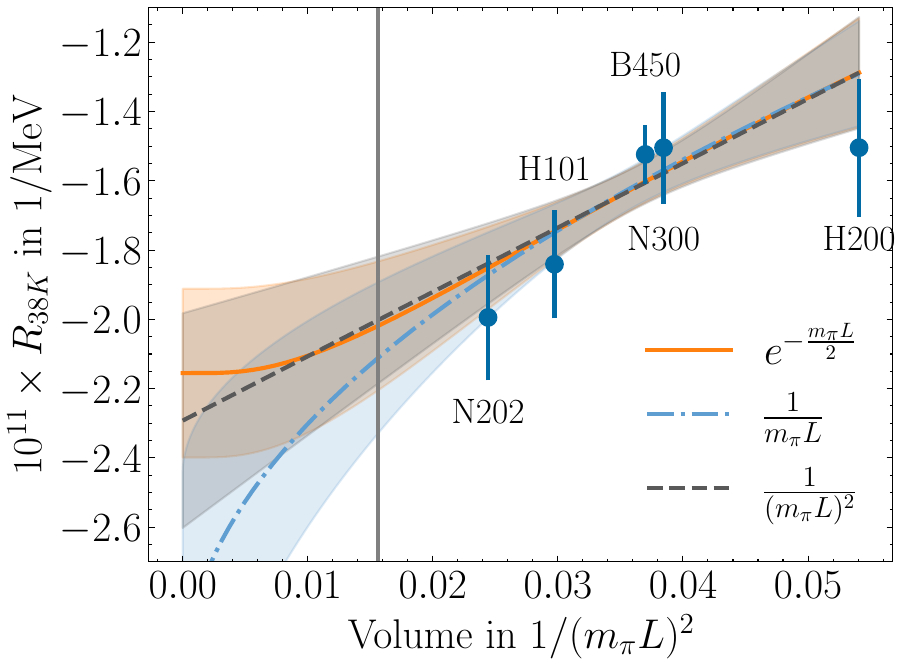}
    \caption{Extrapolation of $R_{38K}$ to infinite volume. The vertical line is located at the reference volume of $m_\pi L=8$. Three different fits using the general ansatz of Eq. \eqref{equ::R_Ratio_Fit_Func} are shown. The labels give the volume term for each fit respectively. The $\chi^2/DOF$ of the fits toghether with their values at the reference volume are written in Table \ref{tab::R_Ratio_Fits}. }
  \label{fig::HVP_MI_R38K}
 \end{figure}

At fixed physical volume, we do not observe any statistically significant dependence on the lattice spacing (compare in particular the results on ensembles B450 and N300), while the volume dependence is clearly significant: see Fig.~\ref{fig::HVP_MI_R38K}.
We therefore employ a fit ansatz for extrapolating to infinite volume of the form 
\begin{align}
    f_{fit}(m_\pi L) = c_0 + c_1 \, f_{vol}(m_\pi L) \;.\label{equ::R_Ratio_Fit_Func}
\end{align}
Asymptotically at large $L$, one would expect the volume term to be given by $f_{vol}(m_\pi L) = e^{-m_\pi L}$, but this expression does not describe our data well. Instead we use three different ansätze given in the first column of Table \ref{tab::R_Ratio_Fits}. All of them have $\chi^2/
DOF$ close to one. We will use them to extrapolate $R_{38K}$ to a reference volume of $m_\pi L=8$. The last part of the extrapolation to infinite volume is then given as a correction calculated in scalar QED. For the value at the reference volume, we take the mean of the extrapolations, which gives us 
\begin{align}
    R_{38K}(m_\pi L =8)=-2.04(22)_{\rm stat}(10)_{\rm syst} \times 10^{-11}\, {\rm MeV}^{-1}\;.
\end{align}
The correction to get to infinite volume vanishes completely within the error and has a value of  $-0.006 (60) \, {\rm MeV}^{-1} \times 10^{-11}$. The error was chosen in order to have a very conservative estimate. 
With this, the continuum extrapolated result of $R_{38K}$ is given by 
\begin{align}\label{eq:R38Klatresult}
R_{38K}   = -2.05 (22)_{\rm stat}(12)_{\rm syst} \times 10^{-11} \, {\rm MeV}^{-1}. 
\end{align}

\begin{table}[t]
	\centering
	\caption{$\chi^2/DOF$ and value at the reference volume of the different fit ansätze in Fig~\ref{fig::HVP_MI_R38K}. The general form of the fit functions can be found in Eq.~\eqref{equ::R_Ratio_Fit_Func}.}
	\label{tab::R_Ratio_Fits}

 \begin{tabular}{c c c}
		\hline
		
		  $f_{vol}$ Ansatz & $\chi^2/DOF$ & $10^{11} \times R_{38K}(m_\pi L =8)$ [1/MeV] \\ \hline
		  $e^{-\frac{m_\pi L}{2}}$ & 1.08 & $-2.02 (19)$ \\ \hline		
		  $\frac{1}{m_\pi L}$ & 0.95 & $-2.11 (22)$ \\ \hline		
		  $\frac{1}{(m_\pi L)^2}$ & 1.16 & $-2.00 (19)$ \\ \hline		
	\end{tabular}
\end{table}

In the FLAG24 report~\cite{FlavourLatticeAveragingGroupFLAG:2024oxs}, specific values for the reference hadron masses $(M_{\pi^+},M_{K^+},M_{K^0})$ are given in a QCD world with $m_u\neq m_d$, electromagnetism being `switched off'. These choices correspond to a scheme for separating electromagnetic from strong isospin-breaking effects. Since $\dmK=-6$\,MeV per definition in that `world', the strong IB contribution to $\atotvio$ is given by 
\begin{align}
    \atotviosib = -6\,{\rm MeV}\times R_{38K} = 12.30 (1.32)_{\rm stat}(72)_{\rm syst} \times 10^{-11}.
\end{align}

 \subsection{The quantity $\atotvio(\Lambda)$, continuum extrapolated  at fixed $\Lambda$\label{sec::Total}}

\begin{figure}[t]
    \centering
    \begin{subfigure}{0.49\textwidth}
        \includegraphics[width=\textwidth]{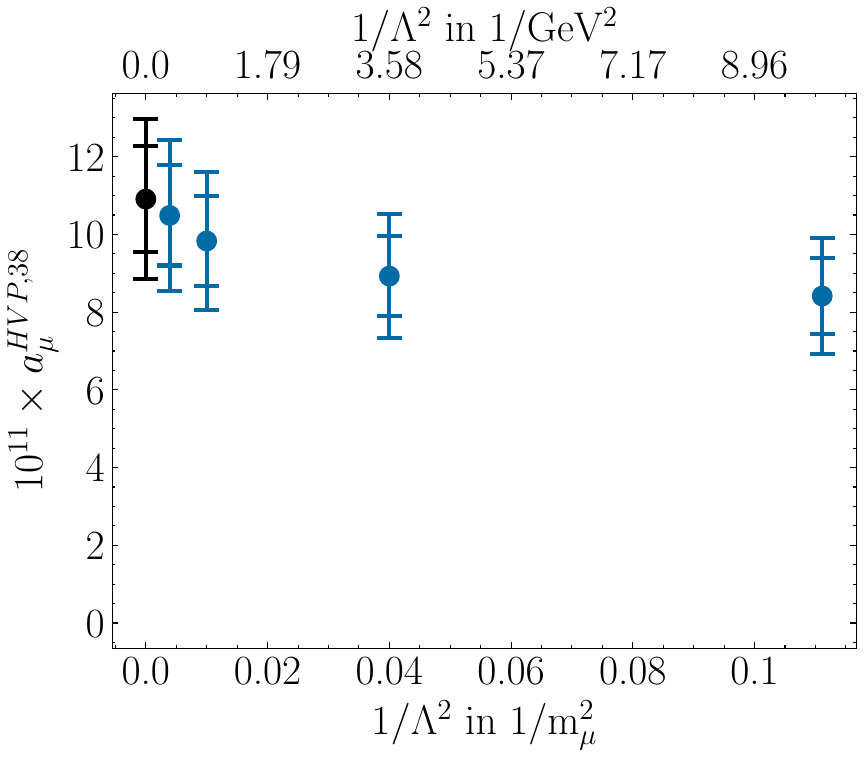}
    \end{subfigure}
\caption{PV-mass extrapolation of $\atotvio$. The inner horizontal bars show the statistical error, while the outer bar shows the additional systematic error. The black dot shows the result of a extrapolation linear in $1/\Lambda^2$ using the two points with the larges PV-masses. A correlation coefficient of one was assumed between them.}
\label{fig::Total_Extrapolations}
\end{figure}

We can now combine the results of the previous sections in order to calculate the total contribution $\atotvio$ based on Eqs.\ (\ref{equ::Master_amu38}--\ref{equ::Master_Counterterm}).
For that we use the continuum extrapolated values of $\atotvio(\Lambda)$, $\dmKem(\Lambda)$ and $R_{38K}$, which are summarized in Table~\ref{tab::Total_Cont_Extra_Values}. It is important to note that $\atotviotpta$ and $R_{38K}$ in rows three and five are independent of the PV-mass. The last two rows show the counterterm and the total contribution $\atotvio$ in dependence of $\Lambda$, respectively. From this one can observe that the result and its error are completely dominated by the counterterm. The bare electromagnetic contributions $\atotvioconn$ and $\atotviotpta$ are only small corrections which vanish within the error. Still, the data provides a hint that the connected part $\atotvioconn$ cancels out the logarithmic divergence of the counterterm, bringing the total result of the smallest PV-mass and of the largest PV-mass closer together. 
Fig.~\ref{fig::Total_Extrapolations} shows a plot of the residual dependence on $\Lambda$ of the total contribution $\atotvio$. The extrapolation to infinite PV-mass is very flat, all of the points being equal to one another within the errors.

\begin{table}[b]
	\centering
	\caption{Continuum extrapolated values of the different contributions to $\atotvio$. All values except $\dmKem$ are multiplied by $10^{11}$.}
	\label{tab::Total_Cont_Extra_Values}

\resizebox{\textwidth}{!}{%
 \begin{tabular}{c c c c c}
		\hline
		
		  PV-mass & $\Lambda = 3\, m_\mu $ & $\Lambda = 5\, m_\mu $ & $\Lambda = 10\, m_\mu $ & $\Lambda = 16\, m_\mu $  \\ \hline
		  $\atotvioconn (\Lambda) \phantom{\Big|}$ & $-0.006(39)_{\rm st.} $ & $0.002(74)_{\rm st.} $ & $-0.068(193)_{\rm st.} $ & $-0.249(321)_{\rm st.} $ \\ \hline
		  $\atotviotpta \phantom{\Big|} $ & $-0.53(17)_{\rm st.}$ & $-0.53(17)_{\rm st.}$ & $-0.53(17)_{\rm st.}$ &  $-0.53(17)_{\rm st.}$ \\ \hline
		  $\dmKem \phantom{\Big|}$[MeV] & $0.432(21)_{\rm st.}$ & $0.678(41)_{\rm st.}$ & $1.153(83)_{\rm st.}$ & $1.562(124)_{\rm st.}$  \\ \hline
		  $R_{38K} \phantom{\Big|}$[1/MeV]& $-2.05(22)_{\rm st.}(12)_{\rm sy.}$ & $-2.05(22)_{\rm st.}(12)_{\rm sy.}$ & $-2.05(22)_{\rm st.}(12)_{\rm sy.}$ &  $-2.05(22)_{\rm st.}(12)_{\rm sy.}$ \\ \hline
		  $C_T (\Lambda)\phantom{\Big|}$ & $ 8.95 (97)_{\rm st.}(53)_{\rm sy.} $ & $ 9.45 (1.02)_{\rm st.}(56)_{\rm sy.} $ & $ 10.43 (1.14)_{\rm st.}(61)_{\rm sy.} $ & $ 11.27 (1.24)_{\rm st.}(66)_{\rm sy.} $ \\ \hline
		  $\atotvio(\Lambda)\phantom{\Big|}$ & $ 8.41 (98)_{\rm st.}(53)_{\rm sy.} $ & $ 8.93 (1.04)_{\rm st.}(56)_{\rm sy.} $ & $ 9.83 (1.16)_{\rm st.}(61)_{\rm sy.} $ & $ 10.49 (1.29)_{\rm st.}(66)_{\rm sy.} $ \\ \hline
	\end{tabular}
    }

\end{table}

We remark that in continuum field theory, typically the approach to the $\Lambda=\infty$ limit occurs with leading O$(1/\Lambda^2)$ corrections. In the future, it could be interesting to evaluate the elastic contribution to the kaon mass splitting directly at $\Lambda=\infty$, maintaining the regulator only for the inelastic part, the idea being to reduce any residual $\Lambda$ dependence in $\atotvio(\Lambda)$ even further.

\subsection{Comparison of lattice results and model estimates \la{sec:compa_lat_model}}

Appendix \ref{app::Pheno} provides  estimates from a hadronic model for the various quantities computed on the lattice. For convenience we have collected them in Table \ref{tab::Total_Pheno_Prediction}, where for the $\Lambda$-dependent quantities we have set $\Lambda=16\, m_\mu$ in the first line. In the second line the PV-mass is infinite. 
The main observations are that the bare electromagnetic correction is very small, below $1\times 10^{-11}$, and the model estimate for $\atotvio$ is consistent with the lattice result. The most important quantity, $R_{38K}$, is somewhat underestimated in magnitude by the model, but nevertheless consistent within the uncertainties. The elastic part, of size 1.53\,MeV when evaluated  at $\Lambda=16\, m_\mu$, appears to saturate the electromagnetic kaon mass splitting computed on the lattice with the same value of $\Lambda$.

\begin{table}[h]
    \centering
    \caption{Hadronic-model estimate for a PV-mass of $\Lambda=16 \, m_\mu$ in the first line, and for infinite PV-mass in the second line. All values except $\dmKel$ are multiplied by $10^{11}$.}
    \label{tab::Total_Pheno_Prediction}
    \begin{tabular}{cccc}
    \hline
        $\atotvioemlow \phantom{\Big|}$ &  $\dmKel\phantom{\Big|}$[MeV] & $R_{38K} \phantom{\Big|}$[1/MeV] & $\atotvio\phantom{\Big|}$ \\ \hline
          $-0.25(0.4)$& $1.53$ & $-1.53(38)$ &  $8.1(2.1)_{R_{38K}}(0.4)_{\rm em}(0.4)_{\rm sd}$ \\         
          $0.01(0.6)$ & $2.4$ & $-1.53(38)$ &   $9.7 (2.4)_{R_{38K}} (0.6)_{\rm em}(0.6)_{\rm sd}$\\ \hline
    \end{tabular}
\end{table}

 \section{Conclusion\label{sec::Concl}}

In this paper, we have performed a lattice QCD calculation of the isospin-violating part of the hadronic vacuum polarization (HVP), which corresponds to the cross-terms in the two-point correlation function of the electromagnetic current $j_\mu^{em}$, when the latter is written as the sum of an isovector ($j_\mu^3$) and an isoscalar ($j_\mu^8$) current.
It would vanish for equal up and down quark masses in the absence of electromagnetic effects.
Our result for the isospin-violating HVP contribution to the anomalous magnetic moment of the muon, at the SU$(3)_{\rm f}$ symmetric point, reads:
\begin{equation}\label{eq:finalresult}
     2\,\atotvio = 21.8 (2.8)_{\rm stat}(1.4)_{\rm syst}\times 10^{-11}\qquad
    (M_\pi=M_K\simeq 416\,{\rm MeV}),
\end{equation}
where the factor of two on the left-hand side accounts for the equally sized (3,8) and (8,3) contributions.

In order to deal with the long-range electromagnetic effects on the lattice, we have used a Pauli-Villars (PV) regulated photon propagator with the cutoff scale $\Lambda$ set above the typical hadronic scale of 1 GeV. Our results, however, exhibit little dependence on the cutoff, compared to the uncertainties. We account for the residual dependence on the PV-mass by performing a linear fit in $1/\Lambda^2$ on the results of the two largest PV-masses to get our result for $\Lambda\to\infty$.  
 
The PV regularization enables us to do crosschecks of our intermediate results. For the connected diagrams, we were able to obtain consistent results between lattice calculations on `gluonless' ensembles and continuum QED calculations. 
For the charged/neutral kaon mass splitting, which was part of the (mass) counterterm, our results are consistent with the expected logarithmic dependence on $\Lambda$. This expectation resulted from an Operator Product Expansion.
In addition, we were able to compare the bare electromagnetic correction $\atotvioem(\Lambda)$ to continuum model predictions based on the kaon loop and pseudoscalar meson exchanges. 

In general, an advantage of decoupling the photon cutoff $\Lambda$ from the lattice QCD cutoff $1/a$ is that intermediate results such as $\atotvioem(\Lambda)$ and $\dmKem(\Lambda)$ no longer depend on the lattice action employed and can be reused, or cross-checked, by other lattice collaborations. By contrast, when $\Lambda$ is identified with $1/a$ by using a lattice photon propagator, as has been the case in most previous lattice calculations of QED effects, the quantities $\atotvioem(1/a)$ and $\dmKem(1/a)$  depend on the details of the lattice QCD action used.

We have also pointed out the importance of precisely determining the quantity $R_{38K}$ defined in Eq.\ (\ref{equ::HVP_MI_R38}), a scheme-independent and renormalization group invariant quantity.
Based on the results presented here, an extension of our lattice calculations towards physical pion and kaon masses seems promising. In this case, the additional disconnected diagrams will become relevant -- see Fig. \ref{fig::Master_feynman}. Still, we expect the elastic contribution to the counterterm to remain the numerically dominant part.

We have complemented our lattice calculations by hadronic-model estimates of $R_{38K}$ and of the low-energy part $\atotvioemlow$ of the vacuum polarization -- see section \ref{sec:compa_lat_model}, and Appendix \ref{app::Pheno} for details. We have found agreement within the uncertainty of $25\%$ of the model with our lattice QCD results. This emboldens us to attempt an estimate at physical pion and kaon masses. Our estimate of $R_{38K}$ is consistent with previous phenomenological results \cite{James:2021sor,Colangelo:2022prz}, and we arrive at an estimate of 
\begin{align}
2\,\atotvio = 32(8)\times10^{-11} \qquad \textrm{($(M_\pi,M_K)$ physical; hadronic model)}
  \end{align}
  for the total isospin-violating part of the hadronic vacuum polarization in the muon $(g-2)$. Clearly it is a relevant effect worth bringing under control, given the current precision ($22\times10^{-11}$) of the world average for the direct measurement of $\amu$ \cite{Aguillard_2023,Muong-2:2006rrc,Muong-2:2021ojo}.

 \acknowledgments We thank Simon Kuberski for providing the data for the calculation of the HVP light-quark mass derivative. We also thank Franziska Hagelstein for an ongoing collaboration on computing electromagnetic corrections to hadronic vacuum polarization. We acknowledge the support of Deutsche Forschungsgemeinschaft (DFG) through the research unit FOR 5327 “Photon-photon interactions in the Standard Model and beyond exploiting the discovery potential from MESA to the LHC” (grant 458854507), and through the Cluster of Excellence “Precision Physics, Fundamental Interactions and Structure of Matter” (PRISMA+ EXC 2118/1) funded within the German Excellence Strategy (project ID 39083149). Calculations for this project were partly performed on the HPC cluster and “HIMster II” at the Helmholtz-Institut Mainz and “Mogon II” at JGU Mainz. We are grateful to our colleagues in the CLS initiative for sharing ensembles.\\
 \appendix
 \appendix
\section{Derivation of the expression for the counterterm to $\atotvio$ \label{sec:CTderivation}}

The expression for the kaon mass splitting $\dmK \equiv M_{K^+}-M_{K^0}$, expanded  to leading order in $\alpha$ and $(m_u-m_d)$ around an isosymmetric point in the parameter space of QCD, reads
\begin{equation} \label{eq:dmK_master}
 \dmKphys =   \dmKem(\Lambda) + (m_u-m_d)(\Lambda) \frac{\partial\; \dmK}{\partial(m_u-m_d)}.
\end{equation}
The partial derivative is taken at fixed $(m_u+m_d)$ and $m_s$, as well as at fixed bare (QCD) coupling $g_0$ and at $\alpha=0$: the right-hand side is an expansion in the bare parameters of QCD+QED.
Since the splitting is isospin breaking, only one counterterm, associated with the quark-mass difference $(m_u-m_d)$, appears.
We thus interpret Eq.\ (\ref{eq:dmK_master}) as a natural renormalization condition that determines the bare parameter $(m_u-m_d)(\Lambda)$.

In the past, the cutoff $\Lambda$ on the photon propagator has been identified with the lattice cutoff $1/a$ by computing this propagator on the lattice. In our approach, the scale $\Lambda$, which appears in the Pauli-Villars regularization terms of the photon propagator (see Eq.\ (\ref{eq:pv_photon_prop})), is decoupled from the lattice spacing. Thus, at fixed $\Lambda$ each term in Eq.\ (\ref{eq:dmK_master}) has a continuum limit.

The analogous expansion for $\atotvio$ is 
\begin{align}
    \atotvio
    = \atotvioem(\Lambda) + (m_u-m_d)(\Lambda) \frac{\partial \alovio }{\partial(m_u-m_d)}.
\end{align}
Eliminating the bare parameter $(m_u-m_d)(\Lambda) $ using Eq.\ (\ref{eq:dmK_master}), we arrive at 
\begin{align}
    \atotvio
    \label{equ::Master_4pt_apdx}
    = \atotvioem(\Lambda) + \frac{\dmKphys - \dmKem(\Lambda)}{\frac{\partial\; \dmK}{\partial(m_u-m_d)}} \frac{\partial \alovio }{\partial(m_u-m_d)}.
\end{align}
Finally, realizing that the Hamiltonian of QCD can be written
\begin{align}
    H = \int d^3x\; \left(  {\textstyle\frac{1}{2}}(m_u-m_d)(\bar u u - \bar d d) + \textrm{isosymmetric terms} \right),
\end{align}
we can trade the partial derivative for a scalar matrix element,
\begin{align} \label{eq:dmKdm}
    \frac{\partial\; \dmK}{\partial(m_u-m_d)} =
    {\textstyle\frac{1}{2}}\left( \langle K^+|\bar u u - \bar d d | K^+\rangle - 
    \langle K^0|\bar u u - \bar d d | K^0\rangle \right)
    = \langle K^+|\bar u u - \bar d d | K^+\rangle,
\end{align}
where in the last step we have exploited the isospin symmetry to simplify the expression.
The kaon states are at rest ($\vec{p}=0$) and their normalization is the non-covariant one, 
$\langle K_{\vec{p}}| K_{\vec{p}'} \rangle = (2\pi)^3 \delta^{(3)}(\vec{p}-\vec{p}')$.
Inserting the simplified expression (\ref{eq:dmKdm}) into Eq.\ (\ref{equ::Master_4pt_apdx}), one arrives at Eq.\ (\ref{equ::Master_amu38}) with the counterterm given by Eq.\ (\ref{equ::Master_Counterterm}).
 \section{QED continuum prediction for the connected contribution \label{app::QED_Cont}}

A formalism for obtaining the two-loop vacuum polarization in QED via the Cottingham formula has been explained in details in \cite{Biloshytskyi:2022ets}.
Here we sketch only the main steps of the vacuum polarization calculation pertinent to a double Pauli-Villars photon regularization and provide its contribution to the muon $(g-2)$.  

Using the Cottingham formula, the four-point (two-loop) vacuum polarization can be written in terms of the forward doubly-virtual LbL scattering amplitude $\mathcal{M}(\nu,K^2,Q^2)$ as
\begin{align}
	\Pi_{{\rm 4pt}}(Q^2,\Lambda) &=  \frac{1}{3 (2\pi)^3 Q^2} \int\limits_0^{\infty} d K^2
 \,K^2\,\left[\frac{1}{K^2}\right]_{\Lambda}
	\,\int\limits_{0}^{1} d x \, \sqrt{1-x^2}\,
	\mathcal{M}( K Q x ,\,K^2,\,Q^2),
\label{eq:CottinghamFormulaFinal}
\end{align}
where $K^2=-k^2$ and $Q^2=-q^2$ are virtualities of the scattered photons with momenta $k$ and $q$, correspondingly, and $\nu = k\cdot q= K Q x$.
Substituting the dispersive representation in the variable $\nu$ of the forward LbL amplitude with one subtraction at $\nu = \bar\nu$, 
\begin{equation}
\mathcal{M}(\nu ,K^2,\,Q^2) = \mathcal{M}(\bar\nu ,K^2,\,Q^2)+\frac{2}{\pi} (\nu^2-\bar\nu^2)\int\limits_{\nu_\mathrm{thr.}}^\infty d \nu'\,
\frac{ \nu' \im \mathcal{M}(\nu', K^2,Q^2)}{(\nu^{\prime\,2}-\bar\nu^2)(\nu^{\prime\,2}-\nu^2)},
\label{eq:Mdisprep}
\end{equation}
and making use of the optical theorem 
\begin{equation}
\im M(\nu,K^2,Q^2) = 2\sqrt{X}\sigma(\nu,K^2,Q^2),
\end{equation}
where $X = \nu^2-Q^2K^2$ and $\sigma = 4\sigma_{TT}-2\sigma_{LT}-2\sigma_{TL}+\sigma_{LL}$ is an unpolarized $\gamma^*\gamma^*$-fusion cross section,
 we arrive at the dispersive representation of the Cottingham formula:
\begin{align}
\Pi_{{\rm 4pt}}(Q^2,\Lambda)
=& \frac{1}{3(2\pi)^3 Q^2}\int\limits_0^\infty d K^2 \,
K^2\,\left[\frac{1}{K^2}\right]_{\Lambda}\,
\Bigg[ \frac{\pi}{4}
\mathcal{M}(\bar\nu ,K^2,\,Q^2)\nn
&+ \int\limits_{\nu_\mathrm{thr.}}^\infty d \nu  \left(\frac{2}{\nu+\sqrt{X}} - \frac{\nu}{\nu^{ 2} -\bar{\nu}^2 }\right) \sqrt{X}\,  \sigma(\nu,K^2,\,Q^2) 
\Bigg].
\label{eq:DRform}
\end{align}
The double Pauli-Villars regularization of the photon propagator \eqref{eq:pv_photon_prop}, which is employed on the lattice, corresponds to the following regulator in continuum calculation:
\begin{equation}
    \left[\frac{1}{K^2}\right]_\Lambda = \frac{\zeta^2\Lambda^4}{K^2(K^2+\Lambda^2)(K^2+\zeta^2\Lambda^2)}, \quad \zeta = \frac{1}{\sqrt{2}}.
\end{equation}
The contribution to the muon $(g-2)$ that stems from the Cottingham formula reads
\begin{equation}
    a^\ast_\mu = -\frac{\alpha_\mathrm{em}}{2\pi}\Pi_\mathrm{4pt}(0,\Lambda)+\frac{\alpha_\mathrm{em}}{\pi}\int_0^\infty dQ^2\mathcal{K}(Q^2)\Pi_\mathrm{4pt}(Q^2,\Lambda),
    \label{eq:AMMCottingham}
\end{equation}
with the QED kernel $\mathcal{K}$ given by
\begin{equation}
\mathcal{K}(Q^2) = \frac{1}{2m_\mu^2}\frac{(v-1)^3}{2v(v+1)},\quad v=\sqrt{1+\frac{4m_\mu^2}{Q^2}}.
\end{equation}
A numerical evaluation of Eq.~\eqref{eq:AMMCottingham}, with the regulator set to $\Lambda = 3\, m_\mu$ and the mass of the lepton inducing vacuum polarization set to $m_\ell=m_\mu$, yields
\begin{equation}
    a^\ast_\mu\approx 7.499\times 10^{-11}.
\end{equation}

\section{Kaon electromagnetic form factor \label{app::Kaon_FF}}
The kaon electromagnetic form factor enters the calculation of the elastic contribution of the electromagnetic kaon-mass splitting in section \ref{sec::Kaon_MS_elastic}. While the form factor is calculable on the lattice directly, we use the vector-meson-dominance (VMD) monopole parametrization (\ref{equ::vmd_papram}) throughout the paper.  In this appendix we argue that the VMD approximation is sufficient for our purposes by comparing the results of the elastic contribution on the H101 ensemble for both methods of obtaining the electromagnetic kaon form factor. 

The functional form of the VMD form factor is given by Eq.\ (\ref{equ::vmd_papram}).\footnote{One way to evaluate the expression in finite volume is to use the Gaussian representation $G_m(x) = \frac{1}{16\pi^2} \int_0^\infty \frac{dt}{t^2} \,e^{-tm^2-x^2/(4t)}$ of the scalar propagator, which allows one to obtain the integrals over the spatial components of $x$ over the interval $[-L/2,L/2]$ in terms of the error function.}
For the H101 ensemble, following \cite{Chao:2020kwq}, a VMD mass of 921 MeV was used; this value comes from a VMD fit to the $\pi^0$ transition form factor on the same ensemble \cite{Gerardin:2019vio}.

In order to obtain the form factor from the lattice, we follow the procedure from \cite{koponen2022isovectoraxialformfactor}, which is a very similar approach to the summation method in section \ref{sec::Kaon_MI}, but instead of a unit matrix $\mathbb{1}$ the gamma-matrices $\gamma_j$ get inserted while calculating the three-point function. Additionally the kaon has to be projected to non-zero momenta. At the SU$(3)_{\rm f}$ symmetric point the propagator of the light quarks and the strange quark are the same, which means the two- and three-point functions for this calculation are 
\begin{align}
    C_2(\vec{p}, x_0)&=-a^3\sum_{\vec{x}} e^{-i\vec{p}\vec{x}} \ \text{Tr}[S(0,x)\gamma_5 S(x,0)\gamma_5]\,,\\
    C^\nu_3(\vec{p}, x_0, y_0)&=-a^6\sum_{\vec{x}, \vec{y}} e^{-i\vec{p}\vec{y}} \  \text{Tr}[S(0,y)\gamma_\nu S(y,x)\gamma_5 S(x,0)\gamma_5]\,.
\end{align}
From these the lattice form factor $F_{\text{lat}}(-P^2)$ can be obtained in the following way:
\begin{align}
\Bar{C}_2(\bar{p},x_0)&=\frac{\sum_{\vec{p}:|\vec{p}|=\bar{p}}C_2(\vec{p}, x_0)}{\sum_{\vec{p}:|\vec{p}|=\bar{p}}1}
\\
    R(\vec{p}, x_0, y_0)&= \frac{C^0_3(\vec{p}, x_0, y_0)}{\Bar{C}_2(0,x_0)}\sqrt{\frac{\Bar{C}_2(|\vec{p}|, x_0-y_0) \Bar{C}_2(0,y_0)\Bar{C}_2(0,x_0)}{\Bar{C}_2(0, x_0-y_0) \Bar{C}_2(|\vec{p}|,y_0)\Bar{C}_2(|\vec{p}|,x_0)}}
    \\
    F^{\text{eff}}(\bar{p}, x_0,y_0)&=\frac{2\sqrt{M_K E_p}}{M_K+E_p}\ \frac{\sum_{\vec{p}:|\vec{p}|=\bar{p}}R(\vec{p}, x_0, y_0)}{\sum_{\vec{p}:|\vec{p}|=\bar{p}}1}
    \\
    \Rightarrow a\sum_{y_0=a}^{x_0-a} F^{\text{eff}}(\bar{p}, x_0,y_0)& \overset{x_0\rightarrow \infty}{=} b(\bar{p})+ x_0\, F_{\text{lat}}(-P^2) + \dots
\end{align}
with $P^2 = 2M_K(\sqrt{\bar p^{\,2}+M_K^2}-M_K)$ in our choice of kinematics.
Following the last step, a linear fit in $x_0$ to the resulting data was performed in order to extract the lattice form factor.

We will use the integrand of the elastic contribution in finite volume in order to compare both calculations. The integrand $f_{elast}(x_0)$ is computed based on Eq.\ (\ref{equ::elast_fin_2}).
Note that the maximal value of $\vec{p}^2$ amounts to about $2.4 \text{ GeV}^2$ in our lattice calculation.
For a like-by-like comparison, two versions of the VMD calculation were performed, one for these restricted values of $\vec{p}^2$ and one without this restriction. 

\begin{figure}[t]
    \centering
    \begin{subfigure}[t]{0.49\textwidth}
        \includegraphics[width=\textwidth]{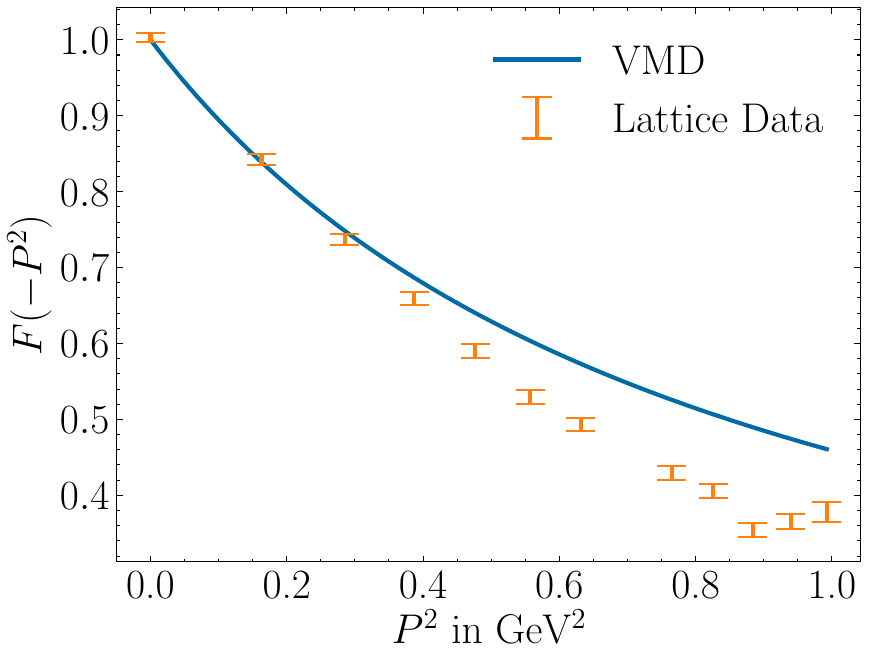}
        \caption{VMD vs.\ lattice form factor.}
        \label{fig::Kaon_FF_Plots_FF}
    \end{subfigure}
    \begin{subfigure}[t]{0.49\textwidth}
        \includegraphics[width=\textwidth]{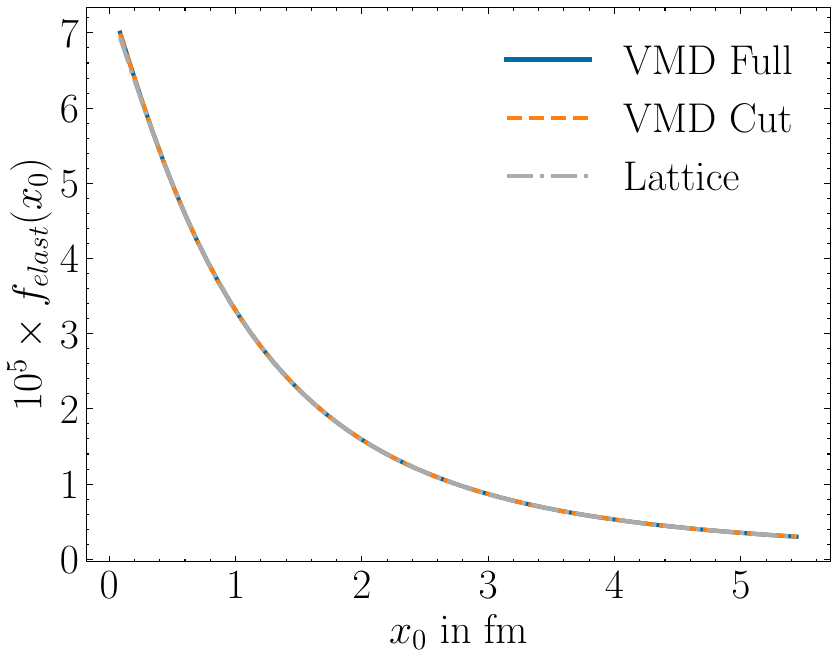}
        \caption{$f_{elast}$ calculated with different form factors.}
        \label{fig::Kaon_FF_Plots_Elast}
    \end{subfigure}

\caption{Comparison of the kaon form factor as obtained on ensemble H101 to the VMD form factor with $M_{\rm vmd}=921\,$MeV. On the right hand side, $f_{elast}(x_0)$ is calculated in the finite volume of ensemble H101 using the discrete momentum values $\vec p^2= (2\pi/L)^2\vec n^2$. For the lattice and the 
``VMD Cut'' curves, the momentum sum is truncated at $\vec p^2=2.4\,{\rm GeV}^2$, while for ``VMD Full'' it extends to significantly higher values.}
\label{fig::Kaon_FF_Plots}
\end{figure}

The comparison between the VMD form factor and lattice data can be seen in Fig.~\ref{fig::Kaon_FF_Plots_FF}. For the first three values, the lattice and the VMD form factor are in perfect agreement. From the fourth point onward, the lattice calculation is decreasing more rapidly. However, as one can see on the right-hand-side in Fig.~\ref{fig::Kaon_FF_Plots_Elast}, this discrepancy results in a completely negligible difference for $f_{elast}(x_0)$ at the Euclidean times of interest. Moreover, there is practically no difference between the limited and full VMD versions. 

As the result, we have decided to forgo the systematic lattice calculation of the kaon form factor and use instead the VMD approximation for the calculations in section \ref{sec::Kaon_MS}.

 \section{Phenomenological estimate of $\atotvio$ \label{app::Pheno}}

In this appendix, we derive various predictions for $\atotvio(\Lambda)$, $\atotvioemlow$ and $R_{38K}$ based on phenomenological models. We refer the reader to Eqs.\ (\ref{equ::Master_amu38}--\ref{equ::HVP_MI_R38}) and  sec.\ \ref{sec::SIB} for the definition of these quantities. After comparing the predictions to our lattice results at the SU(3)$_{\rm f}$ point, we also provide estimates at physical quark masses.

\subsection{The bare electromagnetic contribution at fixed photon cutoff $\Lambda$}

We begin by estimating the bare e.m.\ contribution $\atotvio(\Lambda)$ for a fixed value $\Lambda=16\,m_\mu\simeq 1.69\,$GeV. We take into account the contributions of the pseudoscalar (PS) meson poles $(\pi^0,\eta,\eta')$, as well as of the charged kaon loop. We compute the PS-pole contributions using the master formula given in Ref.\ \cite{Biloshytskyi:2022ets}, which involves the meson's transition form factor for spacelike photon virtualities. We use the vector-meson dominance (VMD) parametrization of the form factor, with the parameters chosen as in Ref.\ \cite{Parrino:2025afq}.
As compared to the full $\ahvp$, we give the pion contribution a factor $1/4$
in $\atotvio$. For the $(\eta,\eta')$ contributions, we proceed as follows\footnote{For the pion, the factor applied follows from the approximation that the TFF with respect to one isovector and one isoscalar current is approximated by ${\cal F}_{\pi^0 38}(q_1^2,q_2^2)\simeq (1/2){\cal F}_{\pi^0 \gamma^*\gamma^*}(q_1^2,q_2^2)$. For the isoscalar PS mesons, we have ${\cal F}_{P\gamma^*\gamma^*}={\cal F}_{P33}+{\cal F}_{P88} $; at the SU(3)$_{\rm f}$ point, this implies ${\cal F}_{P88}=(1/3){\cal F}_{P33}$ for the $\eta'$, and ${\cal F}_{P88}=(-1/3){\cal F}_{P33}$ for the $\eta$.}: at the SU(3)$_{\rm f}$ symmetric point, the flavour symmetry dictates that the $\eta'$ receive a factor $3/16$ and the $\eta$ a factor $(-3/4)$, which leads to a complete cancellation between the $\eta$ and the $\pi^0$ exchanges. At the physical point, we treat the $\eta$ and $\eta'$ as linear superpositions of the octet and singlet states with a mixing angle as in Ref.\ \cite{Parrino:2025afq}. 
As for the kaon loop, we evaluate it in the scalar QED approximation using the expressions given in appendices A and B of Ref.\ \cite{Parrino:2025afq}, applying to it a factor $1/4$ relative to its contribution in $\ahvp$, due to the charge of $K^\pm$ with respect to the currents $j_\lambda^3$ and $j_\lambda^8$. The results are collected in Table~\ref{tab:pheno_bareQED}.

\begin{table}[t]
    \centering
    \caption{Contributions to $\atotvioem(\Lambda=16\, m_\mu)$ at physical quark masses and at the SU(3)$_f$ symmetric point $M_\pi=M_K\simeq 416$\,MeV. All contributions in $10^{-11}$ units.}
    \label{tab:pheno_bareQED}
    \begin{tabular}{c c@{~~~} c}
    \hline
      Contribution: $\phantom{\Big|}$  & $\atotvioem(\Lambda=16\, m_\mu)|_{\rm physpt}$& $\atotvioem(\Lambda=16\, m_\mu)|_{\rm SU(3)_{f}}$ \\
         \hline
      charged kaon loop (sQED)  & $-0.16$ & $-0.41$\\
      $\pi^0$ exchange &  $0.57$  &  $0.21$\\
        $\eta$ exchange & $-0.10$ & $-0.21$\\
          $\eta'$ exchange & $0.14$ &  $0.16$\\
          \hline
          Total &  $0.45$  & $-0.25$ \\
          \hline
    \end{tabular}
\end{table}

At the SU(3)$_{\rm f}$ symmetric point, the charged kaon loop gives the single largest contribution in absolute terms, imposing its negative sign on the total $\atotvio(\Lambda)$.
At physical pion and kaon masses, we find that it is the $\pi^0$ exchange that dominates, resulting in a positive estimate for $\atotvio(\Lambda)$. Still, this estimate is very small, with $2\,\atotvio(\Lambda)$ just below the $1\times10^{-11}$ level.

Since in our lattice QCD calculation of $\atotviotpta$, the photon cutoff $\Lambda$ is directly set to infinity, we cannot directly compare the model predictions of Table \ref{tab:pheno_bareQED} to our lattice results.
We therefore also use our model to estimate $\atotviotpta(\Lambda=16\, m_\mu)$: subtracting this contribution from our estimate of $\atotvioem(\Lambda=16\, m_\mu)$, we obtain a prediction for the connected contribution alone.
In Table~\ref{tab:pheno_2p2}, we provide the predictions\footnote{In $\atotviotpta(\Lambda)$, the combined $\pi_0$ and $\eta$ exchange comes with factor $-3/8$ and the $\eta'$ exchange with factor $3/16$, relative to the size of these contributions in the full $a_{\mu,{\rm em}}^{\rm HVP}(\Lambda)$. The combined pion and kaon loop contribution comes with weight $1/12$ of its size in $a_{\mu,{\rm em}}^{\rm HVP}(\Lambda)$, to which the pion and the kaon loop make equal contributions.} for  $\atotviotpta(\Lambda=16\, m_\mu)$. For the connected contribution, we then obtain the estimate 
\begin{align}
 \atotvioconn(\Lambda) = \atotvioem(\Lambda)-\atotviotpta(\Lambda) \simeq +0.28(21) \times 10^{-11},
 \qquad \Lambda=16\, m_\mu,
\end{align}
to which we attach an uncertainty of the typical size of an individual contribution such as the $\pi^0$ exchange.
This value is to be compared to the continuum-extrapolated lattice QCD result provided in Table \ref{tab::Conn_QED_Results}, $\atotvioconn(\Lambda)=-0.25(32)\times 10^{-11}$.
While this represents a difference at the level of $1.5\sigma$, both the lattice result and the model estimate are very small and essentially compatible with zero. 

For a hint as to where the slight tension comes from, we give the model prediction for $\atotviotpta(\Lambda=\infty)$ in the rightmost column of Table~\ref{tab:pheno_2p2}, which can be compared to the corresponding lattice result, Eq.\ (\ref{eq:tpta_Continuum}). As already visible in figure~\ref{fig:tpta_N300}, the sum of the pseudoscalar exchanges already approximates the integrand quite well. This hints at the magnitude of the pseudoscalar loop contribution to $\atotviotpta(\infty)$ being overestimated by its evaluation using its matching factor of $1/6$ and a VMD form factor, or else at relevant positive contributions missing from our model. Similarly, the evaluation of the pseudoscalar loop contribution to $\atotviotpta(\Lambda=16\, m_\mu)$ using scalar QED likely yields an overestimate of its absolute value.

\begin{table}[t]
    \centering
    \caption{Contributions to the (2+2)a diagrams $\atotviotpta(\Lambda)$ for $\Lambda=16\, m_\mu$ and $\Lambda=\infty$ at the SU(3)$_{\rm f}$ symmetric point $M_\pi=M_K\simeq 416$\,MeV. All contributions in $10^{-11}$ units.}
    \label{tab:pheno_2p2}
    \begin{tabular}{c  c   c}
    \hline
      Contribution at ${\rm SU(3)_{f}}$ point:   &  $\atotviotpta(\Lambda=16\, m_\mu)\phantom{\Big|}$  &  $\atotviotpta(\Lambda=\infty)$\\
         \hline
      charged pion \& kaon loop    & $-0.27$ (sQED) & $-0.37$ (VMD)\\
        $\pi^0$ and $\eta$ exchange & $-0.42$ & $-0.96$\\
          $\eta'$ exchange &  $+0.16$ &  $+0.58 $\\
          \hline
          Total  & $-0.53$ &  $-0.75$ \\
          \hline
    \end{tabular}
\end{table}

\subsection{Low-energy contribution $\atotvioemlow$ }

The pseudoscalar meson exchanges, as evaluated in the previous subsection with a transition form factor, remain UV-finite when $\Lambda\to\infty$. Upon introducing a VMD form factor in the evaluation of the kaon loop, the latter contribution also remains finite. Together, the PS meson exchanges and the kaon loop form the basis of our estimate of the low-energy contributions to $\atotvioemlow$, which is evaluated without a UV-cutoff on the photon propagator.
The results at the SU(3)$_{\rm f}$ and at the physical point are collected in Table~\ref{tab:pheno_renormd}.

At the SU(3)$_{\rm f}$ symmetric point, the kaon loop tends to cancel very strongly against the $\eta'$ exchange, while the $\pi^0$ and $\eta$ exchanges cancel each other exactly. In total, $\atotvioemlow$ is practically zero.
At the physical point, as noted in the previous subsection, the $\pi^0$ exchange dominates, but nevertheless remains a small contribution in comparison to the strong isospin-breaking contribution analyzed in the two coming subsections.

\subsection{Elastic part of the kaon mass splitting \label{app::Pheno_Kaon_MS}}

First of all, we use $\dmKphys=-3.9\,$MeV in all cases. 
Secondly, we defined $\dmKel$ as resulting from the elastic contribution to the forward Compton amplitude on the kaon.
At the physical point, we borrow the (finite) estimate $\dmKel =+2.1\,$MeV based on the elastic contribution from~\cite{Stamen:2022uqh}. 
At the SU(3)$_{\rm f}$ point, the same master formula with a VMD form factor and a VMD mass of 860\,MeV yields $\dmKel=+2.4\,$MeV.

For comparison with the lattice, we also compute $\dmKel$ with a regulated photon propagator ($\Lambda=16\, m_\mu$). The value of 1.53\,MeV is reported in Table \ref{tab::Total_Pheno_Prediction} and appears to saturate the lattice result for $\dmKem(\Lambda)$, suggesting that the inelastic contribution is still very small for this value of $\Lambda$.

The most important quantity to complete the prediction for the renormalized $\atotvio$ is then $R_{38K}$.

\subsection{Estimate of $R_{38K}$ and $\atotvio$ at the SU(3)$_{f}$ point}
In order to estimate $R_{38K}$ at the SU(3)$_{f}$ point, we note that the quark-mass derivative of $\atotvio$  consists only of the three-point connected diagram, which can also be interpreted as a similar derivative within isosymmetric QCD:
\begin{align}
    \frac{\partial \atotvio}{\partial(m_u-m_d)}
    = -\frac{3}{2}  \frac{\partial \atotisoS}{\partial(m_l-m_s)}\bigg|_{2m_l+m_s}
    = \frac{\partial\atotisoS}{\partial m_s}\bigg|_{m_l} -\frac{1}{2} \frac{\partial\atotisoS}{\partial m_l}\bigg|_{m_s}\;.
\end{align}
In the second equality, we have only made use of the chain rule to change variables. At this point, we make the approximation that only the valence quarks of the relevant vector meson matter for estimating the quark-mass derivatives. Thus we arrive at 

\begin{align}
     \frac{\partial \atotvio}{\partial(m_u-m_d)}
     \simeq \frac{3}{4} \, \frac{\partial \astrange}{\partial m_s}\bigg|_{m_l}.
\end{align}
Note that the strangeness contribution $\astrange$ is defined to contain the charge factor of $(-1/3)^2$.
We can now relate the expression to a mass derivative already modeled in~\cite{Ce:2022kxy}, Appendix B.2,
\begin{align}
  R_{38K} &\simeq \frac{3}{2} M_K\, \,\frac{\partial \astrange}{\partial M_K^2}\bigg|_{M_\pi^2}.
\end{align}
In particular, we estimate the kaon mass dependence of the $\bar s s$ vector meson mass from
\begin{align}
    M^V_{ss} = M_\phi^{\rm phys} + 
    \left[\frac{M_\phi-M_\omega}{M_K^2-M_\pi^2}\right]_{\rm phys}
    \left(M_K^2-{\textstyle\frac{1}{2}}M_\pi^2 - (M_K^2-{\textstyle\frac{1}{2}}M_\pi^2 )_{\rm phys}\right).
\end{align}
This expression returns the physical $\phi$ meson mass for physical $(M_K,M_\pi)$ values, and the physical $\omega$ mass for $M_K=M_\pi=(M_\pi)_{\rm phys}$.
It delivers a successful prediction for the vector meson mass at the SU(3)$_{\rm f}$ point as compared to a direct lattice calculation~\cite{Chao:2020kwq}, in which the mass splitting between the octet and the singlet states is found to be small.
Secondly, we neglect the dependence of the electronic width of vector mesons on the valence-quark mass~\cite{Sakurai:1978xb,Ce:2022kxy}. 
As in~\cite{Ce:2022kxy}, we do however take into account the dependence of the perturbative threshold on the valence quark mass.
Setting $M_\phi= 0.860$\,GeV and $(9\pi/\alpha^2)\Gamma_{ee}(\phi)= 0.65\,$GeV at our SU(3)$_{\rm f}$ point, we obtain
\begin{align}
 R_{38K} = -1.53(38)\times 10^{-11}\;{\rm MeV}^{-1}.
\end{align}
We have assigned a conservative uncertainty of 25\% to the model (as compared to 15\% in~\cite{Ce:2022kxy}), since we are applying it a distance away from the physical point, on which its parameter values are based.
With this estimate of $R_{38K}$, we obtain 
\begin{align}
   \atotvio = 9.7(2.4)_{R_{38K}}(0.6)_{\rm em}(0.6)_{\rm sd}\times 10^{-11}.
\end{align}
Our estimates of the bare e.m.\ contribution to $\atotvio$ cancel almost completely, so that the final result arises solely from the counterterm.
The uncertainties come, as indicated in the equation, from the estimate of $R_{38K}$; from $\atotvioem$, to which we assign the size of the kaon loop  as uncertainty to account for missing effects such as scalar, axial-vector and tensor meson exchanges; and from the short-distance contributions in the bare e.m.\ contribution, which must cancel against the short-distance contribution in the inelastic part of the e.m.\ kaon mass splitting. We have assigned an uncertainty to these short-distance parts given again by the size of the kaon loop, which has been computed with form factors.

\subsection{Predicting $R_{38K}$ and $\atotvio$ at the physical point}
The quantity $R_{38K}$ is more subtle to estimate at the physical point, where one cannot make use of the SU(3)$_f$ symmetry.
In that regime, the disconnected mass insertion plays an important role, canceling the dominant part of the connected three-point function~\cite{Lehner:2020crt}.
In the derivative ${\partial \atotvio}/{\partial(m_u-m_d)}$ at constant $(m_u+m_d)$ and $m_s$, due to the narrowness of the $\omega$ resonance, we expect a strong enhancement at $s\simeq M_\omega^2$. Since the mass operator $\bar u u - \bar d d$ carries no energy or momentum, of all the isovector final states the $\rho$ resonance contribution is enhanced, since it has practically the same invariant mass as the $\omega$ resonance. Now, on resonance we expect the large-$N_c$ behavior that the properties of the resonances depend dominantly on their valence quarks. In that regime, we neglect the disconnected diagrams 
and write for the contributions from centre-of-mass energies ${s}\gtrsim M_\omega^2$,
\begin{align}
    \frac{\partial \atotvio}{\partial(m_u-m_d)}\bigg|_{m_u+m_d}
    \simeq \frac{3}{2}\; \frac{\partial}{\partial m_l}\bigg|_{m_s} \atotisoS\,,
\end{align}
where on the right-hand side the contribution of the strangeness current is neglected for the same large-$N_c$ reasons. 
We then estimate by the same method as for the SU(3)$_{\rm f}$ point above 
\begin{align}
    m_l \; \frac{\partial\atotisoS}{\partial m_l}\bigg|_{m_s} \simeq (-1.35-0.19)\times 10^{-10}= - 1.54\times 10^{-10}.
\end{align}
In particular, the first term follows from the approximate relation
\begin{align}
    M^V_{ll} = M_\omega^{\rm phys} 
    + \frac{1}{2}\left[\frac{M_\phi-M_\omega}{M_K^2-M_\pi^2}\right]_{\rm phys} (M_\pi^2 - (M_\pi^{\rm phys})^2)
\end{align}
for the pion-mass dependence of the $\omega$ meson,
while the second term corresponds to the quark-mass dependence of the threshold to a quasi-perturbative $\bar q q$ continuum.
In this way, we arrive at 
\begin{align}
    R_{38K} \simeq -2.50(63)\times 10^{-11}\;{\rm MeV}^{-1}.
\end{align}
The quoted uncertainty of 25\% comes from combining quadratically a 15\% uncertainty for the estimate of the connected diagram and a 20\% uncertainty for the non-accounted for contribution of the $\omega$ transition to the $\pi\pi$ continuum at energies below $M_\rho$. This continuum, whose relevance has already been pointed out in Ref.\ \cite{James:2021sor}, also presents a certain challenge for lattice QCD, since it leads to a long-range contribution and presumably sizeable finite-volume effects in the calculation of $R_{38K}$.

The resulting counterterm $C_T$ in the elastic approximation and the final model estimate of $\atotvio$ are given in Table \ref{tab:pheno_bareQED}. Since the counterterm dominates, the uncertainty on $\atotvio$ is again $25\%$. The total isospin-violating part of $\ahvp$ is thus estimated at 
\begin{align}
   2\, \atotvio = 31.6(7.9) \times 10^{-11}.
\end{align}

The strong isospin-breaking contribution to $\ahvp$ in the FLAG24 scheme~\cite{FlavourLatticeAveragingGroupFLAG:2024oxs} amounts to
\begin{align}
\ahvpsib = 2\,\atotviosib =  -6\,{\rm MeV}\times 2\,R_{38K} = 30.0(7.5)\times 10^{-11}.
\end{align}
This estimate is consistent within the errors with the previous result $33.2(8.9)$~\cite{James:2021sor}, based on SU(3) chiral perturbation theory as well as with the phenomenological estimate
$\ahvpsib=14.7(16.7)\times10^{-11}$~\cite{Hoferichter:2023sli}.

\begin{table}[t]
    \centering
    \caption{
    UV-finite contributions to $\atotvioem$ at physical quark masses, where $M_K=494.6$\,MeV, and at the SU(3)$_f$ symmetric point $M_\pi=M_K\simeq 416$\,MeV. All contributions in $10^{-11}$ units.} 
    \label{tab:pheno_renormd}
    \begin{tabular}{c c c @{~~~}c}
    \hline
    &  Contribution:   &  phys.\;point & $\rm SU(3)_{f}$ point \\
         \hline
$\atotvioemlow\phantom{^\big|}$: & $K^+K^-$ (VMD) &  $-0.31$ & $-0.56$ \\
   &   $\pi^0$ exchange &  0.93  & 0.48\\
    &    $\eta$ exchange & $-0.24$ & $-0.48$ \\
     &     $\eta'$ exchange & 0.44  & 0.58\\
     & Total       & $0.81$  & $0.01$ \\
     $\atotviosib  \phantom{\Big|}$: &    &         15.0  &   9.7 \\
          \hline
     $\atotvio\phantom{^\big|}$:   &   &  15.8  & 9.7  \\
          \hline
    \end{tabular}
\end{table}

\newpage
\bibliographystyle{JHEP}
\bibliography{references}
\end{document}